\documentclass[11pt,a4paper]{article}
\pdfoutput=1
\usepackage{jheppub}
\usepackage[dvipsnames]{xcolor}
\usepackage{colortbl}
\usepackage{tabularx}
\usepackage{arydshln}
\usepackage{custom_commands}
\usepackage{bbm} 
\usepackage{subcaption}
\usepackage{soul}
\usepackage{cleveref}

\definecolor{Gray}{gray}{0.9}
\newcolumntype{g}{>{\columncolor{Gray}} Y}
\newcolumntype{h}{>{\columncolor{Gray}} c}

\newcolumntype{Y}{>{\centering\arraybackslash}X}

\title{Holomorphic D-brane embeddings in D-brane backgrounds}
\author[1]{James Ratcliffe,}
\author[2]{Ronnie Rodgers,}
\author[3]{and Sangsoo Ryu}
\affiliation[1]{STAG Research Centre, Physics and Astronomy, University of Southampton,\\
    Highfield, Southampton SO17 1BJ, United Kingdom}
\affiliation[2]{Nordita, Stockholm University and KTH Royal Institute of Technology,
Hannes Alfv\'{e}ns v\"{a}g 12, SE-106 91 Stockholm, Sweden}
\affiliation[3]{Department of Physics, Freie Universitat Berlin, 14195 Berlin, Germany}
\emailAdd{J.Ratcliffe@soton.ac.uk}
\emailAdd{ronnie.rodgers@su.se}
\emailAdd{sangsoo.ryu@fu-berlin.de}
\preprint{NORDITA 2026-008}
\abstract{We describe families of probe D$q$-brane embeddings in the extremal black D$p$-brane backgrounds of type IIA and type IIB supergravity, specified by an arbitrary holomorphic function of a complex coordinate on the worldvolume of the D$q$-branes. These embeddings preserve one-quarter of the supersymmetry of the D$p$-brane background, or sometimes one-half of the supersymmetry when $p = q$. We discuss the holography of two example families of holomorphic probe branes in the near-horizon limit of the D3-brane background. The first is probe D5-branes, dual to defect hypermultiplets with a holomorphic mass, which in the infrared flow to Wilson lines located at the zeros of the mass. The second is probe D3-branes, holographically dual to states in the presence of Gukov--Witten surface defects in the dual $\mathcal{N}=4$ supersymmetric Yang--Mills theory.}

\begin{document}

\setcounter{tocdepth}{2}
\maketitle

\section{Introduction and summary of results}
\label{sec:intro}

A family of embeddings of probe D7-branes in the extremal black D3-brane background of type IIB supergravity has recently been introduced~\cite{holomorphic_branes}, in which a complex coordinate \(y\), formed from the two directions orthogonal to the D7-branes, may be any holomorphic or antiholomorphic function of another complex coordinate \(z\) formed from two directions parallel to both the D7-branes and the D3-branes sourcing the background. These embeddings are similar to brane embeddings in flat space found in ref.~\cite{Gauntlett:1997ss} and, like the flat space embeddings, their energy saturates a Bogomol'nyi--Prasad--Sommerfield (BPS) bound and they preserve a fraction of supersymmetry.
    
The near-horizon limit of the D3-brane background is \(\ads[5] \times \sph[5]\), holographically dual to four-dimensional \(\mathcal{N}=4\) supersymmetric Yang--Mills (SYM) theory~\cite{Maldacena:1997re,Gubser:1998bc,Witten:1998qj}. In this limit, introducing probe D7-branes corresponds to coupling \(\mathcal{N}=4\) SYM to four-dimensional \(\mathcal{N}=2\) hypermultiplets~\cite{Karch:2002sh}. Non-trivial holomorphic \(y\) corresponds to giving these hypermultiplets a complex mass proportional to \(y\), which therefore depends holomorphically on position~\cite{holomorphic_branes}. Holomorphic D7-branes in \(\ads[5]\times \sph[5]\) thus provide an analytically tractable holographic description of a strongly coupled quantum field theory (QFT) with explicitly broken translational symmetry.

There is nothing particular about D3- or D7-branes that implies that the embeddings of ref.~\cite{holomorphic_branes} should be the only examples of such holomorphic embeddings of D-branes in extremal D-brane backgrounds. In this work we will perform the natural generalisation, considering probe D\(q\)-branes embedded in the extremal D\(p\)-brane backgrounds of type IIA and type IIB supergravity, for other values \(p\) and \(q\). We determine the conditions under which the embedding of the D\(q\)-branes may be specified by a holomorphic function in a manner similar to the D7-brane embeddings described above. We will restrict to \(p < 7\), as the supergravity solutions for larger values of \(p\) are more subtle, see for example ref.~\cite{Blumenhagen:2013fgp}, and require separate analysis.

The result that we find is what one might intuitively expect. Starting from an intersection between flat D\(p\)- and D\(q\)-branes in Minkowski space, upon replacing the D\(p\)-branes by their corresponding extremal type II supergravity background, the D\(q\)-brane equations of motion admit embeddings specified by an arbitrary holomorphic or antiholomorphic \(y\) if the original intersection preserves a fraction of supersymmetry. This occurs when the number of Neumann--Dirichlet (ND) directions for strings connecting the D\(p\)- and D\(q\)-branes in the original intersection, which we denote \(d\), is a multiple of four~\cite{Polchinski:1998rr,Skenderis:2002vf}. ND directions are the directions spanned by the D\(p\)- but not the D\(q\)-branes, or by the D\(q\)- but not the D\(p\)-branes. We will show that when \(d\) is a multiple of four, holomorphic embeddings have energy saturating a BPS bound similar to that of ref.~\cite{Gauntlett:1997ss} and preserve a fraction of the supersymmetry of the D\(q\)-brane background; typically one-half for \(d=0\) or one-quarter for \(d=4\) or \(8\).

In all of the embeddings that we construct, \(y\) is a complex coordinate formed from two directions orthogonal to the probe D\(q\)-branes, while \(z\) is a complex coordinate formed from two of the directions along the D\(q\)-branes. In general, one can choose to form each of \(y\) and \(z\) from directions \(x_\parallel^\m\) parallel to the D\(p\)-branes sourcing the supergravity background, or directions \(x_\perp^i\) orthogonal to them. We classify the holomorphic embeddings that we construct according to these choices, as summarised in table~\ref{tab:embedding_classification}. The D7-branes of ref.~\cite{holomorphic_branes} are of the type we call class 1, in which \(y\) is formed from the \(x_\perp^i\) directions and \(z\) from the \(x_\parallel^\m\) directions. We will also construct embeddings in which \(y\) and \(z\) are both formed from \(x_\perp^i\) or \(x_\parallel^\m\) directions, that we will refer to as class 2 and class 3, respectively. The final possibility, that \(y\) is formed from \(x_\parallel^\m\) directions and \(z\) from \(x_\perp^i\) directions and which we label class 1\('\), is related to class 1 by a reparameterisation of the D\(q\)-branes, in a sense discussed in section~\ref{sec:class1_embedding}.

\begin{table}
\begin{center}
\begin{tabular}{| l | l l |}
    \hline 
    Class &\,\(y\) & \(z\)
    \\ \hline
    ~~\,1 &\,\(x_\perp\) & \(x_\parallel\)
    \\
    ~~\,1\('\) &\,\(x_\parallel\) & \(x_\perp\)
    \\
    ~~\,2 &\,\(x_\perp\) & \(x_\perp\)
    \\
    ~~\,3 &\,\(x_\parallel\) & \(x_\parallel\)
    \\[0.25em] \hline
\end{tabular}
\caption{
    We will construct embeddings of probe D\(q\)-branes in extremal D\(p\)-brane backgrounds, specified by a complex embedding function \(y\) that is a holomorphic or antiholomorphic function of a complex coordinate \(z\) on the D\(q\)-branes. We classify these embeddings into four different types, depending on whether \(y\) and \(z\) are built from coordinates \(x_\parallel^\m\) parallel to the D\(p\)-branes sourcing the background, or coordinates \(x_\perp^i\) perpendicular to them.}
    \label{tab:embedding_classification}
\end{center}
\end{table}

Extremal D\(p\)-brane backgrounds have a decoupling limit, in which they are holographically dual to maximally supersymmetric \((p+1)\)-dimensional supersymmetric Yang--Mills (SYM) theory~\cite{Maldacena:1997re,Gubser:1998bc,Witten:1998qj,Itzhaki:1998dd,Boonstra:1998mp,Skenderis:1998dq,Kanitscheider:2008kd,Wiseman:2008qa}. Embedding probe D\(q\)-branes into the D\(p\)-brane background is typically holographically dual to coupling SYM to additional degrees of freedom, as with the hypermultiplets described above, or the insertion of defect operators into the path integral~\cite{Karch:2000ct,Karch_2001a}. The holomorphic embeddings that we construct each have holographic duals to explore. In this article we will examine two. We will focus on the most interesting case of the D3-brane background, holographically dual to four-dimensional \(\cN=4\) SYM~\cite{Maldacena:1997re,Gubser:1998bc,Witten:1998qj} and consider embeddings of \(d=4\) D5-branes and \(d=0\) D3-branes. As will be shown in section~\ref{sec:D3_background}, the probe D5-branes are dual to three-dimensional \(\cN=4\) hypermultiplets with mass depending holomorphically on position, while the probe D3-branes are dual to certain states in the presence of Gukov--Witten surface defects~\cite{Gukov:2006jk,Witten:2007td}.

\paragraph{Outline.}

The structure of this paper is as follows. In section~\ref{sec:embeddings} we will construct the different classes of holomorphic D-brane embeddings described above, and show that their energy saturates a BPS bound. We tabulate all supersymmetric holomorphic embeddings of classes 1, 1\('\), 2, and 3 in tables~\ref{tab:class1_holomorphic},~\ref{tab:class1prime_holomorphic},~\ref{tab:class2_holomorphic}, and~\ref{tab:class3_holomorphic}, respectively. In section~\ref{sec:susy} we compute the fraction of the supersymmetry of the extremal D\(p\)-brane backgrounds preserved by probe D\(q\)-branes with holomorphic embeddings. In section~\ref{sec:D3_background} we analyse the holography of class 1 D5- and D3-brane embeddings in the extremal D3-brane background. We close with discussion and outlook for future work in section~\ref{sec:discussion}.

We include two appendices which contain different generalisations of the embeddings that appear in the main body of the text. In appendix~\ref{sec:multiple_coords} we show that it is possible to construct embeddings in which \(y\) is a holomorphic function of \emph{multiple} complex coordinates \(z_1\), \(z_2\), etc. In appendix~\ref{sec:m_brane} we construct holomorphic embeddings of probe M2- and M5-branes in the M2- and M5-brane backgrounds of eleven-dimensional supergravity.

\section{Holomorphic embeddings}
\label{sec:embeddings}

In this section we demonstrate the existence of the holomorphic embeddings described in section~\ref{sec:intro}. Our starting point is the extremal black D\(p\)-brane background in type IIA or IIB supergravity, for \(p\) even or odd respectively. We will restrict to cases where \(p<7\), for which the string frame metric, the dilaton \(\f\), and the \((p+1)\)-form Ramond--Ramond (RR) field \(C_{p+1}\) of this background may be written as~\cite{Horowitz:1991cd}\footnote{For the case \(p=3\), the RR field \(C_4\) has additional terms with legs in the \(x_\perp^i\) directions, in order to make its field strength \(F_5 = \diff C_4\) self-dual. These terms will play no role in our discussion.}
\begin{equation}\begin{aligned} \label{eq:Dp_brane_background}
    \diff s^2 &= H(r)^{-1/2} \h_{\m\n} \diff x_\parallel^\m \diff x_\parallel^\n + H(r)^{1/2} \d_{ij} \diff x_\perp^i \diff x_\perp^j \, ,
    \\
    e^{\f(r)} &= g_s H(r)^{(3-p)/4} \, ,
    \\
    C_{p+1} &= \le[H(r)^{-1} - 1 \ri] \diff x_\parallel^0 \wedge \diff x_\parallel^1 \wedge \cdots \wedge \diff x_\parallel^p \, ,
\end{aligned}\end{equation}
with all other supergravity fields vanishing. In equation~\eqref{eq:Dp_brane_background}, \(\h_{\m\n}\) is the \((p+1)\)-dimensional Minkowski metric in mostly-plus signature, \(\d_{ij}\) is the Kronecker delta, \(g_s\) is the closed string coupling, \(r^2 = \d_{ij} x_\perp^i x_\perp^j\), and \(H(r)\) is the harmonic function
\begin{equation}
    H(r) = 1 + \le(\frac{L}{r}\ri)^{7-p} \, .
\end{equation}
The parameter \(L\), which has dimensions of length, is related to the number of D\(p\)-branes \(N\), the string coupling, and the Regge parameter \(\a'\), through 
\begin{equation} \label{eq:curvature_radius}
    L^{7-p} = (4\pi)^{(5-p)/2} \, \Gamma \le(\frac{7 - p}{2} \ri) g_s N {\a'}^{(7-p)/2} \, .
\end{equation}

We will embed \(k\) coincident D\(q\)-branes into the geometry in equation~\eqref{eq:Dp_brane_background}. We work in the probe limit, in which \(k\) is sufficiently small compared to \(N\) that we can neglect the backreaction of the D\(q\)-branes on the metric and other supergravity fields. We will assume that the D\(q\)-branes' worldvolume gauge field \(A\) vanishes. We will always work in a static gauge, in which we parameterise the D\(q\)-branes by \((q+1)\) of the coordinates in the background~\eqref{eq:Dp_brane_background}, which we denote \(\xi\). In a slight abuse of terminology we will often refer to the \(\xi\) directions as \emph{spanned} by the D\(q\)-branes. The embedding of the D\(q\)-branes is specified by how the directions orthogonal to the D\(q\)-branes depend on \(\xi\).\footnote{In general the world-volume scalars are \(k \times k\) matrices, valued in the adjoint representation of the Lie algebra \(\mathfrak{u}(k)\). We always consider abelian configurations in which the scalars are proportional to the identity matrix.} Allowed embeddings extremise the bosonic part of the D\(q\)-brane action \(S\), which for vanishing \(A\) is
\begin{equation} \label{eq:Dq_action}
    S = - k T_q \int \diff^{q+1}\xi \, e^{-\bar{\f}} \sqrt{\lvert \det g \rvert} + k T_q \int P[C_{q+1}],
\end{equation}
where \(e^{-\bar{\f}} \equiv g_s e^{-\f}\), and \(g\) is the induced metric on the D\(q\)-branes' worldvolume, i.e. the pullback of the metric~\eqref{eq:Dp_brane_background} to the worldvolume of the D\(q\)-branes. Further, \(P[C_q]\) is the pullback of \(C_q\), which in the D\(p\)-brane background~\eqref{eq:Dp_brane_background} vanishes unless \(p=q\). The D\(q\)-brane tension \(T_q\) is given by
\begin{equation}
    T_q = \frac{1}{(2\pi)^q \a'^{(q+1)/2} g_s}\, .
\end{equation}

Combining two of the directions perpendicular to the D\(q\)-branes into a complex coordinate \(y\), we will show in this section that there are combinations of \(p\) and \(q\) for which the D\(q\)-brane equations of motion that follow from extremisation of the action~\eqref{eq:Dq_action} allow \(y\) to be any holomorphic or antiholomorphic function of another complex coordinate \(z\) formed from two of the directions along the D\(q\)-branes. We will also show that when this happens, the energy of the D\(q\)-branes saturates a BPS bound. As discussed in section~\ref{sec:intro}, we will classify the embeddings we construct into four different classes, depending on whether the complex coordinates \(y\) and \(z\) are formed from \(x_\parallel^\m\) or \(x_\perp^i\) directions of the background~\eqref{eq:Dp_brane_background}, as summarised in table~\ref{tab:embedding_classification}. We will discuss the different classes of embeddings in the next three subsections, but first we will introduce some notation that will be common to all three embeddings.

Our D\(q\)-branes will always span time \(t = x_\parallel^0\) and the complex \(z\) plane. Two of the directions not spanned by the D\(q\)-branes will be used to form the complex coordinate \(y\). Depending on the class of embedding under consideration, \(z\) and \(y\) may be formed either from \(x_\parallel^\m\) directions or \(x_\perp^i\) directions. Any remaining \(x_\parallel^\m\) or \(x_\perp^i\) directions spanned by the D\(q\)-branes will be denoted by vectors \(\vec{x}\) and \(\vec{v}\), respectively, so that in total the D\(q\)-branes are parameterised by \(\xi = (t,z,\zb,\vec{x},\vec{v})\). Any remaining \(x_\parallel^\m\) or \(x_\perp^i\) directions orthogonal to the D\(q\)-branes will be denoted by vectors \(\vec{U}\) and \(\vec{W}\), respectively. Thus, a probe D\(q\)-brane embedding is specified by how \((y,\yb,\vec{U},\vec{W})\) depend on \(\xi\). These coordinates are summarised in table~\ref{tab:notation}. Since there must be at least two worldvolume scalar fields \((y,\yb)\), our holomorphic embeddings exist only for D\(q\)-branes with \(q \leq 7\).

\begin{table}[t]
\begin{center}
    \begin{tabular}{|c | l|}
    \hline
    Coordinate & Meaning
    \\ \hline
    \(t\) & Time, \(x_\parallel^0\)
    \\
    \((z,\zb)\) & Complex coordinates on worldvolume of D\(q\)-branes
    \\
    \(\vec{x}\) & \(x_\parallel^\m\) directions spanned by brane, excluding \(t\) and \((z,\zb)\)
    \\
    \(\vec{v}\) & \(x_\perp^i\) directions spanned by brane, excluding \((z,\zb)\)
    \\ \hdashline
    \((y,\yb)\) & Complex coordinates orthogonal to D\(q\)-branes
    \\
    \(\vec{U}\) & \(x_\parallel^\m\) directions orthogonal to D\(q\)-branes, excluding \((y,\yb)\)
    \\
    \(\vec{W}\) & \(x_\perp^i\) directions orthogonal to D\(q\)-branes, excluding \((y,\yb)\)
    \\ \hline
    \end{tabular}
    \end{center}
    \caption{Summary of the notation we use for the different types of coordinates in sections~\ref{sec:class1_embedding},~\ref{sec:class2_embedding}, and~\ref{sec:class3_embedding}. The first four rows are the coordinates \(\xi = (t,z,\zb,\vec{x},\vec{v})\) with which we parameterise the D\(q\)-branes. The remaining rows denote the transverse directions, which act as worldvolume scalars on the D\(q\)-branes. Whether \((z,\zb)\) and \((y,\yb)\) are formed from \(x_\parallel^\m\) or \(x_\perp^i\) directions depends on the class of embedding under consideration, as indicated in table~\ref{tab:embedding_classification}.}
    \label{tab:notation}
\end{table}

We will denote the number of \(x_\parallel^\m\) and \(x_\perp^i\) directions spanned by the D\(q\)-branes as \(a\) and \(b\), respectively. Since a D\(q\)-brane is \((q+1)\)-dimensional, \(b = q+1-a\). As discussed in the introduction, the number \(d\) of ND directions will be an important quantity. The ND directions are the \((p+1-a)\) \(x_\parallel^\m\) directions not spanned by the D\(q\)-branes and the \(b\) \(x_\perp^i\) directions spanned by the D\(q\)-branes, so that
\begin{equation}\begin{aligned} \label{eq:number_ND}
    d &= p + 1 - a + b 
    \\
    &= p + q + 2 ( 1 - a )\, .
\end{aligned}\end{equation}
Since \(p\) and \(q\) are both even or both odd in type IIA or type IIB supergravity, respectively, \(d\) is always even. We will see that the value of \(d\) determines whether or not holomorphic embeddings can exist as stable, supersymmetric solutions of the D\(q\)-brane equations of motion.

\subsection{Class 1 and class 1\texorpdfstring{\('\)}{'}}
\label{sec:class1_embedding}

\subsubsection{Class 1} 
\label{sec:class1_embedding_1}

We begin by constructing the class 1 embeddings. As indicated in table~\ref{tab:embedding_classification}, for such embeddings \(z\) is formed from \(x_\parallel^\m\) directions and \(y\) from \(x_\perp^i\) directions. Thus, in this section we define our complex coordinates as
\begin{equation} \label{eq:class1_z}
\begin{aligned}
    z = x_\parallel^1 + i x_\parallel^2 \, ,
    \qquad
    y = x_\perp^1 + i x_\perp^2 ,
\end{aligned}
\end{equation}
with \(\zb\) and \(\yb\) the complex conjugates of \(z\) and \(y\), respectively. Since the D\(q\)-branes span \((t,z,\zb)\), the number of \(x_\parallel^\m\) directions, \(a\), spanned by the D\(q\)-branes satisfies \(a \geq 3\). The D\(q\)-branes span a further \((a-3)\) \(x_\parallel^\m\) directions which, as discussed above and indicated in table~\ref{tab:notation}, we denote \(\vec{x}\). Any remaining \(x_\parallel^\m\) directions orthogonal to the D\(q\)-branes are denoted \(\vec{U}\). When \(b = q+1 -a >0\), the D\(q\)-branes span \(b\) of the \(x_\perp^i\) directions, denoted \(\vec{v}\). Apart from \((y,\yb)\), any remaining \(x_\perp^i\) directions are denoted \(\vec{W}\). Counting the number of each of these directions, the lengths of the vectors $(\vec{x}, \vec{U}, \vec{v}, \vec{W})$ are
\begin{equation} \label{eq:class1_dimensions}
\begin{aligned}
	\dim \vec{x} &= a - 3 \, ,
	& \qquad
	\dim \vec{U} &= p + 1 - a \, ,
	\\
	\dim \vec{v} &= q + 1 - a \, ,
	&
	\dim \vec{W} &= 6 - p - q + a \ .
\end{aligned}
\end{equation}
When \((p,q,a)\) are chosen such that any of these lengths are zero, the corresponding coordinates should be ignored from subsequent equations. Since both the D\(p\)- and D\(q\)-branes span at least three \(x_\parallel^\m\) directions \((t,z,\zb)\), the ansatz for class 1 embeddings requires \(p,q \geq 2\). The ND directions are \(\vec{U}\) and \(\vec{v}\), so equation~\eqref{eq:class1_dimensions} implies that there are \(d=p+q+2(1-a)\) of them, consistent with equation~\eqref{eq:number_ND}. Since five out of the ten dimensions, \((t,z,\zb,y,\yb)\), cannot be ND directions, the numbers of possible ND directions consistent with our ansatz for class 1 embeddings are \(d=0\), \(2\), or \(4\).

After relabelling the coordinates in this way, the blocks in the ten-dimensional metric in equation~\eqref{eq:Dp_brane_background} become
\begin{equation}\begin{aligned} \label{eq:metric_component_decomposition}
    \h_{\m\n}\diff x_\parallel^\m \diff x_\parallel^\n &= -  \diff t^2 + \diff z \diff \zb + \diff \vec{x}^{\,2} + \diff\vec{U}^{\,2} \, ,
    \\
    \d_{ij} \diff x_\perp^i \diff x_\perp^j &=   \diff y \diff \yb + \diff \vec{v}^{\,2} + \diff \vec{W}^{\,2} \, ,
\end{aligned}\end{equation}
where \(\diff \vec{x}^{\,2}\) denotes the flat metric \(\diff \vec{x}^{\,2} = \d_{\a\b} \diff x_\a \diff x_\b\), and similar for \(\diff \vec{U}^{\,2}\), \(\diff \vec{v}^{\,2}\), and \(\diff \vec{W}^{\,2}\). The radial distance \(r\) appearing in the harmonic function \(H(r)\) is determined by \(r^2 = |y|^2 + v^2 + W^2\), where \(v^2 = \vec{v}\cdot\vec{v} = \d_{ij}v_i v_j\) and similar for \(W^2\).

In table~\ref{tab:example_class_1_embeddings} we provide two examples to illustrate our notation. Table~\ref{tab:example_class_1_embeddings_1} shows the directions in the D4-brane background \((p=4)\) spanned by probe D4-branes \((q=4)\) when \(a=4\), and consequently \(b=1\). The shaded columns in the table indicate the \(x_\parallel^\m\) directions and the crosses indicate the directions spanned by the probe branes. In accordance with equation~\eqref{eq:class1_dimensions}, for these values of \((p,q,a)\) there is one each of \(\vec{x}\), \(\vec{U}\), and \(\vec{v}\) directions, and two \(\vec{W}\) directions. We chose \(p=q=a=4\) as an example since most other choices of these parameters leads to at least one of \((\vec{x},\vec{U},\vec{v},\vec{W})\) having zero length. For example, in table~\ref{tab:example_class_1_embeddings_2} we show the directions in the D3-brane background \((p=3)\) spanned by probe D5-branes \((q=5)\) when \(a=3\). Again in accordance with equation~\eqref{eq:class1_dimensions}, there are no \(\vec{x}\) directions when \(a=3\). For both examples we indicate the number \(d\) of ND directions, which correspond to the shaded columns without crosses plus the unshaded columns with crosses.

\begin{table}
\setlength{\tabcolsep}{3pt}
    \begin{subtable}{0.5\textwidth}\begin{tabularx}{0.98\textwidth}{| c | g g g g g Y Y : Y Y Y | c |}
            \hline
            & \(t\) & \(z\) & \(\zb\) & \(x_1\) & \(U_1\) & \(y\) & \(\yb\) & \(v_1\) & \(\!W_1\) & \(\!W_2\) & \(d\)
            \\ \hline
            D4 & \(\times\) & \(\times\) & \(\times\) & \(\times\) & & & & \(\times\) & & & 2
            \\
            \hline
    \end{tabularx}
    \caption{\(p = q = a = 4\,\).}
    \label{tab:example_class_1_embeddings_1}
    \end{subtable}\begin{subtable}{0.5\textwidth}
    \hfill
    \begin{tabularx}{0.98\textwidth}{| c | g g g g Y Y : Y Y Y Y | c |}
            \hline
            D\(q\) & \(t\) & \(z\) & \(\zb\) & \(U_1\) & \(y\) & \(\yb\) & \(v_1\) & \(v_2\) & \(v_3\)& \(\!W_1\) & \(d\)
            \\ \hline
            D5 & \(\times\) & \(\times\) & \(\times\) &  & & & \(\times\) & \(\times\) & \(\times\) & & \(4\)
            \\
            \hline
    \end{tabularx}
    \caption{\(p = a = 3\,\), \(q = 5\,\).}
    \label{tab:example_class_1_embeddings_2}
    \end{subtable}
    \caption{Two examples to illustrate the coordinate system defined by equation~\eqref{eq:metric_component_decomposition}. In each example, the shaded columns correspond to the \(x_\parallel^\m\) directions while the crosses indicate the directions \(\xi\) spanned by the probe branes. The ND directions are therefore the shaded columns without crosses, and the unshaded columns with crosses. The number \(d\) of ND directions is indicated in the final column of each sub-table. \textbf{(a):} Probe D4-branes in the extremal black D4-brane background, such that they span four of the five \(x_\parallel^\m\) directions. \textbf{(b):} Probe D5-branes in the extremal black D3-brane background, such that they span three of the four \(x_\parallel^\m\) directions. The analysis of section~\ref{sec:class1_embedding} shows that the example in (a) does not admit holomorphic embeddings while the example in (b) does, due to their respective values of \(d\).}
    \label{tab:example_class_1_embeddings}
\end{table}

The embedding of the coincident probe D\(q\)-branes in the D\(p\)-brane background is specified by how the transverse directions \((y,\yb,\vec{U},\vec{W})\) depend on \(\xi = (t,z,\zb,\vec{x},\vec{v})\). Following refs.~\cite{holomorphic_branes,Gauntlett:1997ss}, we will seek solutions to the D\(q\)-brane equations of motion where \(\vec{U}\) and \(\vec{W}\) are constant, while \(y\) and \(\yb\) depend only on \(z\) and \(\zb\),
\begin{equation} \label{eq:ansatz}
    y = y(z,\zb) \, ,
    \qquad
    \yb = \yb(z,\zb) \, .
\end{equation}
With this ansatz, the induced metric on the D\(q\)-branes' worldvolume is \(\diff s_{\mathrm{D}q}^2 \equiv g_{mn} \diff \xi^m \diff \xi^n\) given by
\begin{multline}  \label{eq:class1_metric}
    \diff s_{\mathrm{D}q}^2 = H(r)^{-1/2} \le(- \diff t^2 + \diff z \diff \zb + \diff\vec{x}^2 \ri) + H(r)^{1/2} \diff \vec{v}^2
    \\
    + H(r)^{1/2} \le( \p y \diff z + \pb y \diff \zb\ri)\le( \p \yb \diff z + \pb \yb \diff \zb\ri),
\end{multline}
where \(\p \equiv \p/\p z\) and \(\pb \equiv \p /\p \zb\). Using the numbers of \(\vec{x}\) and \(\vec{v}\) directions from equation~\eqref{eq:class1_dimensions}, we find that the determinant of the induced metric is
\begin{equation}  \label{eq:class1_metric_determinant}
    \lvert\det g \rvert = \frac{H(r)^{(q+1-2a)/2}}{4} \le(\le[1 + H(r) \le(|\p y|^2 + |\pb y|^2 \ri)\ri]^2 - 4 H(r)^2 |\p y|^2 |\pb y|^2 \ri) .
\end{equation}

For generic \(p\), \(q\), and \(a\), the pullback of \(C_{q+1}\) to the D\(q\)-branes' worldvolume will vanish and not contribute to the equations of motion evaluated on our ansatz. This happens when \(p \neq q\), because then \(C_{q+1}=0\) in the D\(p\)-brane background~\eqref{eq:Dp_brane_background}, and also when \(p=q\) with \(a\neq p+1\), as then the D\(q\)-branes do not span all the \(x_\parallel\) directions and hence the pullback vanishes. Thus, \(P[C_{q+1}]\) only contributes to the D\(q\)-brane action when \(p = q = a-1\), which from equation~\eqref{eq:number_ND} corresponds to \(d=0\) ND directions. This is the only way to obtain \(d=0\), since if \(q > p\) the D\(q\)-branes must span some \(x_\perp^i\) directions, while if \(q < p\) and/or \(a \leq p-1\) there must be some \(x_\parallel^\m\) directions not spanned by the D\(q\)-branes. This allows us to compactly write the pullback of \(C_{q+1}\) as
\begin{equation} \label{eq:class1_pullback_Cq}
    P[C_{q+1}] = \frac{i}{2} \d_{d,0}  \le[ H(r)^{-1} - 1 \ri] \diff t \wedge \diff z \wedge \diff \zb \wedge \diff x_1 \wedge \diff x_2 \wedge \cdots \wedge \diff x_{a-3}\,,
\end{equation}
where \(\d_{d,0}\) is the Kronecker delta.

Substituting equations~\eqref{eq:class1_metric_determinant} and~\eqref{eq:class1_pullback_Cq} into the D\(q\)-brane action~\eqref{eq:Dq_action} and using the expression for the dilaton in equation~\eqref{eq:Dp_brane_background}, we find that the action evaluated on our ansatz takes the form
\begin{equation}\begin{aligned} \label{eq:class1_action}
    S_1 &= - \frac{k T_q}{2}  \int \diff t  \diff z \diff \zb \diff \vec{x} \diff \vec{v} \, \cL_1\,,
    \\
    \cL_1 &=
        H(r)^{(d-4)/4} \sqrt{\le[1 + H(r) \le(|\p y|^2 + |\pb y|^2 \ri)\ri]^2 - 4 H(r)^2 |\p y|^2 |\pb y|^2} - \d_{d,0} \le[H(r)^{-1} - 1\ri],
\end{aligned}\end{equation}
with \(r^2 = |y|^2 + v^2 + W^2\). We have added the subscript ``\(1\)'' to indicate that this action is evaluated on the ansatz corresponding to class \(1\) embeddings. Notice that the Lagrangian density \(\cL_1\) depends on \(p\), \(q\), and \(a\) only through the number of ND directions \(d\).

In order to write the Euler--Lagrange equations that follow from the action~\eqref{eq:class1_action} in a relatively compact form, it is useful to define a quantity \(\cA_1\) and a differential operator \(\cD_1\),
\begin{equation}\begin{aligned}
    \cA_1 &= H(r)^{-1} + |\p y|^2 + |\pb y|^2 \, ,
    \\
    \cD_1[\bullet] &= \pb y \, \pb \yb \, \p^2 \bullet + \p y \, \p \yb \, \pb^2 \bullet - \cA_1 \, \p \pb \bullet \, .
\end{aligned}\end{equation}
Crucially for our purposes, \(\cD_1[y] = \cD_1[\yb] = 0\) if \(y\) is any holomorphic or antiholomorphic function of \(z\). The Euler--Lagrange equation for \(y(z,\zb)\) that follows from equation~\eqref{eq:class1_action} is
\begin{equation}\begin{aligned} \label{eq:class1_eom}
    0 &= \p y \, \pb y \,  \cD_1[y] - \frac{\cA_1}{2} \cD_1[\yb] + \frac{\p_r H}{4r H^2} \le(\cA_1 y  - 2 \, \yb \, \p y \, \pb y\ri) \p \yb \pb \yb
    \\ &\phantom{=}
    - \frac{d-4}{32} \frac{\p_r H}{r H} \le(\cA_1 \yb - 2 y \, \p \yb \, \pb \yb \ri) \le(\cA_1^2 - 4 |\p y|^2 |\pb y|^2 \ri)
    \\ &\phantom{=}
    - \d_{d,0} \frac{\p_r H}{8 r H^{(d+4)/4}} \yb  \le(\cA_1^2 - 4 |\p y|^2 |\pb y|^2 \ri)^{3/2} \,.
\end{aligned}\end{equation}
The Euler--Lagrange equation for \(\yb(z,\zb)\) is the complex conjugate of equation~\eqref{eq:class1_eom}.

The first line of equation~\eqref{eq:class1_eom} vanishes when \(y\) is a holomorphic or antiholomorphic function of \(z\), since then \(\cD_1[y]=\cD_1[\yb]=\p\yb \, \pb\yb=0\). The second and third lines each vanish when \(d=4\), and cancel against each other for holomorphic or antiholomorphic \(y\) when \(d=0\). Thus, equation~\eqref{eq:class1_eom} admits solutions with arbitrary holomorphic or antiholomorphic \(y\) when \(d=0\) or \(d=4\), but not when \(d=2\). We will collectively refer to any solution with \(y = y(z)\) or \(y = y(\zb)\) as a \emph{holomorphic embedding}.

Notice that the Wess--Zumino term in the D\(q\)-brane action, which gives rise to the third line of the Euler--Lagrange equation~\eqref{eq:class1_eom}, plays a crucial role in the existence of holomorphic embeddings for \(d=0\), since in this case holomorphic embeddings only exist because the second and third lines of equation~\eqref{eq:class1_eom} cancel each other. Physically, then, holomorphic embeddings only exist for \(d=0\) due to a stabilising force present thanks to the D\(q\)-branes' coupling to \(C_{q+1}\). Relatedly, we will shortly show that the energy of D\(q\)-branes with (anti)holomorphic \(y\) saturates a BPS bound for \(d=0\) and \(d=4\), but not for \(d=2\). In the latter case, the failure to saturate a BPS bound is presumably due to the lack of a stabilising Wess--Zumino coupling.

Recall that in our ansatz we took the worldvolume scalars \(\vec{U}\) and \(\vec{W}\) to be constant, and we should confirm that this choice extremises the action. Any constant \(\vec{U}\) solves the Euler--Lagrange equations, since \(\vec{U}\) is a cyclic coordinate. This follows from translational invariance of the D\(p\)-brane background in the \(\vec{U}\) directions. On the other hand, the action in equation~\eqref{eq:class1_action} depends explicitly on \(\vec{W}\) through its dependence on \(r\). The Euler--Lagrange equation for \(\vec{W}\), \(\vec{\nabla}_W \cL_1 = 0\), evaluates to
\begin{equation} \label{eq:class1_w_eom}
     \frac{\vec{W}}{r} H^{(d-8)/4} \p_r H \le[d \sqrt{H^2\cA_1^2 - 4 |\p y|^2 |\pb y|^2} - \frac{4 H \cA_1}{\sqrt{H^2\cA_1^2 - 4 |\p y|^2 |\pb y|^2}} +4 \d_{d,0}\ri]   = 0\, .
\end{equation}
The left-hand side vanishes for any \(\vec{W}\) for \(d=0\) and holomorphic or antiholomorphic \(y\), since then the term in the square brackets vanishes. On the other hand, for \(d=4\) the term in the square brackets is non-zero for (anti)holomorphic but non-constant \(y\), so in general the only way to solve equation~\eqref{eq:class1_w_eom} is to set \(\vec{W} = 0\).

In summary, for \(d=0\) or \(d=4\) the D\(q\)-brane equations of motion admit solutions where \(y\) is a holomorphic or antiholomorphic function of \(z\), sitting at constant \(\vec{W} = 0\) for \(d=4\) or arbitrary constant \(\vec{W}\) for \(d=0\), and at arbitrary constant \(\vec{U}\). All possible class 1 holomorphic embeddings are listed in table~\ref{tab:class1_holomorphic}. They correspond to the values of \(2 \leq p < 7\), \(2 \leq q \leq 7\), and \(3 \leq a \leq \mathrm{max}(p+1,q+1)\) that yield \(d=0\) or \(d=4\). The fact that holomorphic embeddings solve the D\(q\)-branes' equations of motion~\eqref{eq:class1_eom} and~\eqref{eq:class1_w_eom} is independent of the form of the function \(H(r)\), so holomorphic embeddings exist both in the full D\(p\)-brane background in equation~\eqref{eq:Dp_brane_background}, as well as its near-horizon limit obtained by setting \(H(r) = (L/r)^{7-p}\).

\begin{table}
\setlength{\tabcolsep}{2pt}
    \begin{subtable}{0.5\textwidth}
    \begin{tabularx}{0.98\textwidth}{| c | g g g Y Y : Y Y Y Y Y | c |}
            \multicolumn{1}{c}{} & \multicolumn{3}{c}{\textcolor{gray!75!black}{\(\xleftarrow{\hspace{0.25cm}} \hfill x_\parallel^\m \hfill \xrightarrow{\hspace{0.25cm}}\)}} & \multicolumn{7}{c}{\textcolor{gray!75!black}{\(\xleftarrow{\hspace{1.4cm}} \hfill x_\perp^i \hfill \xrightarrow{\hspace{1.4cm}}\)}}
            \\ \hline
            D\(q\) & \(t\) & \(z\) & \(\zb\) & \(y\) & \(\yb\) & \(x_\perp^3\) & \(x_\perp^4\) & \(x_\perp^5\)& \(x_\perp^6\) & \(x_\perp^7\) & \(d\)
            \\ \hline
            D2 & \(\times\) & \(\times\) & \(\times\) & & & & & & & & \(0\)
            \\ 
            D6 & \(\times\) & \(\times\) & \(\times\) & & & \(\times\)  & \(\times\) & \(\times\) & \(\times\) & & \(4\)
            \\\hline
    \end{tabularx}
    \caption{\(p=2\)}
    \end{subtable}\begin{subtable}{0.5\textwidth}
    \hfill
    \begin{tabularx}{0.98\textwidth}{| c | g g g g Y Y : Y Y Y Y | c |}
            \hline
            D\(q\) & \(t\) & \(z\) & \(\zb\) & \(x_\parallel^3\) & \(y\) & \(\yb\) & \(x_\perp^3\) & \(x_\perp^4\) & \(x_\perp^5\)& \(x_\perp^6\) & \(d\)
            \\ \hline
            D3 & \(\times\) & \(\times\) & \(\times\) & \(\times\) & & & & & & & \(0\)
            \\
            D5 & \(\times\) & \(\times\) & \(\times\) &  & & & \(\times\) & \(\times\) & \(\times\) & & \(4\)
            \\ 
            D7 & \(\times\) & \(\times\) & \(\times\) & \(\times\) & & & \(\times\) & \(\times\) & \(\times\) & \(\times\) & \(4\)
            \\
            \hline
    \end{tabularx}
    \caption{\(p=3\)}
    \label{tab:class1_holomorphic_D3}
    \end{subtable}
    \\[1em]
    \begin{subtable}{0.5\textwidth}
    \begin{tabularx}{0.98\textwidth}{| c | g g g g g Y Y : Y Y Y | c |}
            \hline
            D\(q\) & \(t\) & \(z\) & \(\zb\) & \(x_\parallel^3\) & \(x_\parallel^4\) & \(y\) & \(\yb\) &  \(x_\perp^3\) & \(x_\perp^4\) & \(x_\perp^5\)  &\(d\)
            \\ \hline
            D4 & \(\times\) & \(\times\) & \(\times\) & \(\times\) & \(\times\) & & & & & & \(0\)
            \\
            D4 & \(\times\) & \(\times\) & \(\times\) & & & & &  \(\times\) & \(\times\) & & \(4\)
            \\ 
            D6 & \(\times\) & \(\times\) & \(\times\) &  \(\times\) &  & & & \(\times\) & \(\times\) & \(\times\) & \(4\)
            \\
            \hline
    \end{tabularx}
    \caption{\(p=4\)}
    \end{subtable}\begin{subtable}{0.5\textwidth}
    \hfill
    \begin{tabularx}{0.98\textwidth}{| c | g g g g g g Y Y : Y Y | c |}
        \hline
        D\(q\) & \(t\) & \(z\) & \(\zb\) & \(x_\parallel^3\) & \(x_\parallel^4\) & \(x_\parallel^5\) & \(y\) & \(\yb\) & \(x_\perp^3\) & \(x_\perp^4\)  & \(d\)
        \\ \hline
        D5 & \(\times\) & \(\times\) & \(\times\) & \(\times\) & \(\times\) & \(\times\) & & & & & \(0\)
        \\ 
        D3 & \(\times\) & \(\times\) & \(\times\) &  & & & & & \(\times\)& & \(4\)
        \\
        D5 & \(\times\) & \(\times\) & \(\times\) & \(\times\) & & & & & \(\times\) & \(\times\) & \(4\)
        \\
        \hline
    \end{tabularx}
    \caption{\(p=5\)}
    \end{subtable}
    \\[1em]
    \begin{subtable}{\textwidth}\centering
    \begin{tabularx}{0.49\textwidth}{| c | g g g g g g g Y Y : Y | c|}
        \hline
        D\(q\) & \(t\) & \(z\) & \(\zb\) & \(x_\parallel^3\) & \(x_\parallel^4\) & \(x_\parallel^5\) & \(x_\parallel^6\) & \(y\) & \(\yb\) & \(x_\perp^3\) & \(d\)
        \\ \hline
        D6 & \(\times\) & \(\times\) & \(\times\) & \(\times\) & \(\times\) & \(\times\) & \(\times\) & & & & \(0\)
        \\
        D2 & \(\times\) & \(\times\) & \(\times\) & & & & & & & & \(4\)
        \\
        D4 & \(\times\) & \(\times\) & \(\times\) & \(\times\) & & & & & &  \(\times\) & \(4\)
        \\
        \hline
    \end{tabularx}
    \caption{\(p=6\)}
    \end{subtable}
    \caption{All possible holomorphic D\(q\)-brane embeddings of class 1 in extremal black D\(p\)-brane backgrounds with \(p < 7\), as described in section~\ref{sec:class1_embedding_1}, organised by \(p\) and by their number \(d\) of ND directions. Each row of each table shows a possible D\(q\)-brane embedding in the corresponding D\(p\)-brane background, with the crosses indicating the directions spanned by the D\(q\)-branes. The shaded columns indicate the \(x_\parallel^\m\) directions of the D\(p\)-brane background, as indicated explicitly in table (a). The D7-brane in table (b) is the holomorphic embedding of ref.~\cite{holomorphic_branes}. We show in section~\ref{sec:class1_kappa} that holomorphic embeddings with \(d=0\) preserve one-half of the supersymmetry of the D\(p\)-brane background, while holomorphic embeddings with \(d=4\) instead preserve one-quarter.}
    \label{tab:class1_holomorphic}
\end{table}

Although for clarity of presentation we have only presented the equations of motion as derived by first substituting our ansatz into the action~\eqref{eq:class1_action}, we have also checked that the full D\(q\)-brane equations of motion derived from arbitrary variations of the action~\eqref{eq:Dq_action} are satisfied by these holomorphic embeddings when \(d=0\) or \(d=4\).

For completeness, we note that there is a family of cases with \(d=2\) for which our ansatz \(A=0\) for the D\(q\)-branes' worldvolume gauge field is manifestly inconsistent with the equations of motion. When \(A\) is non-zero, the bosonic part of the D\(q\)-brane action in equation~\eqref{eq:Dq_action} contains extra terms, including a Wess--Zumino term
\begin{equation} \label{eq:extra_wz_term}
    S \supset 2\pi\a' k T_q \int F \wedge P[C_{q-1}] \, ,
\end{equation}
where \(F = \diff A\) is the field strength for \(A\). When \(q=p+2\) and \(a=p+1\), i.e. when a probe D\((p+2)\)-brane spans all of the \(x_\parallel^\m\) directions in the D\(p\)-brane background, then the pullback of \(C_{q-1}\) is non-zero in the Wess--Zumino term in equation~\eqref{eq:extra_wz_term}. This term in the action then gives rise to a source term in the Euler--Lagrange equation for \(A\). Substituting \(q=p+2\) and \(a=p+1\) into equation~\eqref{eq:number_ND}, we confirm that such configurations have \(d=2\), so the Wess--Zumino term in equation~\eqref{eq:extra_wz_term} does not spoil the existence of holomorphic embeddings for \(d=0\) or \(d=4\).

\paragraph{BPS bound.} Holomorphic embeddings solve the D\(q\)-brane equations of motion when \(d\) is a multiple of four because their energy saturates a BPS bound. This argument was made for brane embeddings in flat space in ref.~\cite{Gauntlett:1997ss} and adapted to class 1, \(d=4\) D7-brane embeddings in the D3-brane background in ref.~\cite{holomorphic_branes}. We now generalise this argument to arbitrary class 1 D\(q\)-brane embeddings in D\(p\)-brane backgrounds.

For arbitrary integer \(n\) and for \(y = y(z,\zb)\), let us define the quantity
\begin{equation} \label{eq:class1_Y}
    \cY_n = H(r)^{n/4} \le(|\p y|^2 - |\pb y|^2 \ri) \, ,
\end{equation}
in terms of which the Lagrangian density \(\cL_1\) in equation~\eqref{eq:class1_action} can be written in two equivalent forms
\begin{equation}\begin{aligned} \label{eq:class1_equivalent forms}
    \cL_1  &= \sqrt{\le[H(r)^{(d-4)/4} + \cY_d \ri]^2 + 4 H(r)^{(d-2)/2} |\pb y|^2} - \d_{d,0} \le[H(r)^{-1} - 1\ri]
    \\
    &= \sqrt{\le[H(r)^{(d-4)/4} - \cY_d \ri]^2 + 4 H(r)^{(d-2)/2} |\p y|^2}  - \d_{d,0} \le[H(r)^{-1} - 1\ri]\, .
\end{aligned}\end{equation}
Since \(H(r)\), \(|\pb y|^2\), and \(|\p y|^2\) are all non-negative, the square roots appearing in these expressions are bounded from below by the factors in the square brackets,
\begin{equation}\begin{aligned}
    \sqrt{\le[H(r)^{(d-4)/4} + \cY_d \ri]^2 + 4 H(r)^{(d-2)/2} |\pb y|^2} \geq H(r)^{(d-4)/4} + \cY_d \, ,
    \\
     \sqrt{\le[H(r)^{(d-4)/4} - \cY_d \ri]^2 + 4 H(r)^{(d-2)/2} |\p y|^2}  \geq H(r)^{(d-4)/4} - \cY_d \, .
\end{aligned}\end{equation}
Thus, the Lagrangian density for class 1 embeddings in equation~\eqref{eq:class1_equivalent forms} satisfies the bound
\begin{equation} \label{eq:class1_lagrangian_bound}
    \cL_1 \geq \begin{cases}
        1 + |\cY_d| \, , & d=0 \text{ or } 4\,,
        \\
        H(r)^{-1/2} + |\cY_d| \, , & d=2 \, .
    \end{cases}
\end{equation}
This bound is saturated when \(y\) is a holomorphic or antiholomorphic function of \(z\). For example, for holomorphic \(y\) we have that \(|\pb y| = 0\), so that \(\cY_d\) in equation~\eqref{eq:class1_Y} is positive and the square root in the first line of equation~\eqref{eq:class1_equivalent forms} is equal to \(H(r)^{(d-4)/4} + \cY_d\).

Substituting the bound on the Lagrangian density into the action~\eqref{eq:class1_action}, we find that the action is bounded from above. Equivalently, since the D\(q\)-branes are static and so their energy \(E\) is minus the Lagrangian, the energy of the D\(q\)-branes is bounded from below. For \(d=0\) or \(4\) these bounds are
\begin{equation} \label{eq:class1_bps_bound}
    S_1 \leq - \int \diff t \, ( Z + Y_d) \, , \qquad E \geq Z + Y_d \, , \qquad (d = 0\text{ or }4)\,,
\end{equation}
where we have defined the integrals
\begin{equation}\begin{aligned} \label{eq:class1_ZY_integrals}
    Z &= \frac{k T_q}{2}\int \diff z \diff \zb \diff \vec{x} \diff \vec{v} \, ,
    \\
    Y_d &= \frac{k T_q}{2} \int \diff z \diff \zb \diff \vec{x} \diff \vec{v} \, |\cY_d| = \frac{\mathrm{deg}(y) \, k T_q}{2}\int \diff y \diff \yb \diff \vec{x} \diff \vec{v} \, H(r)^{d/4} \, .
\end{aligned}\end{equation}
The second equality in the expression for \(Y_d\) arises because \(\cY_n\) in equation~\eqref{eq:class1_Y} is \(H(r)^{n/4}\) times the Jacobian for a change of integration variables from \((z,\zb)\) to \((y,\yb)\).  The factor \(\mathrm{deg}(y)\) is the degree of the map \(y : \mathbb{C} \to \mathbb{C}\), i.e. how many times we must integrate over the complex \(y\) plane to integrate over the whole of the complex \(z\) plane.

The integrals for \(Z\) and \(Y_d\) in equation~\eqref{eq:class1_ZY_integrals} are divergent due to the infinite extent of the D\(q\)-branes and so require regularisation, for instance by integrating only over a finite extent in each of the coordinates. Provided we maintain consistent regularisation of the integral over the complex \(y\) plane, \(Y_d\) depends only on the topological data of \(y(z,\zb)\), in the form of the degree \(\mathrm{deg}(y)\).

Stable branes in string theory arise as central charges of the target space supersymmetry algebra~\cite{deAzcarraga:1989mza,Sato:1998ax,Sato:1998yu,Callister:2007jy} and, similarly to in refs.~\cite{Gauntlett:1997ss,holomorphic_branes}, the quantities \(Z\) and \(Y_d\) appearing in the BPS bound are precisely such central charges. Concretely, using the general expressions for D-brane central charges in non-trivial supergravity backgrounds in ref.~\cite{Callister:2007jy}, it is straightforward to show that \(Z\) is the central charge corresponding to \(k\) D\(q\)-branes parallel to the \((t,z,\zb,\vec{x},\vec{v})\) directions (i.e. a class 1 embedding with constant \(y\)), while \(Y_d\) is that of \(\mathrm{deg}(y) \, k\) D\(q\)-branes parallel to the directions \((t,y,\yb,\vec{x},\vec{v})\). More generally, these branes minimise their energy because they wrap calibrated manifolds \cite{Harvey:1982xk,Gutowski_1999,Gutowski_1999b}. See for instance refs.~\cite{gauntlett2003branescalibrationssupergravity,Sim_n_2012,townsend2000branetheorysolitons,Townsend_2000} for reviews on calibrated geometry in supergravity.

Holomorphic or antiholomorphic \(y\) saturates the bounds in equation~\eqref{eq:class1_bps_bound}, and therefore extremises the action for fixed regularised central charges \(Z\) and \(Y_d\), providing another perspective on why such holomorphic embeddings solve the D\(q\)-brane equations of motion for \(d=0\) or \(d=4\). The reason why (anti)holomorphic \(y\) does not solve the equations of motion for \(d=2\) is that equation~\eqref{eq:class1_lagrangian_bound} implies that in this case
\begin{equation} \label{eq:class1_d2_bound}
    S_1 \leq - \int \diff t \, (\tilde{Z} + Y_2) \, ,
    \quad \text{where} \quad
    \tilde{Z} \equiv  \frac{k T_q}{2}\int \diff z \diff \zb \diff \vec{x} \diff \vec{v} \, H(r)^{-1/2} \,,  \qquad (d = 2)\,.
\end{equation}
Although this inequality is saturated for holomorphic or antiholomorphic \(y\), the value of \(\tilde{Z}\) depends on the form of \(y(z)\) or \(y(\zb)\), through its dependence on \(r^2 = |y|^2 + \vec{v}^{\,2} + \vec{W}^{\,2}\). Thus, to solve the equations of motion we would still need to extremise the integral of \(\tilde{Z}\), which implies that we must set \(y = 0\).

\subsubsection{Class 1\('\)} 
\label{sec:class_1_prime}

Recall from table~\ref{tab:embedding_classification} that class 1\('\) embeddings were defined in a complementary manner to class 1 embeddings, by interchanging the roles of \(y\) and \(z\). Concretely, for class 1\('\) embeddings \(y\) is formed from the \(x_\parallel^\m\) directions and \(z\) from \(x_\perp^i\) directions.

The ansatz for class 1\('\) embeddings may therefore be obtained from the ansatz for class 1 embeddings by a reparameterisation of the D\(q\)-branes. We begin with a class 1 holomorphic embedding, for which \(y = y(z)\), and then switch to parameterising the D\(q\)-branes by \(z\) rather than \(y\), so that now the embedding is specified by how \(z\) depends on \(y\), \(z = z(y)\). We then relabel the variables \(z \leftrightarrow y\), i.e. the coordinate that we previously called \(z\) we now call \(y\), and vice versa. This effects the change that \(z\) is now built from \(x_\perp^i\) directions and \(y\) from \(x_\parallel^\m\) directions.

Since we are defining the ND directions as the \(x_\perp^i\) directions used to parameterise the D\(q\)-branes plus the \(x_\parallel^\m\) directions \emph{not} used to parameterise the D\(q\)-branes, the class 1\('\) ansatz obtained by applying the above reparameterisation to a class 1 ansatz has four extra ND directions, namely \((z,\zb,y,\yb)\). Thus, in going from the class 1 ansatz to the class 1\('\) ansatz we send \(d \to d + 4\).

\begin{figure}
    \centering
    \begin{subfigure}{0.3\textwidth}
        \includegraphics{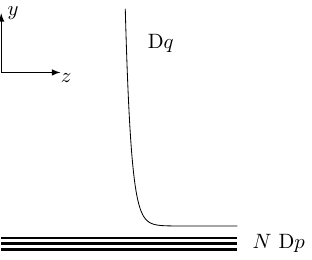}
        \caption{Class 1}
        \label{subfig:class1}
    \end{subfigure}
    \hspace{5em}
    \begin{subfigure}{0.3\textwidth}
        \includegraphics{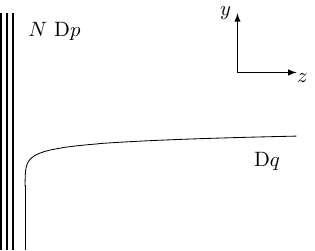}
        \caption{Class 1\('\)}
        \label{subfig:class1'}
    \end{subfigure}
    \caption{The relation between class 1 embeddings and class 1${}^\prime$ embeddings. \textbf{(a):} Cartoon of a class 1 embedding. The thick, horizontal lines represent the D\(p\)-branes sourcing the background~\eqref{eq:Dp_brane_background}. The curve represent the probe D\(q\)-branes, which have embedding specified by how \(y\) depends on \(z\). \textbf{(b):} Cartoon of a class 1\('\) embedding. The thick, vertical lines represent the D\(p\)-branes, and the curve again represents the probe D\(q\)-branes, which again have embedding specified by how \(y\) depends on \(z\). Figures (a) and (b) are the same up to a \(\pi/2\) rotation, representing a reparameterisation of the D\(q\)-branes, and relabelling of variables \(y \leftrightarrow z\). Note however that not every class 1\('\) embedding can be thought of as a simple reparameterisation of class 1 embedding, as discussed in the main text.}
    \label{fig:classVsclass1'}
\end{figure}

At the risk of labouring the point, we illustrate the reparameterisation schematically in figure~\ref{fig:classVsclass1'}. Figure~\ref{subfig:class1} shows a cartoon of a class 1 embedding. The three thick, horizontal lines represent the \(N\) D\(p\)-branes sourcing the supergravity background. The horizontal direction represents the complex coordinate \(z\), which in accordance with table~\ref{tab:embedding_classification} is built from directions parallel to the D\(p\)-branes. Similarly, the vertical direction represents the complex coordinate \(y\), built from directions orthogonal to the D\(p\)-branes. Figure~\ref{subfig:class1'} shows a cartoon of the class 1\('\) embedding obtained by the reparameterisation. It is identical to figure~\ref{subfig:class1}, up to a \(\pi/2\) rotation and the interchange \(y \leftrightarrow z\). The thick, vertical lines again represent the D\(p\)-branes. The rotation represents the change of variables after which we specify the class 1 embedding by \(z(y)\). After the interchange \(y \leftrightarrow z\), we now have that the embedding is specified by \(y = y(z)\), with \(y\) built from \(x_\parallel^\m\) directions and \(x_\perp^i\) directions, as appropriate for class 1\('\).

The punchline is that the action \(S_{1'}\) for class 1\('\) embeddings may be obtained from the action for class 1 embeddings in equation~\eqref{eq:class1_action}, by treating \(z\) and \(\zb\) as functions of \(y\) and \(\yb\), then relabelling the variables \((y,\yb)  \leftrightarrow (z,\zb)\) and sending \(d \to d +4\). This procedure yields the action
\begin{equation}\begin{aligned} \label{eq:class1prime_action}
    S_{1'} &= - \frac{k T_q}{2}  \int \diff t  \diff z \diff \zb \diff \vec{x} \diff \vec{v} \, \cL_{1'}\,,
    \\
    \cL_{1'} &=
        H(r)^{(d-8)/4} \sqrt{\le[H(r) + |\p y|^2 + |\pb y|^2 \ri]^2 - 4  |\p y|^2 |\pb y|^2}
        \\ 
        &\phantom{=} \hspace{4cm} - \d_{d,4} \le| |\p y|^2 - |\pb y|^2 \ri|\le[H(r)^{-1} - 1\ri] ,
\end{aligned}\end{equation}
with \(r^2 = |z|^2 + v^2 + W^2 \). The reparameterisation used to obtain this action immediately implies that the corresponding Euler--Lagrange equations admit solutions with arbitrary holomorphic or antiholomorphic \(y\) when \(d = 4\) or \(8\). This may be verified by direct calculation. All D\(q\)-brane embeddings in D\(p\)-brane backgrounds admitting class 1\('\) holomorphic solutions are listed in table~\ref{tab:class1prime_holomorphic}, which is obtained from table~\ref{tab:class1_holomorphic} by the interchange \(y \leftrightarrow z\) and sending \(d \to d+4\).

\begin{table}
\setlength{\tabcolsep}{2pt}
    \begin{subtable}{0.5\textwidth}
    \begin{tabularx}{0.98\textwidth}{| c | g g g Y Y : Y Y Y Y Y | c |}
            \hline
            D\(q\) & \(t\) & \(y\) & \(\yb\) & \(z\) & \(\zb\) & \(x_\perp^3\) & \(x_\perp^4\) & \(x_\perp^5\)& \(x_\perp^6\) & \(x_\perp^7\) & \(d\)
            \\ \hline
            D2 & \(\times\) & & & \(\times\) & \(\times\) & & & & & & \(4\)
            \\
            D6 & \(\times\) & & & \(\times\)  & \(\times\) & \(\times\) & \(\times\) & \(\times\) & \(\times\) & & \(8\)
            \\\hline
    \end{tabularx}
    \caption{\(p=2\)}
    \end{subtable}\begin{subtable}{0.5\textwidth}
    \hfill
    \begin{tabularx}{0.98\textwidth}{| c | g g g g Y Y : Y Y Y Y | c |}
            \hline
            D\(q\) & \(t\) & \(y\) & \(\yb\) & \(x_\parallel^3\) & \(z\) & \(\zb\) & \(x_\perp^3\) & \(x_\perp^4\) & \(x_\perp^5\)& \(x_\perp^6\) & \(d\)
            \\ \hline
            D3 & \(\times\) & & & \(\times\) & \(\times\) & \(\times\) &  & & & & \(4\)
            \\
            D5 & \(\times\) & & &  & \(\times\) & \(\times\) & \(\times\) &\(\times\) & \(\times\) &  & \(8\)
            \\ 
            D7 & \(\times\) & & & \(\times\) & \(\times\) & \(\times\) & \(\times\) & \(\times\) & \(\times\) & \(\times\) & \(8\)
            \\
            \hline
    \end{tabularx}
    \caption{\(p=3\)}
    \end{subtable}
    \\[1em]
    \begin{subtable}{0.5\textwidth}
    \begin{tabularx}{0.98\textwidth}{| c | g g g g g Y Y : Y Y Y | c |}
            \hline
            D\(q\) & \(t\) & \(y\) & \(\yb\) & \(x_\parallel^3\) & \(x_\parallel^4\) & \(z\) & \(\zb\) & \(x_\perp^3\) & \(x_\perp^4\) & \(x_\perp^5\)  & \(d\)
            \\ \hline
            D4 & \(\times\) & & & \(\times\) & \(\times\) & \(\times\) & \(\times\) &  & & & \(4\)
            \\
            D4 & \(\times\) & & & & &  \(\times\) & \(\times\) & \(\times\) & \(\times\) & & \(8\)
            \\
            D6 & \(\times\) & & &  \(\times\) &  &  \(\times\) & \(\times\) & \(\times\) & \(\times\) & \(\times\) & \(8\)
            \\
            \hline
    \end{tabularx}
    \caption{\(p=4\)}
    \end{subtable}\begin{subtable}{0.5\textwidth}
    \hfill
    \begin{tabularx}{0.98\textwidth}{| c | g g g g g g Y Y : Y Y | c |}
        \hline
        D\(q\) & \(t\) & \(y\) & \(\yb\) & \(x_\parallel^3\) & \(x_\parallel^4\) & \(x_\parallel^5\) & \(z\) & \(\zb\) & \(x_\perp^3\) & \(x_\perp^4\) & \(d\)
        \\ \hline
        D5 & \(\times\) & & & \(\times\) & \(\times\) & \(\times\) & \(\times\) & \(\times\) &  & & \(4\)
        \\
        D3 & \(\times\) & & &  & & & \(\times\) & \(\times\) & \(\times\) &  & \(8\)
        \\ 
        D5 & \(\times\) & & & \(\times\) & &  & \(\times\) & \(\times\) & \(\times\) & \(\times\) & \(8\)
        \\
        \hline
    \end{tabularx}
    \caption{\(p=5\)}
    \end{subtable}
    \\[1em]
    \begin{subtable}{\textwidth}\centering
    \begin{tabularx}{0.49\textwidth}{| c | g g g g g g g Y Y : Y | c |}
        \hline
        D\(q\) & \(t\) & \(y\) & \(\yb\) & \(x_\parallel^3\) & \(x_\parallel^4\) & \(x_\parallel^5\) & \(x_\parallel^6\)  & \(z\) & \(\zb\) & \(x_\perp^1\) & \(d\)
        \\ \hline
        D6 & \(\times\) & & & \(\times\) & \(\times\) & \(\times\) & \(\times\) & \(\times\) & \(\times\) & & \(4\)
        \\
        D2 & \(\times\) & & & & & & & \(\times\) & \(\times\) & & \(8\)
        \\
        D4 & \(\times\) & & & \(\times\) & & & & \(\times\) & \(\times\) & \(\times\) & \(8\)
        \\
        \hline
    \end{tabularx}
    \caption{\(p=6\)}
    \end{subtable}
    \caption{All possible holomorphic D\(q\)-brane embeddings of class 1\('\) in extremal black D\(p\)-brane backgrounds with \(p < 7\), as described in section~\ref{sec:class_1_prime}, organised by \(p\) and by their number \(d\) of ND directions. We show in section~\ref{sec:class1_kappa} that the embeddings with \(d=4\) or \(d=8\) preserve one-half or one-quarter of the supersymmetry of the D\(p\)-brane background, respectively.}
    \label{tab:class1prime_holomorphic}
\end{table}

Although the action and equations of motion for class 1 and 1\('\) embeddings are obtained from each other by a reparameterisation of the D\(q\)-branes, we distinguish these two classes with a prime because this is not always the case for the solutions; the step where we exchange \(y(z)\) for \(z(y)\) only works if \(y(z)\) is invertible. For instance, a class 1 embedding with constant \(y\) cannot be thought of as a class 1\('\) embedding. More subtly, a class 1 embedding for which \(y\) has poles or zeros of degree greater than one would correspond to a class 1\('\) embedding with a branch cut. For example, consider a class 1 embedding for which
\begin{equation}
    y = c z^n \, ,
\end{equation}
for some integer \(n\). Thus \(z = c^{1/n} y^{1/n}\), and after relabelling the variables \(z \leftrightarrow y\), this becomes a class 1\('\) embedding with
\begin{equation}
    y = c^{1/n} z^{1/n} \, ,
\end{equation}
which has a branch point at \(z=0\) for \(|n| > 1\). To properly make sense of this branch cut we would have to use the non-abelian D\(q\)-brane action and introduce non-zero holonomy of the D\(q\)-branes' worldvolume gauge field around the branch point~\cite{Li:1998ce}, which goes beyond the scope of our present work.

\subsection{Class 2}
\label{sec:class2_embedding}

We now describe class 2 embeddings. In accordance with table~\ref{tab:embedding_classification}, for class 2 embeddings we form both of the complex coordinates \(y\) and \(z\) from \(x_\perp^i\) directions. Concretely, in this section we take
\begin{equation} \label{eq:class2_z}
    z = x_\perp^1 + i x_\perp^2 \, ,
    \qquad
    y = x_\perp^3 + i x_\perp^4 ,
\end{equation}
with \(\zb\) and \(\yb\) the complex conjugates of \(z\) and \(y\), respectively. Class 2 embeddings can only exist in the D\(p\)-brane backgrounds with \(p \leq 5\), since they require at least four \(x_\perp^i\) directions to form the complex coordinates in equation~\eqref{eq:class2_z}.

The analysis of class 2 embeddings proceeds almost identically to that performed for class 1 embeddings in section~\ref{sec:class1_embedding_1}. We will therefore be briefer in this section. We again adopt the notation summarised in table~\ref{tab:notation}. We take the D\(q\) branes to span \(a\) of the \(x_\parallel^\m\) directions, \(t\) and \(\vec{x}\), with the remaining \(x_\parallel^\m\) directions denoted as \(\vec{U}\). In addition to \((z,\zb)\), the D\(q\)-branes may span a further \(b = q-1-a\) of the \(x_\perp^i\) directions, which we again denote \(\vec{v}\). There are at least two \(x_\perp^i\) directions transverse to the D\(q\)-branes, \((y,\yb)\). Any further \(x_\perp^i\) directions we denote by \(\vec{W}\). Counting the number of \((\vec{x}, \vec{U}, \vec{v},\vec{W})\) coordinates, we find
\begin{equation} \label{eq:class2_dimensions}
\begin{aligned}
	\dim \vec{x} &= a - 1 \, ,
	& \qquad
	\dim \vec{U} &= p + 1 - a \, ,
	\\
	\dim \vec{v} &= q - 1 - a \, ,
	&
	\dim \vec{W} &= 6 - p - q + a \ .
\end{aligned}
\end{equation}
The ND directions are \((z,\zb,\vec{U},\vec{v})\), so the total number \(d\) of them is again given by equation~\eqref{eq:number_ND}.  Note that for class 2 embeddings \(d\geq 2\), since there are at least two ND directions \((z,\zb)\), while \(d \leq 6\) since there are at least three directions \((t,y,\yb)\) which are not ND.  In terms of the coordinates used in this section, the blocks appearing in the metric in equation~\eqref{eq:Dp_brane_background} are
\begin{equation}\begin{aligned} \label{eq:class2_metric_component_decomposition}
    \h_{\m\n}\diff x_\parallel^\m \diff x_\parallel^\n &= -  \diff t^2 + \diff \vec{x}^{\,2} + \diff\vec{U}^{\,2} \, ,
    \\
    \d_{ij} \diff x_\perp^i \diff x_\perp^j &= \diff z \diff \zb + \diff y \diff \yb + \diff \vec{v}^{\,2} + \diff \vec{W}^{\,2} \, .
\end{aligned}\end{equation}

For the ansatz that \(y = y(z,\zb)\) and \(\yb = \yb(z,\zb)\) with \(\vec{U}\) and \(\vec{W}\) constant, the determinant of the induced metric on the D\(q\)-branes' worldvolume is
\begin{equation} \label{eq:class2_metric_determinant}
    \lvert \det g \rvert = \frac{H(r)^{(q+1-2a)/2}}{4} \le[ \le(1 + |\p y|^2 + |\pb y|^2 \ri)^2 - 4 |\p y|^2 |\pb y|^2\ri].
\end{equation}
The pullback of \(C_{q+1}\) always vanishes on the ansatz for class 2 embeddings since they have \(d \neq 0\), so substituting this expression for \(\lvert \det g \rvert\) into the D\(q\)-brane action~\eqref{eq:Dq_action}, we obtain
\begin{equation}\begin{aligned} \label{eq:class2_action}
    S_2 &= - \frac{k T_q}{2}  \int \diff t  \diff z \diff \zb \diff \vec{x} \diff \vec{v} \, \cL_2\,,
    \\
    \cL_2 &=
        H(r)^{(d-4)/4} \sqrt{\le(1 + |\p y|^2 + |\pb y|^2 \ri)^2 - 4 |\p y|^2 |\pb y|^2} \, ,
\end{aligned}\end{equation}
with \(r^2 = |z|^2 + |y|^2 + v^2 + W^2\), where have added the subscript ``2'' to denote class 2 embeddings.

To write the equations of motion that follow from the action~\eqref{eq:class2_action} in a relatively compact form, we introduce the notation
\begin{equation}\begin{aligned}
    \cA_2 &= 1 + |\p y|^2 + |\pb y|^2 \, ,
    \\
    \cD_2[\bullet] &= \pb y \, \pb \yb  \, \p^2 \bullet + \p y \, \p \yb \, \pb^2 \bullet - \cA_2 \, \p \pb \bullet \, .
\end{aligned}\end{equation}
Notice that \(\cD_2[y] = \cD_2[\bar{y}] = 0\) if \(y\) is either a holomorphic or antiholomorphic function of \(z\). The Euler--Lagrange equation for \(y\) that follows from the action~\eqref{eq:class2_action} is
\begin{multline} \label{eq:class2_eom}
    \p \yb \, \pb \yb \, \cD_2[y] - \frac{\cA_2}{2} \cD_2[\yb] - \frac{d-4}{32} \frac{\p_r H}{r H} \le[\cA_2 \yb - 2 y \, \p \yb \,  \pb \yb - (1-\cY_0) z \, \p \yb - (1+\cY_0) \zb \, \pb \yb \ri] = 0 \, ,
\end{multline}
where \(\cY_0 = |\p y|^2 - |\pb y|^2\) is the central charge density from equation~\eqref{eq:class1_Y}, evaluated for \(n=0\). The Euler--Lagrange equation for \(\yb\) is the complex conjugate of equation~\eqref{eq:class2_eom}. The first two terms in equation~\eqref{eq:class2_eom} vanish when \(y\) is a holomorphic or antiholomorphic function of \(z\), while the rest of the left-hand side only vanishes for non-trivial \(y\) if \(d=4\). Thus, holomorphic embeddings solve the Euler--Lagrange equation for \(y\) if and only if \(d=4\).

As for the class 1 embeddings, \(\vec{U}\) is a cyclic coordinate, so any constant \(\vec{U}\) solves its Euler--Lagrange equation. Moreover, for \(d=4\) the action in equation~\eqref{eq:class2_action} is independent of \(H(r)\) and therefore independent of \(\vec{W}\), and so for \(d=4\) any constant value of \(\vec{W}\) solves its Euler--Lagrange equation.

In sum, holomorphic or antiholomorphic \(y\) solves the D\(q\)-brane equations of motion for \(d=4\) but not for other values of \(d\).\footnote{Note that \(d=4\) is the only possibility for which \(d\) is a multiple of four, since for class 2 embeddings \(2 \leq d \leq 6\) as discussed above.} The holomorphic embeddings can have any constant values of the other worldvolume scalars \(\vec{U}\) and \(\vec{W}\). All possible class 2 holomorphic embeddings are listed in table~\ref{tab:class2_holomorphic}. As for class 1 embeddings, the Wess--Zumino term in equation~\eqref{eq:extra_wz_term} does not spoil the existence of these holomorphic embeddings since when \(A=0\) it only contributes to the equations of motion when \(q=p+2\) and \(a=p+1\), corresponding to \(d=2\).

\begin{table}
\setlength{\tabcolsep}{3pt}
    \begin{subtable}{0.5\textwidth}
    \begin{tabularx}{0.98\textwidth}{| c | g  Y Y : Y Y : Y Y Y Y Y |}
            \hline
            D\(q\) & \(t\) & \(z\) & \(\zb\) & \(y\) & \(\yb\) & \(x_\perp^5\) & \(x_\perp^6\) & \(x_\perp^7\)& \(x_\perp^8\) & \(x_\perp^9\)
            \\ \hline
            D4 & \(\times\) &  \(\times\) &  \(\times\) & & & \(\times\) &  \(\times\) & & &
            \\\hline
    \end{tabularx}
    \caption{\(p=0\)}
    \end{subtable}\begin{subtable}{0.5\textwidth}
    \hfill\begin{tabularx}{0.98\textwidth}{| c | g g Y Y : Y Y : Y Y Y Y |}
            \hline
            D\(q\) & \(t\) & \(x_\parallel^1\) & \(z\) & \(\zb\) & \(y\) & \(\yb\) & \(x_\perp^5\) & \(x_\perp^6\) & \(x_\perp^7\)& \(x_\perp^8\)
            \\ \hline
            D3 & \(\times\) & &  \(\times\) &  \(\times\) & & & \(\times\) & \(\times\) & &
            \\
            D5 & \(\times\) & \(\times\) & \(\times\) & \(\times\) & & & \(\times\) & \(\times\) & & 
            \\ \hline
    \end{tabularx}
    \caption{\(p=1\)}
    \end{subtable}
    \\[1em]
    \begin{subtable}{0.5\textwidth}
    \begin{tabularx}{0.98\textwidth}{| c | g g g Y Y : Y Y : Y Y Y |}
            \hline
            D\(q\) & \(t\) & \(x_\parallel^1\) & \(x_\parallel^2\) & \(z\) & \(\zb\) & \(y\) & \(\yb\) & \(x_\perp^3\) & \(x_\perp^4\) & \(x_\perp^5\)
            \\ \hline
            D2 & \(\times\) & & & \(\times\) & \(\times\) & & & & &
            \\ 
            D4 & \(\times\) & \(\times\) & &  \(\times\) &  \(\times\) & & & \(\times\) & &
            \\
            D6 & \(\times\) & \(\times\) & \(\times\) & \(\times\)  & \(\times\) & & & \(\times\) & \(\times\) &
            \\\hline
    \end{tabularx}
    \caption{\(p=2\)}
    \end{subtable}\begin{subtable}{0.5\textwidth}
    \hfill\begin{tabularx}{0.98\textwidth}{| c | g g g g Y Y : Y Y : Y Y |}
            \hline
            D\(q\) & \(t\) & \(x_\parallel^1\) & \(x_\parallel^2\) & \(x_\parallel^3\) & \(z\) & \(\zb\) & \(y\) & \(\yb\) &  \(x_\perp^3\) & \(x_\perp^4\)
            \\ \hline
            D3 & \(\times\) & \(\times\) &  & & \(\times\) & \(\times\) & & & &
            \\
            D5 & \(\times\) & \(\times\) & \(\times\) & & \(\times\) & \(\times\) & & & \(\times\) &
            \\
            D7 & \(\times\) & \(\times\) & \(\times\) & \(\times\) & \(\times\) & \(\times\) & & & \(\times\) & \(\times\)
            \\\hline
    \end{tabularx}
    \caption{\(p=3\)}
    \end{subtable}
    \\[1em]
    \begin{subtable}{0.5\textwidth}
    \begin{tabularx}{0.98\textwidth}{| c | g g g g g Y Y : Y Y : Y |}
            \hline
            D\(q\) & \(t\) & \(x_\parallel^1\) & \(x_\parallel^2\) & \(x_\parallel^3\) & \(x_\parallel^4\) & \(z\) & \(\zb\) & \(y\) & \(\yb\) & \(x_\perp^3\)
            \\ \hline
            D4 & \(\times\) & \(\times\) & \(\times\) & & & \(\times\) & \(\times\) & & &
            \\
            D6 & \(\times\) & \(\times\) & \(\times\) & \(\times\) &  & \(\times\) & \(\times\) & & & \(\times\)
            \\\hline
    \end{tabularx}
    \caption{\(p=4\)}
    \end{subtable}\begin{subtable}{0.5\textwidth}
    \hfill\begin{tabularx}{0.98\textwidth}{| c | g g g g g g Y Y : Y Y |}
            \hline
            D\(q\) & \(t\) & \(x_\parallel^1\) & \(x_\parallel^2\) & \(x_\parallel^3\) & \(x_\parallel^4\) & \(x_\parallel^5\) & \(z\) & \(\zb\) &  \(y\) & \(\yb\)
            \\ \hline
            D5 & \(\times\) & \(\times\) & \(\times\) & \(\times\) & & & \(\times\) & \(\times\) & & 
            \\\hline
    \end{tabularx}
    \caption{\(p=5\)}
    \end{subtable}
    \caption{All posssible class 2 holomorphic D\(q\)-brane embeddings in extremal black D\(p\)-brane backgrounds, as described in section~\ref{sec:class2_embedding}, organised by \(p\). All have \(d=4\) ND directions. We show in section~\ref{sec:class2_kappa} that holomorphic class 2 embeddings preserve one-quarter of the supersymmetry of the D\(p\)-brane background.}
    \label{tab:class2_holomorphic}
\end{table}

\paragraph{BPS bound.} The energy of class 2 holomorphic embeddings saturates a BPS bound similar to that for class 1 embeddings. To show this, we write the Lagrangian density in equation~\eqref{eq:class2_action} in two equivalent ways, as
\begin{equation}\begin{aligned} \label{eq:class2_action_alternate}
    \cL_2 &= H(r)^{(d-4)/4} \sqrt{(1+\cY_0)^2 + 4 |\pb y|^2}
    \\
    &= H(r)^{(d-4)/4} \sqrt{(1-\cY_0)^2 + 4 |\p y|^2} \, .
\end{aligned}\end{equation}
By similar logic as led to equation~\eqref{eq:class1_bps_bound} for class 1 embeddings, this implies that the action for class 2 embeddings is bounded from above,
\begin{equation} \label{eq:class_2_action_bound}
    S_2 \leq - \frac{k T_q}{2} \int \diff t \diff z \diff \zb \diff \vec{x} \diff \vec{v} \, H(r)^{(d-4)/2}
    - \frac{\mathrm{deg}(y)  \, k T_q}{2} \int \diff t \diff y \diff \yb \diff \vec{x} \diff \vec{v} \, H(r)^{(d-4)/2} \, ,
\end{equation}
with \(r^2 = |z|^2 + |y|^2 + v^2 + W^2\), and where this bound applies after regulating the integrals over the D\(q\)-branes' worldvolume. Holomorphic or antiholomorphic \(y\) saturate the bound in equation~\eqref{eq:class_2_action_bound}. However, only for \(d=4\) does saturation of the bound mean extremisation of the action, since only for \(d=4\) do the integrals in equation~\eqref{eq:class_2_action_bound} become independent of \(r\), and hence independent of the form of \(y(z)\) or \(y(\zb)\) except through topological data in the form of the degree \(\mathrm{deg}(y)\).

Rewriting the bound on the action for \(d=4\) in terms of the energy \(E\) of the D\(q\)-branes, we obtain the BPS bound satisfied by class 2 embeddings,
\begin{equation} \label{eq:class2_bps_bound}
    E \geq Z + Y_0 \, , \qquad (d=4) \, ,
\end{equation}
where we have defined the integrals, as in equation~\eqref{eq:class1_ZY_integrals},
\begin{equation} \label{eq:class2_ZY_integrals}
    Z = \frac{k T_q}{2}\int \diff z \diff \zb \diff \vec{x} \diff \vec{v} \, ,
    \qquad
    Y_0 = \frac{\mathrm{deg}(y) \, k T_q}{2}\int \diff y \diff \yb \diff \vec{x} \diff \vec{v} \,  .
\end{equation}
As for the class 1 embeddings, \(Z\) is the central charge corresponding to \(k\) D\(q\)-branes parallel to the directions \((t,z,\zb,\vec{x},\vec{v})\), while \(Y_0\) is the central charge corresponding to \(\deg(y) \, k\) D\(q\)-branes parallel to the directions \((t,y,\yb,\vec{x},\vec{v})\).

\subsection{Class 3}
\label{sec:class3_embedding}

Finally, we describe the class 3 embeddings. Once again we will be brief. For class 3 embeddings, we form both complex coordinates from \(x_\parallel^\m\) directions, so in this section we take
\begin{equation} \label{eq:class3_z}
    z = x_\parallel^1 + i x_\parallel^2 \, ,
    \qquad
    y = x_\parallel^3 + i x_\parallel^4 ,
\end{equation}
with \(\zb\) and \(\yb\) the complex conjugates of \(z\) and \(y\), respectively. The D\(q\)-branes span \(a\) of the \(x_\parallel^\m\) directions, including \((t,z,\zb)\) but not including \((y,\yb)\), thus \(3 \leq a \leq p-1\). This implies that class 3 embeddings can only exist for \(p \geq 4\). We label the remaining directions as in table~\ref{tab:notation}; the other \(x_\parallel^\m\) directions spanned by the D\(q\)-branes are labelled \(\vec{x}\), the remaining \(x_\parallel^\m\) directions are labelled \(\vec{U}\), and the \(x_\perp^i\) are separated into directions \(\vec{v}\) spanned by the D\(q\)-branes and directions \(\vec{W}\) transverse to them. The number of each of these directions is 
\begin{equation} \label{eq:class3_dimensions}
\begin{aligned}
	\dim \vec{x} &= a - 3\, ,
	& \qquad
	\dim \vec{U} &= p - 1 - a \, ,
	\\
	\dim \vec{v} &= q + 1 - a \, ,
	&
	\dim \vec{W} &= 8 - p - q + a \ .
\end{aligned}
\end{equation}
The blocks appearing in the metric in equation~\eqref{eq:Dp_brane_background} are
\begin{equation}\begin{aligned} \label{eq:class3_metric_component_decomposition}
    \h_{\m\n}\diff x_\parallel^\m \diff x_\parallel^\n &= -  \diff t^2 + \diff z \diff \zb +  \diff y \diff \yb + \diff \vec{x}^{\,2} + \diff\vec{U}^{\,2} \, ,
    \\
    \d_{ij} \diff x_\perp^i \diff x_\perp^j &=  \diff \vec{v}^{\,2} + \diff \vec{W}^{\,2} \, .
\end{aligned}\end{equation}
This ansatz requires \(p\geq4\), so that we have enough \(x_\parallel^\m\) directions to build the complex coordinates in equation~\eqref{eq:class3_z}. The ND directions are \((y,\yb,\vec{U},\vec{v})\) and their number is given by equation~\eqref{eq:number_ND}. Since there are at least two ND directions \((y,\yb)\) and at least three directions \((t,z,\zb)\) which are not ND, we have that \(2 \leq d \leq 6\). Since class 3 embeddings have \(a \leq p-1\), the pullback of \(C_{p+1}\) to the D\(q\)-branes' worldvolume always vanishes.

With the ansatz that \(y = y(z,\zb)\) and \(\yb = \yb(z,\zb)\), and with \(\vec{U}\) and \(\vec{W}\) constant, the determinant of the induced metric on the worldvolume of the D\(q\)-branes is
\begin{equation} \label{eq:class3_metric_determinant}
    \lvert \det g \rvert = \frac{H(r)^{(q+1-2a)/2}}{4} \le[ \le(1 + |\p y|^2 + |\pb y|^2 \ri)^2 - 4 |\p y|^2 |\pb y|^2 \ri]
\end{equation}
Substituting this into the action~\eqref{eq:Dq_action}, we obtain
\begin{equation}\begin{aligned} \label{eq:class3_action}
    S_3 &= - \frac{k T_q}{2}  \int \diff t  \diff z \diff \zb \diff \vec{x} \diff \vec{v} \, \cL_3\,,
    \\
    \cL_3 &=
        H(r)^{(d-4)/4} \sqrt{\le(1 + |\p y|^2 + |\pb y|^2 \ri)^2 - 4 |\p y|^2 |\pb y|^2} \, ,
\end{aligned}\end{equation}
with \(r^2 = v^2 + W^2\), where have added the subscript ``3'' to denote class 3 embeddings. Although the actions for class 2 and class 3 embeddings in equations~\eqref{eq:class2_action} and~\eqref{eq:class3_action} look superficially the same, they differ in how \(r\) depends on \(z\) and \(y\), leading to different equations of motion. Concretely, the Euler--Lagrange equation for \(y\) that follows from equation~\eqref{eq:class3_action} is 
\begin{equation} \label{eq:class3_eom}
    \p \yb \, \pb \yb \, \cD_2[y] - \frac{\cA_2}{2} \cD_2[\yb] = 0 \, ,
\end{equation}
where \(\cD_2\) and \(\cA_2\) are  defined in equation~\eqref{eq:class2_action}. Equation~\eqref{eq:class3_eom} is solved by arbitrary holomorphic or antiholomorphic \(y\), since in either case  \(\cD_2[y] = \cD_2[\yb] = 0\). Equation~\eqref{eq:class3_eom} is independent of the number of ND directions \(d\), so is solved by (anti)holomorphic \(y\) for any \(d\).\footnote{Since equation~\eqref{eq:class3_eom} is independent of \(H(r)\), it is the same as the Euler--Lagrange equation for embedding D\(q\)-branes in Minkowski space, corresponding to \(H(r)=1\), which is known to admit arbitrary (anti)holomorphic solutions~\cite{Gauntlett:1997ss}.} As for class 1 and 2 embeddings, any constant value of \(\vec{U}\) solves its Euler--Lagrange equation, while the Euler--Lagrange equation for \(\vec{W}\) following from equation~\eqref{eq:class3_action} is
\begin{equation} \label{eq:class3_w_eom}
    (d-4) \vec{W} \, \p_r H(r) = 0 \, ,
\end{equation}
which is automatically satisfied for any \(\vec{W}\) if \(d=4\). For other values of \(d\), equation~\eqref{eq:class3_w_eom} requires that the D\(q\)-branes sit at \(\vec{W} = 0\).

\paragraph{BPS bound.} The same reasoning that led us to equation~\eqref{eq:class_2_action_bound} implies that for any \(y = y(z,\zb)\), the action in equation~\eqref{eq:class3_action} satisfies the bound
\begin{equation} \label{eq:class_3_action_bound}
    S_3 \leq - \frac{k T_q}{2} \int \diff t \diff z \diff \zb \diff \vec{x} \diff \vec{v} \, H(r)^{(d-4)/2}
    - \frac{\mathrm{deg}(y)  \, k T_q}{2} \int \diff t \diff y \diff \yb \diff \vec{x} \diff \vec{v} \, H(r)^{(d-4)/2} \, ,
\end{equation}
which is saturated for holomorphic or antiholomorphic \(y\). Equations~\eqref{eq:class_2_action_bound} and~\eqref{eq:class_3_action_bound} again differ due to the different way \(r\) depends on \(y\) and \(z\) for class 2 versus class 3 embeddings. In particular, \(r\) for class 3 embeddings is independent of \(y\), and therefore independent of the form of the function \(y(z,\zb)\). Since for (anti)holomorphic \(y\) the action saturates the bound in equation~\eqref{eq:class_3_action_bound}, this implies that the action is extremised for such \(y\), giving another perspective on why class 3 holomorphic embeddings solve the D\(q\)-brane equations of motion for any \(d\).

Despite the fact that class 3 holomorphic embeddings saturate the bound in equation~\eqref{eq:class_3_action_bound} for any \(d\), we will see in section~\ref{sec:class3_kappa} that they preserve a fraction of the supersymmetry of the D\(p\)-brane background, and are therefore guaranteed to be stable, only for \(d=4\). All possible supersymmetric, \(d=4\) class 3 holomorphic embeddings are listed in table~\ref{tab:class3_holomorphic}.

\begin{table}
\setlength{\tabcolsep}{2pt}
    \begin{subtable}{0.5\textwidth}
    \begin{tabularx}{0.98\textwidth}{| c | g  g g : g g Y Y Y Y Y |}
            \hline
            D\(q\) & \(t\) & \(z\) & \(\zb\) & \(y\) & \(\yb\) & \(x_\perp^1\) & \(x_\perp^2\) & \(x_\perp^3\)& \(x_\perp^4\) & \(x_\perp^5\)
            \\ \hline
            D4 & \(\times\) & \(\times\) & \(\times\) & & & \(\times\) & \(\times\) & & &
            \\ \hline
    \end{tabularx}
    \caption{\(p=4\)}
    \end{subtable}\begin{subtable}{0.5\textwidth}
    \hfill
    \begin{tabularx}{0.98\textwidth}{| c | g  g g : g g : g Y Y Y Y |}
            \hline
            D\(q\) & \(t\) & \(z\) & \(\zb\) & \(y\) & \(\yb\) & \(x_\parallel^5\) & \(x_\perp^1\) & \(x_\perp^2\) & \(x_\perp^3\)& \(x_\perp^4\)
            \\ \hline
            D3 & \(\times\) & \(\times\) & \(\times\) & & & & \(\times\) & & &
            \\
            D5 & \(\times\) & \(\times\) & \(\times\) & & & \(\times\) & \(\times\) & \(\times\) & &
            \\ \hline
    \end{tabularx}
    \caption{\(p=5\)}
    \end{subtable}
    \\[1em]
    \begin{subtable}{\textwidth}\centering
    \begin{tabularx}{0.49\textwidth}{| c | g  g g : g g : g g Y Y Y |}
            \hline
            D\(q\) & \(t\) & \(z\) & \(\zb\) & \(y\) & \(\yb\) & \(x_\parallel^1\) & \(x_\parallel^2\) & \(x_\perp^1\)& \(x_\perp^2\) & \(x_\perp^3\)
            \\ \hline
            D2 & \(\times\) & \(\times\) & \(\times\) & & & & & & &
            \\
            D4 & \(\times\) & \(\times\) & \(\times\) & & & \(\times\) & & \(\times\) & &
            \\
            D6 & \(\times\) & \(\times\) & \(\times\) & & & \(\times\) &  \(\times\) & \(\times\) &  \(\times\) & 
            \\\hline
    \end{tabularx}
    \caption{\(p=6\)}
    \end{subtable}
    \caption{All supersymmetric holomorphic D\(q\)-brane embeddings of class 3 in extremal black D\(p\)-brane backgrounds with \(p<7\), as described in section~\ref{sec:class3_embedding}, organised by \(p\). All have \(d=4\) ND directions. We show in section~\ref{sec:class3_embedding} that each of these embeddings preserve one-quarter of the supersymmetry of the D\(p\)-brane background.}
    \label{tab:class3_holomorphic}
\end{table}

\section{Supersymmetry analysis}
\label{sec:susy}

In this section we will show that the holomorphic embeddings constructed in section~\ref{sec:embeddings} preserve a fraction of the supersymmetry of the extremal D\(p\)-brane background, by checking their kappa symmetry. We begin in subsection~\ref{subsec:susy_notation} by establishing our conventions for spinors and notation for the Killing spinors of the D\(p\)-brane background. In the subsequent subsections we will then perform the kappa symmetry analysis for each of the classes of holomorphic embeddings, in turn. Our analysis will proceed similarly to that for the class 1 D7-brane embedding in the D3-brane background appearing in ref.~\cite{holomorphic_branes}.

\setcounter{subsection}{-1}
\subsection{Spinor conventions and Killing spinors of extremal D-brane backgrounds}
\label{subsec:susy_notation}

For our spinor conventions, we follow ref.~\cite{Freedman_Van_Proeyen_2012}. We adopt the notation that \(\G_A\) are the ten-dimensional Minkowski space Dirac matrices, satisfying
\begin{equation} \label{eq:clifford_10d}
    \{\G_A , \G_B \} = 2 \h_{AB} \id \, , 
\end{equation}
with \(\h_{AB}\) the ten-dimensional Minkowski metric in mostly-plus signature. We take \(\G_0\) to be anti-Hermitian and \(\G_i\) Hermitian for \(i\geq 1\). We use \(\g_m\) to denote the pullback of the \(d=10\) curved space Dirac matrices to the worldvolume of the probe D\(q\)-branes,
\begin{equation}
    \g_m = (\p_m x^\m) e_\m^A \G_A \, ,
\end{equation}
where \(e_\m^A\) are vielbeins for the ten-dimensional metric in equation~\eqref{eq:Dp_brane_background}. When \(\G\) or \(\g\) has multiple indices, this denotes a normalised antisymmetric product, for example
\begin{equation}
    \G_{AB} = \frac{1}{2} \le(\G_A \G_B - \G_B \G_A \ri) .
\end{equation}
We denote the ten-dimensional chirality matrix as \(\G_\sharp= \G_{01\cdots9}\)\,. It is Hermitian. We denote the charge conjugation matrix as \(C\). By definition, it satisfies \(\G_A^T =  - C \G_A C^{-1}\) for all \(A\).

In both type IIA and type IIB supergravities, there are two Majorana--Weyl Killing spinors, \(\veh^{1}\) and \(\veh^2\). Being Majorana spinors, they satisfy the Majorana condition
\begin{equation} \label{eq:majorana_condition}
    (\veh^i)^* = B \veh^i \,
\end{equation}
where \(B = i C \G^0\) is a matrix obeying \(B \G_A B^{-1} = (\G_A)^*\) for all \(A\). Being Weyl spinors, the two Killing spinors satisfy
\begin{equation}  \label{eq:weyl_condition}
    \G_\sharp \veh^1 = \veh^1 \, ,
    \qquad
    \G_\sharp \veh^2 = \mp \veh^2 \, ,
\end{equation}
with the upper and lower signs for the chirality of \(\veh^2\) in type IIA and type IIB supergravity, respectively. It is notationally convenient to package both spinors into a single object~\cite{Bergshoeff:1997kr}. For type IIA, where \(\veh^1\) and \(\veh^2\) have opposite chirality, we package them into a single Majorana spinor \(\veh = \veh^1 + \veh^2\). We can then extract \(\veh^1\) and \(\veh^2\) by applying the appropriate chiral projections. For type IIB we instead package the spinors into a doublet, \(\veh = (\veh^1,\veh^2)\).

For both type IIA and type IIB supergravities, the Killing spinors \(\veh\) of the D\(p\)-brane background of equation~\eqref{eq:Dp_brane_background} take the form~\cite{Bergshoeff:1996wk,Kehagias:1998gn,Grana:2000jj}
\begin{equation} \label{eq:killing_spinor}
    \veh = H(r)^{-1/8} \ve \, ,
\end{equation}
where \(\ve\) is a constant Majorana spinor in type IIA, or a constant doublet of Majorana--Weyl spinors in type IIB, satisfying the projection conditions
\begin{subequations} \label{eq:Dp_brane_projection_conditions}
\begin{align}
    \ve &= \G_{x_\parallel^0 x_\parallel^1 \cdots x_\parallel^p} \le(\G_\sharp\ri)^{(p+2)/2}\ve\, , && \text{(type IIA)},
    \\
    \ve &= \G_{x_\parallel^0 x_\parallel^1 \cdots x_\parallel^p} \otimes (\s_3)^{(p+1)/2} i\s_2 \ve\,, &&\text{(type IIB)},
\end{align}
\end{subequations}
where the Pauli matrices that appear in the type IIB case act on the doublet index of the Killing spinors, and the subscripts on \(\G_{x_\parallel^0 x_\parallel^1 \cdots x_\parallel^p}\) indicate that we should take an antisymmetric product of the Dirac matrices corresponding to all of the \(x_\parallel^\m\) directions. To treat both type IIA and type IIB supergravity in a unified manner, one can define the matrix~\cite{Bergshoeff:1997kr}
\begin{equation} \label{eq:Jp}
     J_{(p)} = \begin{cases}
        \G_\sharp^{(p+2)/2}, &\qquad  \text{(type IIA)},
        \\
        (\s_3)^{(p+1)/2} i \s_2, &\qquad  \text{(type IIB)}.
    \end{cases}
\end{equation}
Then, the conditions in equation~\eqref{eq:Dp_brane_projection_conditions} may be expressed as
\begin{equation} \label{eq:Dp_brane_projection_conditions_unified}
    \ve = \G_{x_\parallel^0 x_\parallel^1 \cdots x_\parallel^p} J_{(p)} \ve \, ,
\end{equation}
where from now on for type IIB we leave implicit the tensor product in any concatenation of Dirac and Pauli matrices.

The supersymmetries preserved by the introduction of our probe D\(q\)-branes correspond to those constant Majorana--Weyl spinors \(\ve\) obeying equation~\eqref{eq:Dp_brane_projection_conditions_unified} that also obey the kappa symmetry condition~\cite{Bergshoeff:1997kr}
\begin{equation} \label{eq:kappa_symmetry_constraint}
    \G \ve = \ve \, ,
\end{equation}
where for D\(q\)-branes with vanishing worldvolume gauge field the kappa symmetry matrix \(\G\) is given by\footnote{In writing equation~\eqref{eq:kappa_matrix_general_cases} we have anticipated that we will use a complex coordinate for two of the directions on the D\(q\)-branes. This is responsible for the prefactor of \(-i\) which does not appear in the expression in ref.~\cite{Bergshoeff:1997kr}.}
\begin{equation} \label{eq:kappa_matrix_general_cases}
    \G = \frac{-i}{\sqrt{\lvert \det g \rvert}} \times \begin{cases}
        \g_{01 \cdots q} \le(\G_\sharp\ri)^{(q+2)/2} \, ,
        &\qquad  \text{(type IIA)},
        \\
        \g_{01 \cdots q} (\s_3)^{(q+1)/2} i \s_2\, ,
        &\qquad  \text{(type IIB)}.
    \end{cases}
\end{equation}
Using equation~\eqref{eq:Jp}, the kappa symmetry matrix may be written in notation that treats type IIA and type IIB supergravities simultaneously~\cite{Bergshoeff:1997kr},
\begin{equation} \label{eq:kappa_matrix_general}
    \G = \frac{-i}{\sqrt{\lvert \det g \rvert}} \g_{01 \cdots q} J_{(q)}.
\end{equation}

\subsection{Class 1}
\label{sec:class1_kappa}

We now determine the supersymmetry preserved by class 1 embeddings. For class 1 embeddings, as described in section~\ref{sec:class1_embedding}, the D\(q\)-branes span \(t\), two of the spatial directions parallel to the D\(p\)-branes parameterised by a complex coordinate \(z\), a further \((a-3)\) directions \(\vec{x}\) parallel to the D\(p\)-branes, and \((q + 1 - a)\) directions \(\vec{v}\) orthogonal to the D\(p\)-branes. Using these directions as the worldvolume coordinates \(\xi\), we will use the following indices to refer to the different components of \(\xi\),
\begin{equation} \label{eq:class1_xi_indices}
    \xi^0 = t \, , \quad
    \xi^1 = z \, , \quad
    \xi^2 = \zb \, , \quad
    \xi^\a = x^{(\a-2)} \, \quad
    \xi^\ell = v^{(\ell+1-a)} \, ,
\end{equation}
where \(\a\) runs from \(3\) to \(a-1\) and \(\ell\) runs from \(a\) to \(q\). The remaining coordinates \((y,\yb,\vec{U},\vec{W})\) act as worldvolume scalars on the D\(q\)-branes. As in section~\ref{sec:class1_embedding}, we make the ansatz that \(y  = y (z,\zb)\) with \(\vec{U}\) and \(\vec{W}\) constant. We can then choose ten-dimensional vielbeins such that the Dirac matrices \(\g_m\) on the worldvolume of the D\(q\)-branes are
\begin{subequations} \label{eq:class1_dirac_matrices}
\begin{align}
    \g_0 &= h^{-1} \G_0 \, ,
    \\
    \g_1 &= \frac{1}{2h} (\G_1 - i \G_2) + \frac{h}{2} \le[\p y \, (\G_8 - i \G_9) + \p \yb \, (\G_8 + i \G_9) \ri] \, ,
    \\
    \g_2 &= \frac{1}{2h} (\G_1 + i \G_2) + \frac{h}{2} \le[\pb y \, (\G_8 - i \G_9) + \pb \yb \, (\G_8 + i \G_9) \ri] \, ,
    \\
    \g_\a &= h^{-1} \G_\a \, ,
    \\
    \g_\ell &= h \G_\ell \, ,
\end{align}
\end{subequations}
where we have introduced the convenient notation
\begin{equation}
    h(r) \equiv H(r)^{1/4} = \le[1 + \le(\frac{L}{r} \ri)^{7-p} \ri]^{1/4} \, .
\end{equation}
In equation~\eqref{eq:class1_dirac_matrices} we have used \(\G_1\) and \(\G_2\) to denote the ten-dimensional flat space Dirac matrices corresponding to the directions forming the real and imaginary parts of the complex coordinate \(z\), and \(\G_8\) and \(\G_9\) to denote those corresponding to the real and imaginary parts of \(y\).

Since the \(\g_m\) in equation~\eqref{eq:class1_dirac_matrices} satisfy the Dirac algebra \(\{\g_m,\g_n\} = 2 g_{mn}\), with \(g_{mn}\) the metric in equation~\eqref{eq:class1_metric}, the only non-zero anticommutator between \(\g_m\) with different indices is
\begin{equation}
    \{\g_1, \g_2\} = 2 g_{12} = \frac{1}{h^2} + h^2 \le(|\p y|^2 + |\pb y|^2\ri) .
\end{equation}
Consequently, we can anticommute \(\g_{12}\) through the other Dirac matrices appearing in the kappa symmetry matrix~\eqref{eq:kappa_matrix_general} to find
\begin{equation} \label{eq:class1_gamma_product_decomposition}
    \g_{01\cdots q} =  \g_{034\cdots q} \g_{12} \, .
\end{equation}
Using equation~\eqref{eq:class1_dirac_matrices}, the first of the two products on the right-hand side is
\begin{equation} \label{eq:class1_gamma_sub_product_1}
    \g_{034 \cdots q} = h^{-(a-2)} h^{q + 1 -a} \G_{034 \cdots q} = h^{q + 3 - 2a} \G_{034 \cdots q} \, .
\end{equation}
The second product is
\begin{equation}\begin{aligned} \label{eq:class1_gamma_sub_product_2}
    \g_{12} &= \frac{i}{2h^2}\G_{12} + \frac{i}{2 h^2} \cY_4 \G_{89} - \frac{\p y - \pb \yb}{4} \le( \G_{18} + \G_{29} \ri)
    \\ &\phantom{=}
    + \frac{\pb y - \p \yb}{4} \le(\G_{18} - \G_{29}\ri)
    + i \frac{\p y + \pb \yb}{4} \le( \G_{19} - \G_{28} \ri)
    - i \frac{\pb y + \p \yb}{4} \le( \G_{19} + \G_{28} \ri),
\end{aligned}\end{equation}
where
\begin{equation} \label{eq:Y4}
    \cY_4 = H(r) \le(|\p y|^2 - |\pb y|^2 \ri),
\end{equation}
is the central charge density appearing in equation~\eqref{eq:class1_Y}, evaluated for \(n=4\).

Substituting equations~\eqref{eq:class1_gamma_sub_product_1} and~\eqref{eq:class1_gamma_sub_product_2} into equation~\eqref{eq:class1_gamma_product_decomposition}, we find that the product \(\g_{01\cdots q}\) appearing in the kappa symmetry matrix is given by
\begin{equation} \label{eq:class1_matrix_product}
\begin{aligned} 
    h^{2a-q-3}\g_{01 \cdots q} &= \frac{i}{2h^2} \le( \G_{01 \cdots q} + \cY_4 \, \G_{034 \cdots q 89} \ri)
    \\ &\phantom{=}
    - \frac{1}{4}\Bigl[ \le(\p y - \pb \yb \ri) \le(\G_{19} - \G_{28} \ri)
    + i \le(\p y + \pb\yb\ri) \le(\G_{18} + \G_{29} \ri)
    \\ &\phantom{=-\frac{1}{4}\Bigl(}
    + \le(\pb y - \p \yb\ri) \le(\G_{19} + \G_{28}\ri) \G_{01 \cdots q}
    + i \le(\pb y + \p \yb\ri) \le(\G_{18} - \G_{29}\ri) \Bigr]\G_{01 \cdots q} \, .
\end{aligned}
\end{equation}
In writing this expression we have made use of the Clifford algebra~\eqref{eq:clifford_10d} satisfied by the \(\G_A\) to rearrange some products, for example \(\G_{034\cdots q} \G_{12} = \G_{01 \cdots q}\) and \(\G_{034 \cdots q} \G_{18} = \G_{28} \G_{01 \cdots q}\)\,. The other ingredient in the kappa symmetry matrix~\eqref{eq:kappa_matrix_general} is the determinant of the induced metric on the D\(q\)-brane world volume, given in equation~\eqref{eq:class1_metric_determinant}. It will be convenient to factorise the determinant as
\begin{equation} \label{eq:class1_metric_determinant_factorised}
    \lvert \det g \rvert= \frac{h^{2q+2-4a}}{4} \D_1 \, ,
    \qquad
    \D_1 \equiv \le[1 + H(r) (|\p y|^2 + |\pb y|^2)\ri]^2 - 4 H(r)^2 |\p y|^2 |\pb y|^2 \, ,
\end{equation}
where the subscript on \(\D_1\) denotes class 1.

Substituting equations~\eqref{eq:class1_matrix_product} and ~\eqref{eq:class1_metric_determinant_factorised} into equation~\eqref{eq:kappa_matrix_general}, we find that the kappa symmetry matrix for class 1 embeddings may be written as
\begin{subequations} \label{eq:class1_kappa_projector}
\begin{equation}\label{eq:class1_kappa_projector_sum}
    \G = \G' + \G'' \, , 
\end{equation}
where we have defined
\begin{align} 
    \G' &= \frac{1}{\sqrt{\D_1}}
        \le(\G_{01 \cdots q} + \cY_4 \, \G_{034 \cdots q 89} \ri) J_{(q)}
    \label{eq:class1_gamma_prime}
    \\ \phantom{=} 
    \G'' &= \frac{\sqrt{H(r)}}{2 \sqrt{\D_1}}\Bigl[
        i\le(\p y - \pb \yb\ri) \le(\G_{19} - \G_{28} \ri) 
        - \le(\p y + \pb\yb \ri) \le(\G_{18} + \G_{29} \ri)
    \nonumber \\ &\phantom{=+ \frac{\sqrt{H(r)}}{2}\Bigl[} 
        + i \le(\pb y - \p \yb\ri) \le(\G_{19} + \G_{28}\ri) 
        -\le(\pb y + \p \yb\ri) \le(\G_{18} - \G_{29}\ri)
    \Bigr]\G_{01 \cdots q} J_{(q)} \, .
    \label{eq:class1_gamma_double_prime}
\end{align}
\end{subequations}
For arbitrary \(y(z,\zb)\) it is not possible to find a constant spinor \(\ve\) satisfying the kappa symmetry condition \(\G\ve = \ve\) with \(\G\) as given in equation~\eqref{eq:class1_kappa_projector}. This is because the different terms in equation~\eqref{eq:class1_kappa_projector} depend non-trivially on \((z,\zb)\), as well as an \(\vec{v}\) through their dependence on \(r\). However, when \(y\) is either a holomorphic or antiholomorphic function of \(z\) it \emph{is} possible to find solutions to the kappa symmetry condition, as we will now show.

The key is that when \(y\) is a holomorphic or antiholomorphic function of \(z\), the factor \(\D_1\) in equation~\eqref{eq:class1_metric_determinant_factorised} satisfies \(\sqrt{\D_1} = 1 \pm \cY_4\), with the plus sign for holomorphic \(y\) and the minus sign for antiholomorphic \(y\). Then, a constant spinor \(\ve\) will obey \(\G' \ve = \ve\) if it satisfies the two conditions
\begin{subequations}\label{eq:parallel_Dq_brane_kappa_combined}
\begin{align}
    \G_{01 \cdots q} J_{(q)} \ve &= \ve \, ,
    \label{eq:parallel_Dq_brane_kappa_1}
    \\
    \G_{034 \dots q 89} J_{(q)} \ve &= \pm \ve \, ,
    \label{eq:parallel_Dq_brane_kappa_2}
\end{align}
\end{subequations}
where the plus or minus sign in equation~\eqref{eq:parallel_Dq_brane_kappa_2} are for holomorphic and antiholomorphic \(y\), respectively. With \(\G = \G' + \G''\), a spinor satisfying \(\G' \ve = \ve\) will satisfy the kappa symmetry condition \(\G \ve = \ve\) if it also obeys \(\G'' \ve = 0\), which occurs if all four of the following conditions hold
\begin{subequations}\label{eq:parallel_Dq_brane_kappa_combined_2}
\begin{align}
    \le(\p y - \pb \yb\ri) \le(\G_{19} - \G_{28} \ri) J_{(q)} \ve &= 0 \,,
    &
    \le(\p y + \pb\yb \ri) \le(\G_{18} + \G_{29} \ri) J_{(q)} \ve &= 0\,,
    \label{eq:parallel_Dq_brane_kappa_3}
    \\
    \le(\pb y - \p \yb\ri) \le(\G_{19} + \G_{28}\ri) J_{(q)} \ve &= 0 \,,
    &
    \le(\pb y + \p \yb\ri) \le(\G_{18} - \G_{29}\ri) J_{(q)} \ve &= 0\,.
    \label{eq:parallel_Dq_brane_kappa_4}
\end{align}
\end{subequations}

We therefore need to determine whether equations~\eqref{eq:parallel_Dq_brane_kappa_combined} and~\eqref{eq:parallel_Dq_brane_kappa_combined_2} can be satisfied simultaneously. Notice that the left-hand sides of equations~\eqref{eq:parallel_Dq_brane_kappa_3} and~\eqref{eq:parallel_Dq_brane_kappa_4} vanish automatically for antiholomorphic and holomorphic \(y\), respectively. Thus, for holomorphic or antiholomorphic solutions, the requirement on \(\ve\) following from equation~\eqref{eq:parallel_Dq_brane_kappa_combined_2} is that
\begin{equation}\label{eq:parallel_Dq_brane_kappa_5}
    \le(\G_{19} \mp \G_{28} \ri) J_{(q)} \ve = 0 \,,
    \qquad
    \le(\G_{18} \pm \G_{29} \ri) J_{(q)} \ve = 0\,,
\end{equation}
where the signs in equations~\eqref{eq:parallel_Dq_brane_kappa_combined} and~\eqref{eq:parallel_Dq_brane_kappa_5} are correlated, i.e. the upper signs are for holomorphic \(y\) and the lower signs for antiholomorphic \(y\). In fact, the two conditions in equation~\eqref{eq:parallel_Dq_brane_kappa_5} are equivalent to each other, since the Clifford algebra implies that \(\G_{18} \pm \G_{29} = \G_{89} \le(\G_{19} \mp \G_{28} \ri)\). Moreover, the left-hand side of the second condition in equation~\eqref{eq:parallel_Dq_brane_kappa_5} may be rewritten using the Clifford algebra as
\begin{equation}
    \le(\G_{18} \pm \G_{29}\ri) J_{(q)}\ve
    = (-1)^{\lfloor\frac{3q-1}{2}\rfloor} \G_0 \G_{23 \cdots q} \G_8 \le( \G_{01 \cdots q} \mp \G_{034 \cdots q 89}  \ri) J_{(q)} \ve \, ,
\end{equation}
where \(\lfloor \frac{3q-1}{2} \rfloor\) denotes the integer part of \(\frac{3q-1}{2}\). The right-hand side of this expression vanishes for any \(\ve\) satisfying equation~\eqref{eq:parallel_Dq_brane_kappa_combined}, so any such \(\ve\) satisfies equation~\eqref{eq:parallel_Dq_brane_kappa_5}.

It is therefore sufficient to consider only the conditions in equation~\eqref{eq:parallel_Dq_brane_kappa_combined}. We need to know when these conditions are compatible with equation~\eqref{eq:Dp_brane_projection_conditions_unified} coming from the supergravity background. Since the D\(q\)-branes span \(a\) of the \(x_\parallel^\m\) directions, and therefore there are \((p+1-a)\) of the \(x_\parallel^\m\) directions orthogonal to the D\(q\)-branes, equation~\eqref{eq:Dp_brane_projection_conditions_unified} may be written as
\begin{equation}\label{eq:parallel_Dq_brane_kappa_Dp}
    \G_{01 \cdots (a-1)} \G_{(q+1)(q+2) \cdots (q + p + 1 - a)} J_{(p)} \ve = \ve \, .
\end{equation}
We need to know when this condition is compatible with equation~\eqref{eq:parallel_Dq_brane_kappa_combined}.

First, note that equation~\eqref{eq:parallel_Dq_brane_kappa_1} is the kappa symmetry condition for a flat D\(q\)-brane along the directions \((0,1,\cdots,q)\) while equation~\eqref{eq:parallel_Dq_brane_kappa_Dp} is the kappa symmetry condition for a flat D\(p\)-brane, both in Minkowski space. They are compatible if the number of ND directions between these branes, which is the number we have been denoting by \(d\), is a multiple of four~\cite{Polchinski:1998rr,Skenderis:2002vf}. Similarly, the condition in equation~\eqref{eq:parallel_Dq_brane_kappa_2} is the kappa symmetry condition for a flat D\(q\)-brane in Minkowski space along the directions \((0,3,4,\cdots,q,8,9)\). Such a D\(q\)-brane has 4 ND directions, \((1,2,8,9)\), relative to the D\(q\)-brane giving rise to equation~\eqref{eq:parallel_Dq_brane_kappa_1}, and \((d+4)\) ND directions relative to the D\(p\)-brane giving rise to equation~\eqref{eq:parallel_Dq_brane_kappa_Dp}.\footnote{The four extra ND directions are again \((1,2,8,9)\): the D\(p\)-branes span \((1,2)\) since \(a\geq 3\) and, relatedly, since the complex coordinates \((z,\zb)\) are formed from \(x_\parallel^\m\) directions for class 1 embeddings, see table~\ref{tab:embedding_classification}. The D\(p\)-branes do not span \((8,9)\) since these are the \(x_\perp^i\) directions used to form the complex coordinates \((y,\yb)\).} Thus, if \(d\) is a multiple of four, all of the conditions in equations~\eqref{eq:parallel_Dq_brane_kappa_combined} and~\eqref{eq:parallel_Dq_brane_kappa_Dp} are compatible, in which case holomorphic or antiholomorphic \(y\) preserves a fraction of the supersymmetry of the D\(p\)-brane background.

The fraction of preserved supersymmetry depends on \(d\). As shown in section~\ref{sec:class1_embedding}, class 1 holomorphic embeddings can exist for \(d=0\) or \(d=4\), while \(d=8\) is incompatible with the ansatz for class 1 embeddings. Obtaining \(d=0\) is only possible for \(p=q=a-1\), in which case the conditions in equations~\eqref{eq:parallel_Dq_brane_kappa_1} and~\eqref{eq:parallel_Dq_brane_kappa_Dp} are identical. Thus the only non-trivial kappa symmetry condition for a \(d=0\) holomorphic embedding is equation~\eqref{eq:parallel_Dq_brane_kappa_2}. This condition reduces the number of independent components of \(\ve\) by one-half. Thus, \(d=0\) holomorphic embeddings preserve one-half of the supersymmetry of the D\(p\)-brane background. For \(d=4\), both of the conditions in equation~\eqref{eq:parallel_Dq_brane_kappa_combined} are non-trivial. Since these conditions are independent from each other, and each reduces the number of independent components of \(\ve\) by one-half, in total \(d=4\) holomorphic embeddings preserve only one-quarter of the supersymmetry of the D\(p\)-brane background. These conclusions may be checked explicitly for each case in table~\ref{tab:class1_holomorphic} by choosing a basis for the Dirac matrices. We have done so using the ``really real'' basis given in ref.~\cite{Freedman_Van_Proeyen_2012}.

Since class 1\('\) and class 1 are related by a reparameterisation of the D\(q\)-branes in the sense described in section~\ref{sec:class1_embedding}, the kappa symmetry analysis described in this section also applies to class 1\('\): a class 1\('\) embedding with \(y\) a holomorphic or antiholomorphic function of \(z\) will preserve a fraction of the supersymmetry of the D\(p\)-brane background. Since under the reparameterisation that takes class 1 to 1\('\) the number of ND directions changes as \(d \to d + 4\), a \(d=4\) class 1\('\) embedding preserves one-half of the supersymmetries of the D\(p\)-brane background, while a \(d=8\) class 1\('\) embedding preserves one-quarter.

\subsection{Class 2}
\label{sec:class2_kappa}

We now check the kappa symmetry of class 2 embeddings. This proceeds almost identically to that for class 1 embeddings, so we will be brief. Recall from section~\ref{sec:class2_embedding} that for class 2 embeddings the D\(q\)-branes span \(t\), a further \((a-1)\) directions \(\vec{x}\) parallel to the D\(p\)-branes, two of the spatial directions \(z\) and \(\zb\) orthogonal to the D\(p\)-branes, and \((q - 1 - a)\) directions \(\vec{v}\) orthogonal to the D\(p\)-branes. Using these directions as the worldvolume coordinates \(\xi\), in this section we will use the following indices to refer to the different components of \(\xi\),
\begin{equation}
    \xi^0 = t \, , \quad
    \xi^1 = z \, , \quad
    \xi^2 = \zb \, , \quad
    \xi^\a = x^{(\a-2)} \, \quad
    \xi^\ell = v^{(\ell-1-a)} \, ,
\end{equation}
with \(\a\) running from 3 to \(a+1\) and \(\ell\) running from \(a+2\) to \(q\). We choose vielbeins such that the curved space Dirac matrices on the worldvolume of the D\(q\)-branes are
\begin{subequations} \label{eq:class2_dirac_matrices}
\begin{align}
    \g_0 &= h^{-1}\G_0 \, ,
    \\
    \g_1 & = \frac{h}{2}(\G_1 - i \G_2) + \frac{h}{2}[\p y \, (\G_8 - i \G_9) + \p\yb \, (\G_8 + i \G_9)] \, ,
    \\
    \g_2 & = \frac{h}{2}(\G_1 + i \G_2) + \frac{h}{2}[\pb y \, (\G_8 - i \G_9) + \pb\yb \, (\G_8 + i \G_9)] \, ,
    \\
    \g_\a &= h^{-1}\G_\a \, ,
    \\
    \g_\ell &= h\G_\ell \, .
\end{align}
\end{subequations}
As in the class 1 case in  equation~\eqref{eq:class1_dirac_matrices}, we have used \((\G_1,\G_2)\) and \((\G_8,\G_9)\) to denote the ten-dimensional flat space Dirac matrices corresponding to the real and imaginary parts of \(z\) and \(y\), respectively. 

There are two differences between the \(\g_m\) for class 1 and class 2 embeddings, in  equations~\eqref{eq:class1_dirac_matrices} and~\eqref{eq:class2_dirac_matrices} respectively: the ranges of the \(\a\) and \(\ell\) indices, and the prefactors of \(\G_1\) and \(\G_2\), which are proportional to \(h^{-1}\) in equation~\eqref{eq:class1_dirac_matrices} and to \(h\) in equation~\eqref{eq:class2_dirac_matrices}. Performing the same manipulations as led to equation~\eqref{eq:class1_matrix_product}, accounting for these differences, one finds that for class 2 embeddings the antisymmetric product of Dirac matrices appearing in the kappa symmetry matrix is given by
\begin{equation} \label{eq:class2_matrix_product}
\begin{aligned} 
    h^{2a-q-1}\g_{01 \cdots q} &= \frac{i}{2} \le( \G_{01 \cdots q} + \cY_0 \, \G_{034 \cdots q 89} \ri)
    \\ &\phantom{=}
    - \frac{1}{4} \Bigl[ \le(\p y - \pb \yb\ri) \le(\G_{19} - \G_{28} \ri)
    + i \le(\p y + \pb\yb\ri) \le(\G_{18} + \G_{29} \ri)
    \\ &\phantom{= - \frac{1}{4} \Bigl[}
    + \le(\pb y - \p \yb\ri) \le(\G_{19} + \G_{28}\ri)
    + i \le(\pb y + \p \yb\ri) \le(\G_{18} - \G_{29}\ri) \Bigr] \G_{01 \cdots q} \, ,
\end{aligned}
\end{equation}
where
\begin{equation}
    \cY_0 =|\p y|^2 - |\pb y|^2\, ,
\end{equation}
is the central charge density appearing in equation~\eqref{eq:class1_Y}, evaluated for \(n=0\). The determinant of the induced metric on the D\(q\)-branes is written in equation~\eqref{eq:class2_metric_determinant}. It will again be convenient to factorise this determinant, this time as
\begin{equation} \label{eq:class2_metric_determinant_factorised}
    \lvert \det g \rvert= \frac{h^{2q+2-4a}}{4} \D_2 \, ,
    \qquad
    \D_2 \equiv \le(1 + |\p y|^2 + |\pb y|^2\ri)^2 - 4 |\p y|^2 |\pb y|^2 \, ,
\end{equation}
where the subscript on \(\D_2\) denotes class 2.

Substituting equations~\eqref{eq:class2_matrix_product} and~\eqref{eq:class2_metric_determinant_factorised} into the kappa symmetry matrix~\eqref{eq:kappa_matrix_general}, we find that it takes the form
\begin{subequations} \label{eq:class2_kappa_projector}
\begin{align}\label{eq:class2_kappa_projector_sum}
    \G &= \G' + \G'' \, , 
    \\
    \G' &= \frac{1}{\sqrt{\D_2}}
        \le(\G_{01 \cdots q} + \cY_0 \, \G_{034 \cdots q 89} \ri) J_{(q)}
    \\ \phantom{=} 
    \G'' &= \frac{1}{2 \sqrt{\D_2}}\Bigl[
        i\le(\p y - \pb \yb\ri) \le(\G_{19} - \G_{28} \ri) 
        - \le(\p y + \pb\yb \ri) \le(\G_{18} + \G_{29} \ri)
    \nonumber \\ &\phantom{=+ \frac{\sqrt{H(r)}}{2}\Bigl[} 
        + i \le(\pb y - \p \yb\ri) \le(\G_{19} + \G_{28}\ri) 
        -\le(\pb y + \p \yb\ri) \le(\G_{18} - \G_{29}\ri)
    \Bigr]\G_{01 \cdots q} J_{(q)} \, .
\end{align}
\end{subequations}
When \(y\) is a holomorphic or antiholomorphic function of \(z\) we have that \(\sqrt{\D_2} = 1 + \cY_0\) or \(\sqrt{\D_2} = 1 - \cY_0\), respectively. The same reasoning as used in section~\ref{sec:class1_kappa} then implies that we can find constant spinors \(\ve\) satisfying \(\G' \ve = \ve\) and \(\G'' \ve = 0\), and hence satisfying the kappa symmetry condition. These spinors are precisely those \(\ve\) satisfying the conditions in equation~\eqref{eq:parallel_Dq_brane_kappa_combined}.

Consequently, the kappa symmetry condition is again only compatible with equation~\eqref{eq:Dp_brane_projection_conditions_unified} obeyed by the Killing spinors of the D\(p\)-brane background when the number of ND directions \(d\) is a multiple of four. As explained in section~\ref{sec:class2_embedding}, the class 2 ansatz requires \(2 \leq d \leq 6\), so that the only possibility consistent with supersymmetry is \(d=4\), which is also the value of \(d\) for which the equations of motion of the D\(q\)-brane admit holomorphic embeddings. Each of the two conditions in equation~\eqref{eq:parallel_Dq_brane_kappa_combined} reduce the number of independent components of \(\ve\) by one-half, so that every class 2 holomorphic embedding preserves one-quarter of the supersymmetry of the D\(p\)-brane background.

\subsection{Class 3}
\label{sec:class3_kappa}

Finally, we check the kappa symmetry of class 3 embeddings, which again proceed similarly. Recall from section~\ref{sec:class3_embedding} that for class 3 embeddings the D\(q\)-branes span \(t\), the complex directions \(z\) and \(\zb\) parallel to the D\(p\)-branes, a further \((a-3)\) directions \(\vec{x}\) parallel to the D\(p\)-branes, and \((q + 1 - a)\) directions \(\vec{v}\) orthogonal to the D\(p\)-branes. Using these directions as the worldvolume coordinates \(\xi\) and indexing \(\xi\) as in equation~\eqref{eq:class1_xi_indices}, we can take the worldvolume Dirac matrices \(\g_m\) to be
\begin{subequations} \label{eq:class3_dirac_matrices}
\begin{align}
    \g_0 &= h^{-1}\G_0 \, ,
    \\
    \g_1 & = \frac{1}{2h}(\G_1 - i \G_2) + \frac{1}{2h}[\p y \, (\G_8 - i \G_9) + \p\yb \, (\G_8 + i \G_9)] \, ,
    \\
    \g_2 & = \frac{1}{2h}(\G_1 + i \G_2) + \frac{1}{2h}[\pb y \, (\G_8 - i \G_9) + \pb\yb \, (\G_8 + i \G_9)] \, ,
    \\
    \g_\a &= h^{-1}\G_\a \, ,
    \\
    \g_\ell &= h\G_\ell \, .
\end{align}
\end{subequations}
Once more, we use \((\G_1,\G_2)\) and \((\G_8,\G_9)\) as the ten-dimensional flat space Dirac matrices corresponding to the directions forming the real and imaginary parts of \(z\) and \(y\), respectively.

The only difference between the Dirac matrices in equation~\eqref{eq:class3_dirac_matrices} and those for the class 1 case in equation~\eqref{eq:class1_dirac_matrices} are the prefactors of terms involving \(\G_8\) or \(\G_9\), which are proportional to \(h\) in the class 1 case and \(h^{-1}\) in the class 2 case. We can therefore immediately obtain the antisymmetric product of Dirac matrices appearing in the kappa symmetry matrix by making the appropriate adjustments to equation~\eqref{eq:class1_matrix_product}, which results in
\begin{equation} \label{eq:class3_matrix_product}
\begin{aligned} 
    h^{2a-q-1}\g_{01 \cdots q} &= \frac{i}{2} \le( \G_{01 \cdots q} + \cY_0 \, \G_{034 \cdots q 89} \ri)
    \\ &\phantom{=}
    - \frac{1}{4} \Bigl[ \le(\p y - \pb \yb \ri) \le(\G_{19} - \G_{28} \ri)
    + i \le(\p y + \pb\yb\ri) \le(\G_{18} + \G_{29} \ri) 
    \\ &\phantom{= - \frac{1}{4} \Bigl[}
    + \le(\pb y - \p \yb \ri) \le(\G_{19} + \G_{28}\ri) 
    + i \le(\pb y + \p \yb\ri) \le(\G_{18} - \G_{29}\ri) \Bigr]\G_{01 \cdots q} \, .
\end{aligned}
\end{equation}
We factorise the determinant of the metric induced on the D\(q\)-branes, given in equation~\eqref{eq:class3_metric_determinant}, in a similar manner to before
\begin{equation} \label{eq:class3_metric_determinant_factorised}
    \lvert \det g \rvert= \frac{h^{2q+2-4a}}{4} \D_2 \, ,
\end{equation}
with \(\D_2\) as in equation~\eqref{eq:class2_metric_determinant}.

Substituting equations~\eqref{eq:class3_matrix_product} and~\eqref{eq:class3_metric_determinant_factorised} into equation~\eqref{eq:kappa_matrix_general}, we find that the kappa symmetry matrix for class 3 embeddings takes the same form as for class 2 embeddings written in equation~\eqref{eq:class2_kappa_projector}. The same reasoning as in section~\ref{sec:class2_kappa} therefore implies that when \(y\) is a holomorphic or antiholomorphic function of \(z\), a class 3 embedding preserves one-quarter of the supersymmetry of the D\(p\)-brane background when the number \(d\) of ND directions is a multiple of four, but not for other values of \(d\). As explained in section~\ref{sec:class3_embedding}, the values of \(d\) consistent with the ansatz for a class 3 embedding satisfy \(2 \leq d \leq 6\). Thus, for class 3 embeddings the only value of \(d\) that preserves any supersymmetry is \(d=4\). This is true despite the fact that holomorphic or antiholomorphic \(y\) solves the equations of motion for class 3 embeddings for any value of \(d\).

\section{Class 1 embeddings in \texorpdfstring{\(\ads[5] \times \sph[5]\)}{AdS5 x S5} and holography}
\label{sec:D3_background}

In the near-horizon limit \(r \ll L\), the extremal black D\(3\)-brane background becomes \(\ads[5] \times \sph[5]\), which has metric and \(C_4\) which may be written as
\begin{equation}\begin{aligned} \label{eq:ads5_x_s5}
    \diff s^2 &= \frac{r^2}{L^2} \h_{\m\n} \diff x_\parallel^\m \diff x_\parallel^\n + \frac{L^2}{r^2} \d_{ij} \diff x_\perp^i \diff x_\perp^j \, ,
    \\
    C_4 &= \frac{r^4}{L^4} \diff x_\parallel^0 \wedge \diff x_\parallel^1 \wedge \diff x_\parallel^2 \wedge \diff x_\parallel^3 + \cdots \; ,
\end{aligned}\end{equation}
with \(r^2 = \d_{ij} x_\perp^i x_\perp^j\), and the dots denote additional terms in \(C_4\) with legs in the \(x_\perp^i\) directions, needed to make \(F_5 = \diff C_4\) self-dual. The dilaton is constant \(e^\f = g_s\). Equation~\eqref{eq:ads5_x_s5} is obtained from the extremal black D\(p\)-brane background~\eqref{eq:Dp_brane_background} by setting \(p=3\) and \(H(r) = L^4/r^4\), except for two modifications of \(C_4\): the introduction of the terms required for self-duality of \(F_5\), and a gauge transformation that shifts the coefficient of \(\diff x_\parallel^0 \wedge \diff x_\parallel^1 \wedge \diff x_\parallel^2 \wedge \diff x_\parallel^3\) by a constant so that it vanishes at \(r=0\).

Type IIB supergravity in \(\ads[5] \times \sph[5]\) is holographically dual to four-dimensional \(\cN=4\) SYM with gauge group \(\SU(N)\) and gauge coupling \(g_\mathrm{YM}\), in the limit of large \(N\) followed by large `t Hooft coupling \(\la = g_\mathrm{YM}^2 N\)~\cite{Maldacena:1997re,Witten:1998qj,Gubser:1998bc}. The rank \(N\) is related to \(L\) as in equation~\eqref{eq:curvature_radius} and the gauge coupling is determined by the string coupling through \(g_\mathrm{YM}^2 = 4\pi g_s\). Embedding probe D-branes into \(\ads[5] \times \sph[5]\) typically corresponds to deforming \(\cN=4\) in some way. For example, introducing probe D7-branes that span \ads[5]\ and wrap an \(\sph[3] \subset \sph[5]\) corresponds to coupling \(\cN=4\) SYM to four-dimensional \(\cN=2\) hypermultiplets~\cite{Karch:2002sh}. Ref.~\cite{holomorphic_branes} analysed the holography of class 1 holomorphic D7-branes in \(\ads[5] \times \sph[5]\) in detail, arguing that they are holographically dual to \(\cN=2\) hypermultiplets with a mass that depends holomorphically on position, as we review in section~\ref{sec:d7_review}.

We will extend the analysis of ref.~\cite{holomorphic_branes} by studying the holographic duals of the other two class 1 D-brane embeddings in the near-horizon limit of the D3-brane background, listed in table~\ref{tab:class1_holomorphic}, namely \(d=4\) D5-branes and \(d=0\) D3-branes. The D5-branes are discussed in section~\ref{sec:d5} and the D3-branes in section~\ref{sec:d3}.

We specialise to class 1 embeddings since, as we will see, their holographic duals have relatively simple interpretations in terms of position-dependent sources or states. The holographic duals of class 2 embeddings, which depend on directions orthogonal to the D\(3\)-branes sourcing the background, are more intricate, while there are no class 3 embeddings in the D3-brane background as there are not enough \(x_\parallel^\m\) directions to make the class 3 ansatz. Throughout this section we will specialise to holomorphic embeddings with \(y = y(z)\) for simplicity of discussion, commenting on the differences with the antiholomorphic case \(y = y(\zb)\) where appropriate.

In the near-horizon limit, the D3-brane geometry has 16 further supercharges in addition to those discussed in section~\ref{sec:susy}~\cite{Maldacena:1997re}. These additional supercharges are dual to the superconformal symmetries of the dual \(\cN=4\) SYM theory. For the holomorphic embeddings that we discuss, any non-zero \(y\) introduces at least one dimensionful scale and thus breaks superconformal symmetry. We will therefore neglect these additional supercharges in our discussion.

\subsection{Review: D7-branes}
\label{sec:d7_review}

Ref.~\cite{holomorphic_branes} studied the holographic dual of class 1 holomorphic D7-branes in detail, and we will briefly summarise some of their findings. As can be seen in table~\ref{tab:class1_holomorphic}, the class 1 holomorphic D7-branes span \(a=4\) of the \(x_\parallel^\m\) directions, so from equation~\eqref{eq:class1_dimensions} we find that there is a single \(\vec{x}\) direction, four \(\vec{v}\) directions, and no \(\vec{U}\) or \(\vec{W}\) directions. Thus, decomposing \(\h_{\m\n} \diff x_\parallel^\m \diff x_\parallel^\n\) and \(\d_{ij} \diff x_\perp^i \diff x_\perp^j\) as in equation~\eqref{eq:metric_component_decomposition}, the \(\ads[5] \times \sph[5]\) metric in equation~\eqref{eq:ads5_x_s5} becomes
\begin{equation}
    \diff s^2 = \frac{r^2}{L^2} \le(-\diff t^2 + \diff z \diff \zb + \diff x^2 \ri) + \frac{L^2}{r^2} \le(\diff y \diff \yb + \diff \vec{v}^{\,2}\ri) ,
\end{equation}
with \(r^2 = |y|^2 + v^2\), where \(x\) is the single component of \(\vec{x}\). As usual, we think of the \(x_\parallel^\m\) coordinates, which in this case are \((t,z,\zb,x)\), as the coordinates in the dual \(\cN=4\) SYM theory.

The introduction of \(k\) D7-branes that span \(\xi = (t,z,\zb,x,\vec{v})\) is holographically dual to coupling \(\cN=4\) SYM to \(k\) four-dimensional \(\cN=4\) hypermultiplets~\cite{Karch:2002sh}. The embedding of the D7-branes is specified by how the remaining directions \((y,\yb)\) depend on \(\xi\). When \(y\) is non-zero, the dual hypermultiplets have a complex mass \(m\) which, in a weak coupling description, is equal to the minimum energy of strings stretched between the D7-branes and the D3-branes sourcing the background. This in turn is equal to the separation between the D3- and D7-branes multiplied by the string tension, so that the hypermultiplets have mass~\cite{Karch:2002sh}
\begin{equation} \label{eq:mass_y}
    m = \frac{y}{2\pi \a'} \, .
\end{equation}
Thus, holomorphic embeddings with \(y = y(z)\) are dual to hypermultiplets with a mass that depends holomorphically on position in the dual QFT.

When \(y\) is not constant, the position-dependent hypermultiplet mass explicitly breaks translational symmetry in the complex \(z\) plane. A holomorphic D7-brane embedding preserves one-quarter of the supersymmetries of the D3-brane background~\cite{holomorphic_branes}, consistent with the analysis in section~\ref{sec:class1_kappa} with \(p=3\) and \(q=7\). Correspondingly, a position-dependent hypermultiplet mass preserves one-quarter of the supersymmetries of \(\cN=4\) SYM, amounting to four supercharges. Ref.~\cite{holomorphic_branes} showed that the preserved supersymmetries all have the same two-dimensional chirality in the directions \((t,x)\) with unbroken translational symmetry, corresponding to two-dimensional \(\cN=(4,0)\) supersymmetry. For antiholomorphic \(y = y(\zb)\) the supercharges have opposite two-dimensional chirality, corresponding to two-dimensional \(\cN = (0,4)\) supersymmetry~\cite{holomorphic_branes}.

The index theorem of ref.~\cite{weinberg_1981} implies that if the holomorphic hypermultiplet mass \(m(z)\) has \(n\) zeros, then there are \(n k\) two-dimensional chiral fermion zero modes in the dual QFT. Ref.~\cite{holomorphic_branes} showed holographically that in the infrared (IR) these zero modes form the field content of the two-dimensional \(\cN=(8,0)\) defects holographically dual to D7-branes spanning \(\ads[3] \times \sph[5]\) of refs.~\cite{Harvey:2007ab,Buchbinder:2007ar,Harvey:2008zz}. The defects are located at the zeros of the mass. We will similarly show that for class 1 holomorphic D5- and D3-branes, zeros of the embedding function \(y(z)\) correspond in the IR to defects.

To obtain defects preserving two-dimensional \(\cN=(8,0)\) supersymmetry in the IR requires a low-energy enhancement of the \(\cN=(4,0)\) supersymmetry preserved by the holomorphic hypermultiplet mass, which ref.~\cite{holomorphic_branes} argued could be seen holographically as follows. The two kappa symmetry conditions in equation~\eqref{eq:parallel_Dq_brane_kappa_combined}, which each reduce the number of supersymmetries preserved by the embedding by one-half, follow from the condition \(\G' \ve = \ve\), with \(\G'\) given in equation~\eqref{eq:class1_gamma_prime}. For holomorphic or antiholomorphic \(y\), where \(\sqrt{\D_1} = 1 \pm \cY_4\), equation~\eqref{eq:class1_gamma_prime} becomes
\begin{equation} \label{eq:class1_gamma_prime_2}
    \G' = \frac{1}{1 \pm \cY_4} \le(\G_{01\cdots q} + \cY_4 \G_{034\cdots q 89}\ri) J_{(q)} \, ,
\end{equation}
while for \(p=3\) and in the near-horizon limit, \(\cY_4\) in equation~\eqref{eq:Y4} is
\begin{equation}
    \cY_4 = \frac{L^4}{r^4} \le(|\p y|^2 - |\pb y|^2 \ri).
\end{equation}
Since \(\cY_4\) diverges in the IR limit \(r \to 0\), the coefficient of \(\G_{01\cdots q}\) in equation~\eqref{eq:class1_gamma_prime_2} vanishes in that same limit. Meanwhile, the coefficient of \(\G_{034\cdots q 89}\) remains finite. Consequently, of the two kappa symmetry conditions in equation~\eqref{eq:parallel_Dq_brane_kappa_combined}, only the one in equation~\eqref{eq:parallel_Dq_brane_kappa_2} survives in the IR, leading to the doubling of supersymmetry at low energies.

This argument applies for any \(q \neq 3\) in equation~\eqref{eq:class1_gamma_prime_2}. The case in ref.~\cite{holomorphic_branes} corresponds to \(q = 7\). We will make use of the \(q=5\) case in the next subsection. For \(q=3\), the kappa symmetry condition in equation~\eqref{eq:parallel_Dq_brane_kappa_1}, \(\G_{0123} J_{(3)} \ve = \ve\) is satisfied by all of the Killing spinors of the \(\ads[5] \times \sph[5]\) background (this is the near-horizon limit equation~\eqref{eq:Dp_brane_projection_conditions_unified}), so the fact that the coefficient of \(\G_{0123}\) in \(\G'\) for \(q=3\) vanishes at \(r=0\) does not lead to supersymmetry enhancement at low energies. In other words, class 1 holomorphic embeddings of D3-branes preserve one-half of the Poincar\'e supersymmetries of \(\ads[5] \times \sph[5]\) for all \(r\), not just at \(r \to 0\).

\subsection{D5-branes}
\label{sec:d5}

From table~\ref{tab:class1_holomorphic} we see that the class 1 embeddings in the D3-brane background span \(a = 3\) of the \(x_\parallel^\m\) directions. Since \(p=3\) and \(q=5\), from equation~\eqref{eq:class1_dimensions} we see that there are no \(\vec{x}\) directions, a single \(\vec{U}\) direction, three \(\vec{v}\) directions, and one \(\vec{W}\) direction. In the notation used in section~\ref{sec:embeddings}, the \(\ads[5] \times \sph[5]\) metric in equation~\eqref{eq:ads5_x_s5} therefore becomes
\begin{equation} \label{eq:ad5_x_s5_d5}
    \diff s^2 = \frac{r^2}{L^2} \le(-\diff t^2 + \diff z \diff \zb + \diff U^2 \ri) + \frac{L^2}{r^2} \le(\diff y \diff \yb + \diff \vec{v}^{\,2} + \diff W^2\ri) ,
\end{equation}
where we use \(U\) and \(W\) to denote the single components of \(\vec{U}\) and \(\vec{W}\), respectively, while \(\diff \vec{v}^{\,2} = (\diff v_1)^2 + (\diff v_2)^2 + (\diff v_3)^2\). Further, \(r^2 = |y|^2 + v^2 + W^2\). We think of \((t,z,\zb,U)\) as the coordinates in the dual \(\cN=4\) SYM theory.

Class 1 holomorphic D5-brane embeddings in the background~\eqref{eq:ad5_x_s5_d5} span \(\xi = (t,z,\zb,\vec{v})\), and sit at constant \(W=0\) and constant \(U\). Using the symmetry of the background~\eqref{eq:ad5_x_s5_d5} under translations in the \(U\) direction, we will always take the D5-branes to be located at \(U=0\). The introduction of \(k\) D5-branes spanning these directions is holographically dual to coupling \(\cN=4\) SYM to \(k\) three-dimensional \(\cN=4\) hypermultiplets transforming in the fundamental representation of the gauge group, located on a codimension-one defect at \(U=0\)~\cite{Karch_2001a,DeWolfe:2001pq,Erdmenger_2002}. Similarly to the D7-brane case, the defect hypermultiplets have a mass \(m\) given by equation~\eqref{eq:mass_y}. Thus, for our holomorphic embeddings, with \(y\) a non-trivial function of \(z\) the mass of the defect hypermultiplets depends on position on the defect in a holomorphic manner.

From the analysis in section~\ref{sec:class1_kappa} we know that holomorphic D5-branes preserve one-quarter of the sixteen Poincar\'e supersymmetries of the \(\ads[5] \times \sph[5]\) background. This implies, via holography, that giving the defect hypermultiplets in the dual QFT a mass \(m(z)\) that depends holomorphically on \(z\) preserves four supercharges. That this is so can also be seen directly in the QFT. The action for three-dimensional hypermultiplets coupled to \(\cN=4\) SYM is given in ref.~\cite{DeWolfe:2001pq}, where it can be seen that a non-zero hypermultiplet mass arises from coupling the hypermultiplets to a non-zero vacuum expectation value (VEV) of the scalar component of a background four-dimensional \(\cN=2\) vector multiplet. Ref.~\cite{holomorphic_branes} showed that if the VEV of such a scalar field depends holomorphically on \(z\), then four supercharges are preserved.

Just as for the D7-branes, the index theorem of ref.~\cite{weinberg_1981} implies that if \(y(z)\), and therefore \(m(z)\), has \(n\) zeros (counted with their multiplicity), then there will be \(n k\) fermion zero modes. We expect these zero modes to be the degrees of freedom associated to the D5-branes that survive to the IR in the dual QFT, and it is natural to expect that the zero modes associated to a given zero of \(y(z)\) at some \(z = z_0\) will be localised to \(z_0\). In other words, at each zero of \(y(z)\) we expect to find a codimension-three defect in the IR, located at \(z = z_0\) and \(U = 0\).

We will argue holographically that this is indeed the case, and that the defect associated to each zero is a half-BPS Maldacena--Wilson line (hereafter referred to simply as a Wilson line) in the totally antisymmetric representation of \(\SU(N)\) with \(N/2\) indices. To do so, we examine the geometry of the worldvolume of the D\(q\)-branes in the region of \(\ads[5] \times \sph[5]\) at \(r \to 0\), holographically dual to the IR of \(\cN=4\) SYM. In the near-horizon limit \(r \ll L\), the induced metric on the D5-branes' worldvolume given in equation~\eqref{eq:class1_metric} becomes, for holomorphic \(y\),
\begin{equation} \label{eq:D5_induced_metric}
    \diff s_{\text{D5}}^2 = \frac{r^2}{L^2} \le[
        -\diff t^2 + \le( 1 + \frac{L^4}{r^4}|\p y|^2\ri) \diff z \diff \zb
    \ri] + \frac{L^2}{r^2} \diff \vec{v}^{\,2} \, .
\end{equation}
To approach the IR we wish to take the limit \(r \to 0\). Since on the worldvolume of the D5-branes \(r^2 = |y|^2 + v^2\), this requires that we send both \(y \to 0\) and \(v \to 0\), so that in particular we must approach a zero of the holomorphic function \(y(z)\). In the \(r \to 0\) limit, the induced metric becomes
\begin{equation}\begin{aligned} \label{eq:D5_ir_metric_1}
    \diff s_{\text{D5}}^2  &\approx -\frac{r^2}{L^2} \diff t^2 +  \frac{L^2}{r^2} \le( |\p y|^2 \diff z \diff \zb + \diff \vec{v}^2 \ri)
    \\
    &= -\frac{r^2}{L^2} \diff t^2 + \frac{L^2}{r^2} \le( \diff y \diff \yb + \diff \vec{v}^{\, 2}\ri).
\end{aligned}\end{equation}
We then define polar coordinates \((r,\q_1,\q_2,\q_3,\q_4)\) in the directions \((y,\yb,\vec{v})\) through the coordinate transformation
\begin{equation}\begin{aligned}
    v^1 &= r \cos \q_1 \, ,
    \\
    v^2 &= r \sin \q_1 \cos \q_2 \, , 
    \\
    v^3 &= r \sin \q_1 \sin \q_2 \cos \q_3 \, ,
    \\
    y &= r \sin \q_1 \sin \q_2 \sin \q_3 \, e^{i \q_4} \, ,
\end{aligned}\end{equation}
in terms of which the induced metric in equation~\eqref{eq:D5_ir_metric_1} becomes
\begin{equation}\begin{aligned} \label{eq:D5_ir_metric_2}
     \diff s_{\text{D5}}^2 &= -\frac{r^2}{L^2} \diff t^2 + \frac{L^2}{r^2} \diff r^2 + L^2 \diff \W_4^2 \, ,
     \\
     \diff \W_4^2 &\equiv \diff \q_1^2 + \sin^2 \q_1 \diff \q_2^2 + \sin^2 \q_1 \sin^2 \q_2 \diff \q_3^2 + \sin^2 \q_1 \sin^2 \q_2 \sin^3 \q_3 \diff \q_4^2 \, .
\end{aligned}\end{equation}
We recognise \(\diff s_\mathrm{D5}^2\) in equation~\eqref{eq:D5_ir_metric_2} as the metric of \(\ads[2] \times \sph[4]\), where both the \ads[2] and \sph[4] factors have curvature radius \(L\).

The holographic dual of a D5-brane in \ads[5] spanning an \(\ads[2] \subset \ads[5]\) and wrapping an \(\sph[4] \subset \sph[5]\) is well known: it is a Wilson line in an antisymmetric representation of \(\SU(N)\)~\cite{Yamaguchi:2006tq,Gomis:2006sb}. The dimension of the antisymmetric representation is encoded in the radius of the wrapped \(\sph[4]\). The radius \(L\) that we read off from equation~\eqref{eq:D5_ir_metric_2} is maximal, and corresponds to an antisymmetric representation with \(N/2\) indices. The fact that the wrapped \sph[4] is maximal presumably follows from the fact that our ansatz for the D5-branes has vanishing worldvolume gauge field strength \(F\); \(\ads[2] \times \sph[4]\) D5-branes wrapping a non-maximal \sph[4] require non-zero \(F\) in order to stabilise a slipping mode~\cite{Camino:2001at}.

Since we have \(k\) coincident D5-branes, we expect to find \(k\) insertions of the antisymmetric representation Wilson line at each zero of the mass. If \(m(z)\) has \(n\) zeros, then in total we should find \(n k\) Wilson line insertions. This is the same as the number of fermion zero modes, which has a natural interpretation. An antisymmetric representation Wilson line has an alternative description in terms of coupling \(\cN=4\) SYM to a one-dimensional fermion --- integrating out the fermion reproduces the usual Wilson line insertion in the path integral~\cite{Gomis:2006sb}. We expect that the Wilson lines that we find in the IR arise from integrating out the fermion zero modes associated to zeros of \(m(z)\). We leave a detailed analysis to future work.

D5-branes with \(\ads[2] \times \sph[4]\) worldvolume preserve one-half of the Poincar\'e supersymmetries of \(\ads[5]\times \sph[5]\), corresponding to eight supercharges~\cite{Yamaguchi:2006tq,Gomis:2006sb}. This is twice as many as preserved by holomorphic D5-branes. However, as argued in section~\ref{sec:d7_review}, at \(r \to 0\) there is an enhancement of the supersymmetry preserved by the D5-branes,\footnote{This follows from to \(r \to 0\) limit of equation~\eqref{eq:class1_gamma_prime_2} with \(q=5\).} doubling the number of supercharges to eight, matching the number of supersymmetries preserved by \(\ads[2] \times \sph[4]\) D5-branes.

Having dealt with what happens at the zeros of \(y(z)\), it is natural to wonder what happens to the worldvolume geometry in the opposite regime, namely close to points where \(y(z)\) diverges. Such points will always exist if \(y(z)\) is not constant, since by Liouville's theorem any non-constant holomorphic function in the complex plane must be unbounded~\cite{ahlfors}. If \(y(z)\) is holomorphic on the whole complex plane, then \(y \to \infty\) happens at \(z \to \infty\). On the other hand, we can allow \(y(z)\) to have poles if we demand only that it is holomorphic on the complex plane minus isolated points, in which case we can send \(y \to \infty\) by approaching a pole. See ref.~\cite{holomorphic_branes} for detailed discussion of the subtleties of allowing poles in \(y(z)\).

In either case, since \(r^2 = |y|^2 + v^2\), sending \(|y| \to \infty\) also sends \(r \to \infty\), approaching the boundary of \ads[5], dual to the ultraviolet (UV) of the dual QFT. Suppose for example, that \(y(z)\) has a pole of order \(n\) at infinity, so that at large \(|z|\) we have that \(y(z) \approx c z^n\) for some complex constant \(c\). Then at large \(z\) and fixed \(v\), the D5-branes' induced metric in equation~\eqref{eq:D5_induced_metric} becomes
\begin{equation}\begin{aligned} \label{eq:D5_induced_metric_pole}
    \diff s_{\text{D5}}^2 &\approx \frac{|c|^2 |z|^{2n}}{L^2} \le(- \diff t^2 + \a_n \diff z \diff \zb \ri) + \frac{L^2}{|c|^2 |z|^{2n}} \diff \vec{v}^{\,2} 
    \\
    &= - \frac{\r^2}{L^2} \diff t^2 + \frac{\a_n \r^{2/n}}{n^2 L^2 |c|^{2/n} } \le(  \diff \r^2 + \r^2 \diff \y^2 \ri) + \frac{L^2}{\r^2} \diff \vec{v}^{\,2} \, ,
\end{aligned}\end{equation}
where \(\a_n = 1 + \d_{n,1}(n/|c|)^2\) is a constant coefficient, and in the second line we introduced polar coordinates \((\r,\y)\) in the complex \(y\) plane by defining \(c z^n = \r e^{i \y}\). Similarly, if we instead consider \(y(z)\) with a pole of order \(n\) at some \(z = z_*\), near which \(y(z) \approx c / (z - z_*)^n\), then close to \(z_*\) the D5-branes' induced metric again approximately takes the form in the second line of equation~\eqref{eq:D5_induced_metric_pole}, this time after the substitution \(c / (z - z_*)^n = \r e^{i\y}\). The induced metric in equation~\eqref{eq:D5_induced_metric_pole} has a similar form to that of holomorphic D7-branes close to poles, given in ref.~\cite{holomorphic_branes}, and is unfortunately rather hard to interpret.

For completeness, we note that there is a second way to approach the boundary of \ads[5] along the worldvolume of the D5-branes, by sending \(v \to \infty\) with \(|y|\) fixed. In this limit \(r \approx v\), and the induced metric in equation~\eqref{eq:D5_induced_metric} becomes
\begin{equation}\begin{aligned} \label{eq:D5_induced_metric_uv1}
    \diff s_{\text{D5}}^2 &\approx \frac{v^2}{L^2} \le(- \diff t^2 + \diff z \diff \zb \ri) + \frac{L^2}{v^2} \diff \vec{v}^{\,2} 
    \\
    &=  \frac{v^2}{L^2} \le(- \diff t^2 + \diff z \diff \zb \ri) + \frac{L^2}{v^2} \diff v^2 + L^2 \diff \W_2^2 \, ,
\end{aligned}\end{equation}
where \(\diff \W_2^2\) is the metric on a unit, round \sph[2], and the second line follows from the first after adopting polar coordinates in the \(\vec{v}\) hyperplane. The metric in equation~\eqref{eq:D5_induced_metric_uv1} is that of \(\ads[4] \times \sph[2]\), with radial coordinate \(v\) and the boundary of \(\ads[4]\) at \(v \to \infty\). This is the worldvolume geometry of probe D5-branes dual to massless three-dimensional hypermultiplets~\cite{Karch_2001a,DeWolfe:2001pq,Erdmenger_2002}, which has a straightforward interpretation: except at points where \(m(z)\) diverges, at extremely high energy scales the hypermultiplets with holomorphic mass are indistinguishable from massless hypermultiplets. 

\subsection{D3-branes}
\label{sec:d3}

We now consider class 1 D3-brane embeddings. Aspects of holomorphic D3-brane embeddings in \(\ads[5] \times \sph[5]\) have been studied previously. For example, ref.~\cite{Constable:2002xt} introduced probe D3-brane embeddings in \(\ads[5] \times \sph[5]\). In our language, these embeddings would correspond to class 1 holomorphic D3-branes for which \(y\) has the simple pole form \(y = c/z\) for some complex constant \(c\). This choice is particularly physically interesting as it preserves scale invariance in the dual QFT. In ref.~\cite{Drukker:2008wr} the superconformal surface defects dual to holomorphic D3-branes with \(y=c/z\) were identified as disorder operators, also known as Gukov--Witten defects~\cite{Gukov:2006jk}. The existence and supersymmetry of embeddings with \(y = c/z^n\) for exponents \(n \neq 1\), breaking scale invariance, is also discussed in ref.~\cite{Constable:2002xt}.

Relatedly, ref.~\cite{Koh:2008kt} considered probe D3-brane embeddings in \(\ads[5] \times \sph[5]\) that are specified by a holomorphic function of \emph{two} complex coordinates, again focusing on configurations that preserve scale invariance. We discuss generalisations of D\(q\)-brane embeddings specified by holomorphic functions of multiple complex coordinates in appendix~\ref{sec:multiple_coords}.

In this section we will describe other aspects of class 1 D3-brane embeddings, with a particular focus on choices of the holomorphic function \(y(z)\) that break scale invariance, triggering a renormalisation group (RG) flow. We will argue that in the IR of this RG flow one finds Gukov--Witten defects located at the zeros of \(y(z)\). We begin in section~\ref{sec:holomorphic_scalars} with a discussion in \(\cN=4\) SYM at weak coupling, showing that holomorphic scalar field configurations solve the classical equations of motion and preserve one-half of the supersymmetry. We expect the holomorphic D3-brane embeddings to provide a large-\(N\), strongly coupled description of such configurations. We will also discuss salient features of Gukov--Witten defects. Then in section~\ref{sec:holomorphic_d3} we discuss the holomorphic D3-brane embeddings from the gravity side of the AdS/CFT correspondence.

\subsubsection{Holomorphic scalars in \texorpdfstring{$\cN=4$}{N=4} SYM}
\label{sec:holomorphic_scalars}

The fields of four-dimensional \(\cN=4\) SYM are a gauge field \(\cA_\m\), six real scalar fields \(\f^i\), and four Weyl fermions \(\y_a\), all valued in the adjoint representation of the gauge group's Lie algebra. A compact way to write the action for the theory is to treat it as a dimensional reduction from ten-dimensional \(\cN=1\) SYM (see for example refs.~\cite{Polchinski:1998rr,DHoker:2002nbb} for further details). The bosonic fields are packaged into the ten-dimensional gauge field \(\mathbb{A}_A = (\cA_\m, \f^i)\), while the fermions are packaged into a single ten-dimensional Majorana--Weyl spinor \(\Y\). The action for \(\cN=4\) SYM is then
\begin{equation} \label{eq:N=4_sym_action}
    S = \int \diff^4 x \tr \le( - \frac{1}{2} \mathbb{F}_{AB} \mathbb{F}^{AB} + i \bar{\Y} \G^A \mathbb{D}_A \Y \ri),
\end{equation}
where \(\mathbb{F}_{AB} = \p_A \mathbb{A}_B - \p_B \mathbb{A}_A + i g_\mathrm{YM} [\mathbb{A}_A,\mathbb{A}_B]\) is the ten-dimensional gauge field strength, \(\mathbb{D}_A \Y = \p_A \Y + i g_\mathrm{YM} [\mathbb{A}_A,\Y] \) is the covariant derivative, \(\G^A\) is a ten-dimensional Dirac matrix as in section~\ref{sec:susy}, and indices are contracted with the ten-dimensional Minkowski metric. The dimensional reduction means that the fields should be taken as depending on only the four coordinates \(x^\m\) corresponding to the \(\cA_\m\) components of the gauge field.

We will now seek holomorphic solutions to the classical equations of motion of \(\cN=4\) SYM. We look for solutions with all fields vanishing except for two of the scalar fields, \(\f^5\) and \(\f^6\), which we package into a complex scalar field \(\F\),
\begin{equation} \label{eq:N=4_complex_scalar}
    \F = \f^5 + i \f^6 \, .
\end{equation}
The equation of motion for \(\F\) which follows from the action in equation~\eqref{eq:N=4_sym_action} is
\begin{equation} \label{eq:complex_scalar_eom}
    \Box \F + \frac{g_\mathrm{YM}^2}{2} \bigl[\F, [\F, \F^\dagger] \bigr] = 0 \, ,
\end{equation}
where \(\Box = - \h^{\m\n} \p_\m \p_\n\) is the d'Alembertian.
Equation~\eqref{eq:complex_scalar_eom} is manifestly solved by any \(\F\) that solves the wave equation \(\Box \F = 0\) and commutes with its Hermitian conjugate, \([\F,\F^\dagger] = 0\), or equivalently \([\f^5,\f^6] = 0\). One way to solve the wave equation is to take all components of \(\F\) to be holomorphic or antiholomorphic functions of a complex coordinate \(z\) defined by
\begin{equation}
    z = x^1 + i x^2 \, .
\end{equation}
Such a \(\F\) was shown to preserve one-half of the supersymmetry of \(\cN=4\) SYM in ref.~\cite{holomorphic_branes}. Here we will repeat this calculation in a format that makes for easy comparison with the supergravity calculation in section~\ref{sec:class1_kappa}.

Under a supersymmetry transformation, the fermions of \(\cN=4\) SYM transform as
\begin{equation} \label{eq:10D_SYM_SUSY_transformation}
    \d \Y = \bF^{AB} \G_{AB} \ve \, ,
\end{equation}
where \(\ve\) is a ten-dimensional Majorana--Weyl spinor supersymmetry parameter and \(\G_{AB}\) denotes a normalised antisymmetric product of Dirac matrices, as in section~\ref{sec:susy}. Some supersymmetry will be preserved by the field configuration if there exist choices of \(\ve\) for which \(\d \Y = 0\). Suppose the only non-vanishing components of \(\bA_A\) are \(\f^{5,6}\), that these components are mutually commuting \([\f^5,\f^6] = 0\), and that they depend only on the coordinates \((z,\zb)\). The non-zero components of the field strength are then \(\bF_{\m(i+3)} = - \bF_{(i+3)\m} = \p_\m \f^i\) for \(\m=1,2\) and \(i=5,6\). In terms of the complex scalar field \(\F\) defined in equation~\eqref{eq:N=4_complex_scalar}, the transformation in equation~\eqref{eq:10D_SYM_SUSY_transformation} is then
\begin{equation}\begin{aligned}
    \d \Y = - \Bigl[&
        i (\p \F - \pb \F^\dagger) (\G_{19} - \G_{28}) - i (\p \F + \pb \F^\dagger) (\G_{18} + \G_{29})
    \\ &
        + i (\pb \F - \p \F^\dagger) (\G_{19} + \G_{28}) -  (\pb \F + \p \F^\dagger) (\G_{18} - \G_{29})
    \Bigr] \ve \, .
\end{aligned}\end{equation}
The factor in square brackets takes the same form as that in the matrix \(\G''\) in equation~\eqref{eq:class1_gamma_double_prime}, under the interchange \((\F,\F^\dagger) \leftrightarrow (y,\yb)\). Thus, the analysis of section~\ref{sec:class1_kappa} applies here and any \(\F\) which is a holomorphic or antiholomorphic function of \(z\), preserves one-half of the Poincar\'e supersymmetries of \(\cN=4\) SYM. The preserved supersymmetries correspond to \(\ve\) satisfying
\begin{equation}
    (\G_{18} \pm \G_{28})\ve = 0 \, ,
\end{equation}
with the plus or minus sign for holomorphic or antiholomorphic \(\F\), respectively.

For example, consider a holomorphic \(\F\) which is diagonal in its gauge indices (in a particular gauge). We write this in block diagonal form as
\begin{equation} \label{eq:N=4_holomorphic_scalar}
    \F = \begin{pmatrix}
        \F_1(z) \otimes \id_{N_1} & 0 & \cdots & 0
        \\
        0 & \F_2(z) \otimes \id_{N_2}  & \cdots & 0
        \\
        \vdots & \vdots & \ddots & \vdots
        \\
        0 & 0 & \cdots & ~ \F_M(z) \otimes \id_{N_M}
    \end{pmatrix},
\end{equation}
for some integer \(M\), where \(\F_l(z) \neq \F_k(z)\) for \(l \neq k\), \(\id_{N_l}\) is the \(N_l\)-dimensional identity matrix, and \(\sum_{l=1}^M N_l = N\). Such a field configuration breaks the gauge group from \(\SU(N)\) to \(\mathrm{S}[\prod_{l=1}^M \mathrm{U}(N_l)]\).

When the \(\F_l\) are all constant, the field configuration in equation~\eqref{eq:N=4_holomorphic_scalar} corresponds to a point on the Coulomb branch of \(\cN=4\) SYM.  Another well-known choice is to take each \(\F_l(z) = (\b_l + i \g_l)/z\) where \(\b_l\) and \(\g_l\) are constant, real parameters, so that the scalar field is holomorphic in \(\bC \backslash \{0\}\). This gives the scalar field in the presence of a Gukov--Witten surface defect at \(z=0\)~\cite{Gukov:2006jk}. We will now briefly review some aspects of these defects. We follow the discussion in refs.~\cite{Gomis:2007fi,Drukker:2008wr}.

A Gukov--Witten defect is in general defined by singular boundary conditions for the bosonic fields of \(\cN=4\) SYM. In addition to imposing that \(\F_l = (\b_l + i \g_l)/z\) close to \(z=0\), one can prescribe singular boundary conditions for the gauge field \(\cA\) at \(z=0\) that preserve the same \(\mathrm{S}[\prod_{l=1}^M \mathrm{U}(N_l)]\) subgroup of the gauge group, of the form\footnote{As an aside, introducing a gauge field of the form in equation~\eqref{eq:gukov_witten_gauge_field} does not spoil the fact that \(\F\) in equation~\eqref{eq:N=4_holomorphic_scalar} solves the classical equations of motion and preserves some supersymmetry, since \(\cA\) in equation~\eqref{eq:gukov_witten_gauge_field} is exact on \(\mathbb{C}\backslash\{0\}\) and commutes with itself, so has vanishing field strength, and also commutes with \(\F\).}
\begin{equation} \label{eq:gukov_witten_gauge_field}
    \cA = \begin{pmatrix}
        \a_1 \otimes \id_{N_1} & 0 & \cdots & 0
        \\
        0 & \a_2 \otimes \id_{N_2}  & \cdots & 0
        \\
        \vdots & \vdots & \ddots & \vdots
        \\
        0 & 0 & \cdots & ~ \a_M \otimes \id_{N_M}
    \end{pmatrix} \frac{1}{2i} \le(\frac{\diff z}{z} - \frac{\diff \zb}{\zb} \ri) ,
\end{equation}
the \(\a_l\) are \(2\pi\)-periodic real parameters. In general, a Gukov--Witten defect also has a matrix \(\h\) of two-dimensional theta angles \(\h_l\) on the surface \(\S\) at \(z=0\),
\begin{equation}
    \h = \begin{pmatrix}
        \h_1 \otimes \id_{N_1} & 0 & \cdots & 0
        \\
        0 & \h_2 \otimes \id_{N_2}  & \cdots & 0
        \\
        \vdots & \vdots & \ddots & \vdots
        \\
        0 & 0 & \cdots & ~ \h_M \otimes \id_{N_M}
    \end{pmatrix},
\end{equation}
implemented by an insertion of \(\exp \le(i \sum_l \h_l \int_\S \cF_l\ri)\) into the path integral for \(\cN = 4\) SYM, where \(\cF_l\) is the \(l\)th block in the field strength for \(\cA\). In total, a Gukov--Witten defect is specified by the choices of the parameters \((\a_l,\b_l,\g_l,\h_l)\). Generalisations of Gukov--Witten defects with higher-order poles in \(\F\) and \(\cA\), thus breaking scale invariance, are also possible~\cite{Witten:2007td}.

The holographic dual description of Gukov--Witten defects at large \(N\) and strong `t Hooft coupling was developed in refs.~\cite{Gomis:2007fi,Drukker:2008wr}. They are ``bubbling'' supergravity solutions of type IIB supergravity which are asymptotically \(\ads[5] \times \sph[5]\) and include flux of the Ramond--Ramond field strength \(F_5\) around various five-spheres, encoding the sizes \(N_l\) of the blocks. If only the \(l=1\) block has non-zero \((\a_l,\b_l,\g_l,\h_l)\) and \(N_1 \ll N\), this is holographically dual to a probe limit in which the corresponding bubbling geometry is replaced by a stack of coincident probe D3-branes in \(\ads[5] \times \sph[5]\)~\cite{Drukker:2008wr}. The probe D3-branes have an \(\ads[3] \times \sph[1]\) worldvolume, and in our language correspond to a class 1 holomorphic embedding with \(y \propto (\b_1 + i \g_1)/z\)~\cite{Constable:2002xt}.

\subsubsection{Holomorphic D3-branes}
\label{sec:holomorphic_d3}

Now we return to supergravity and consider class 1 holomorphic D3-branes in \(\ads[5] \times \sph[5]\). We adopt the coordinate system described in section~\ref{sec:class1_embedding}, in which we form two complex coordinates \(z = x_\parallel^1 + i x_\parallel^2\) and \(y = x_\perp^1 + i x_\perp^2\), denote the remaining two parallel directions as \(x_\parallel^0 =t\) and \(x_\parallel^3 = x\), and denote the remaining four \(x_\perp^i\) directions as \(\vec{W} = (W_1,W_2,W_3,W_4)\). In this notation, the \(\ads[5] \times \sph[5]\) metric in equation~\eqref{eq:ads5_x_s5} becomes
\begin{equation}
    \diff s^2 = \frac{r^2}{L^2} \le(-\diff t^2 + \diff x^2 + \diff z \diff \zb \ri)
    + \frac{L^2}{r^2} \le( \diff y \diff \yb + \diff \vec{w}^{\,2} \ri),
\end{equation}
with \(r^2 = |y|^2 + W^2\). The probe D3-branes span the \(x_\parallel^\m\) directions \((t,x,z,\zb)\), as indicated in the first row of table~\ref{tab:class1_holomorphic_D3}, sit at constant \(\vec{W}\), and have \(y\) a holomorphic function of \(z\), \(y = y(z)\).  The metric induced on the D3-branes' worldvolume, given in equation~\eqref{eq:class1_metric}, becomes in the near-horizon limit and for holomorphic \(y\),
\begin{equation} \label{eq:d3_metric}
    \diff s_\mathrm{D3}^2  = \frac{r^2}{L^2} \le[- \diff t^2 + \le(1 + \frac{L^4}{r^4} |\p y|^2 \ri) \diff z \diff \zb + \diff x^2 \ri].
\end{equation}

A holomorphic D3-brane embedding with non-constant \(y = y(z)\) breaks translational symmetry in the \((z,\zb)\) directions and, as shown in section in section~\ref{sec:class1_embedding}, preserves one-half of the supersymmetries of the D3-brane background, amounting to eight supercharges. It turns out that half of the supercharges have positive two-dimensional chirality in the directions \((t,x)\) with unbroken translational symmetry, while the other half have negative chirality. Thus, holomorphic D3-brane embeddings preserve two-dimensional \(\cN=(4,4)\) supersymmetry in the \((t,x)\) directions.

To show that there are equal numbers of preserved supercharges with positive and negative two-dimensional chirality, we first build projectors \(P_1\) and \(P_2\) onto spinors satisfying the conditions in equation~\eqref{eq:parallel_Dq_brane_kappa_combined},
\begin{equation}
    P_1 = \frac{1}{2} \le(\id_{32} \otimes \id_2 + \G_{0123} \otimes i \s_2\ri) \, ,
    \qquad
    P_2 = \frac{1}{2} \le(\id_{32} \otimes \id_2 \pm \G_{0389} \otimes i \s_2 \ri) \, ,
\end{equation}
where we have substituted the explicit form of \(J_{(3)} = i \s_2\), and for clarity of presentation we have restored the tensor product symbol on products of matrices acting on spinor and doublet indices. The plus or minus sign in \(P_2\) is for holomorphic or antiholomorphic \(y\), respectively. The two components of the doublet \(\ve\) of Killing spinors are Majorana--Weyl spinors, both with positive ten-dimensional chirality. We build a third projector \(P_3\) onto doublets where both components positive ten-dimensional chirality
\begin{equation}
    P_3 = \frac{1}{2} \le(\id_{32} \otimes \id_2 + \G_\sharp \otimes \id_2\ri) .
\end{equation}
These projectors mutually commute, \([P_1,P_2] = [P_3,P_1] = [P_2,P_3]  =0\).

The easiest way to work with the Majorana condition is to adopt a ``really real'' basis in which Majorana spinors and the \(\G_A\) are real. In such a basis, we define the projector \(\cP = P_1 P_2 P_3\) onto the space of Killing spinor doublets satisfying the kappa symmetry conditions in equation~\eqref{eq:parallel_Dq_brane_kappa_combined}. It is straightforward to check in an explicit really real basis that the trace of the two-dimensional chirality matrix \(\G_{03}\) vanishes on this space,
\begin{equation} \label{eq:chirality_trace}
    \tr \le(\cP^T \G_{03} \cP\ri) = 0 \, .
\end{equation}
We checked this in the basis given in ref.~\cite{Freedman_Van_Proeyen_2012}. Since \(\G_{03}\) has eigenvalues \(\pm 1\), equation~\eqref{eq:chirality_trace} implies that its restriction to the space of doublets of Majorana--Weyl spinors satisfying the kappa symmetry conditions has equal numbers of positive and negative eigenvalues. Hence, the eight supercharges preserved by the holomorphic D3-branes correspond to two-dimensional \(\cN=(4,4)\) supersymmetry.

As mentioned already, certain choices of the function \(y(z)\) correspond to well-known D3-brane embeddings in \(\ads[5] \times \sph[5]\). When \(y(z)\) is constant, the D3-branes sit at a constant value of the radial coordinate \(r = \sqrt{|y|^2 + W^2}\). This corresponds to putting the dual \(\cN=4\) SYM theory at a point on the Coulomb branch where the gauge group \(\SU(N)\) is spontaneously broken to \(\mathrm{S}[\U(N-k) \times U(k)]\) by a non-zero vacuum expectation value \(\langle \F \rangle \propto r\) for one of the adjoint-valued scalar fields \(\F\)~\cite{Klebanov:1999tb}.\footnote{For constant \(y\) translational symmetry is unbroken, and the supersymmetry is enhanced by a factor of two to four-dimensional \(\cN=4\). The supersymmetry enhancement arises because the kappa symmetry condition in equation~\eqref{eq:parallel_Dq_brane_kappa_2} does not apply as the coefficient of \(\G_{03}\) in the kappa symmetry matrix~\eqref{eq:class1_kappa_projector} vanishes for constant \(y\).}

Alternatively, suppose we choose \(\vec{W} = 0\) and
\begin{equation} \label{eq:d3_simple_pole}
    y(z) = \frac{L^2 \k}{z},
\end{equation}
for some complex constant \(\k\). This D3-brane embedding in \(\ads[5] \times \sph[5]\) is well known~\cite{Constable:2002xt,Drukker:2008wr}. The choice \(y \propto z^{-1}\) is special because the probe D3-branes have an \(\ads[3] \times \sph[1]\) worldvolume, and consequently the dual QFT has two-dimensional defect conformal invariance. To see this, we substitute the solution in equation~\eqref{eq:d3_simple_pole} into the induced metric in equation~\eqref{eq:d3_metric} with \(\vec{W}\) to find
\begin{equation} \begin{aligned} \label{eq:D3_gukov_witten_metric}
    \diff s_\mathrm{D3}^2 &= \frac{L^2|\k|^2 }{|z|^2} \le[-\diff t^2 + \le(1 + |\k|^{-2}\ri) \diff z \diff \zb + \diff x^2 \ri]
    \\
    &= L^2 \le(1 + |\k|^2 \ri) \le[\frac{1}{\s^2} \le(-\diff t^2 + \diff x^2 + \diff \s^2 \ri) + \diff \y^2\ri] ,
\end{aligned}\end{equation}
where the second line is obtained by defining new coordinates \((\s,\y)\) through
\begin{equation}
    z = \frac{\s e^{i\y}}{\sqrt{1 + |\k|^2}} \, .
\end{equation}
The metric equation~\eqref{eq:D3_gukov_witten_metric} is indeed that of \(\ads[3] \times \sph[1]\), where both \ads[3] and \sph[1] have curvature radius \(L \sqrt{1 + |\k|^2}\). Consequently, holomorphic probe D3-branes with \(y(z)\) in equation~\eqref{eq:d3_simple_pole} are holographically dual to a two-dimensional conformal defect in \(\cN=4\) SYM. The defect is located at the surface \(z=0\), where the probe branes meet the boundary of \(\ads[5]\), and is superconformal, as the probe branes preserve two-dimensional \(\cN=(4,4)\) supersymmetry.

As discussed in section~\ref{sec:holomorphic_scalars}, the superconformal surface defect dual to \(k\) D3-branes with \(y(z) \propto z^{-1}\) is a Gukov--Witten defect~\cite{Drukker:2008wr}. The singular boundary conditions on the \(\cN=4\) SYM fields described in section~\ref{sec:holomorphic_scalars} have a single non-zero block of size \(N_1 = k\), and with non-zero parameters \((\b_1,\g_1)\) related to \(\k\) in equation~\eqref{eq:d3_simple_pole} by~\cite{Drukker:2008wr}
\begin{equation} \label{eq:beta_gamma}
   \b_1 + i \g_1 = \frac{L^2}{2\pi \a'} \k \, .
\end{equation}
Although in our ansatz in section~\ref{sec:class1_embedding} we took the D3-branes' worldvolume gauge field \(A\) to vanish, in the presence of a pole one has the freedom to turn on a non-zero holonomy of \(A\) around \(z=0\)~\cite{Drukker:2008wr},
\begin{equation} \label{eq:worldvolume_gauge_field_holonomy}
    A = \frac{\a_1}{2i} \le( \frac{\diff z}{z} - \frac{\diff \zb}{\zb} \ri) = \a_1 \diff \y \, .
\end{equation}
Since the corresponding field strength vanishes everywhere away from \(z=0\), this still solves the D3-branes' equations of motion and preserves supersymmetry. The parameter \(\a_1\) corresponds to the parameter appearing in the singular boundary conditions on the \(\cN=4\) SYM gauge field in equation~\eqref{eq:gukov_witten_gauge_field}~\cite{Drukker:2008wr}. Similarly, one can obtain non-zero \(\h_1\) by turning on non-zero holonomy of the dual gauge field \(\tilde{A}\) on the D3-branes' worldvolume.

We expect that solutions with higher-order poles, of the form
\begin{equation} \label{eq:D3_higher_pole}
    y = \frac{L^2 \k}{z^n} \, ,
\end{equation}
with integer \(n > 1\) and complex constant \(\k\), should be dual to the surface operators considered in ref.~\cite{Witten:2007td}, for which the fields of \(\cN=4\) SYM have boundary conditions with higher-order poles at the location of the defect. Such surface operators break scale invariance. Correspondingly, the induced metric on the D3-branes does not contain an \(\ads[3]\) factor. Concretely, substituting \(y\) in equation~\eqref{eq:D3_higher_pole} into the induced metric in equation~\eqref{eq:d3_metric}, with \(\vec{W}=0\) we find
\begin{equation}\begin{aligned} \label{eq:D3_higher_pole_metric}
    \diff s^2 &= \frac{L^2 |\k|^2}{|z|^{2n}} \le[- \diff t^2 + \le(1 + \frac{n^2}{|\k|^2} |z|^{2(n-1)}\ri)\diff z \diff \zb + \diff x^2\ri]
    \\
    &= \frac{L^2 |\k|^2}{\s^2} \le(-\diff t^2 + \diff x^2 \ri) + L^2 \le( 1 + \frac{|\k|^2}{n^2 \s^{2(n-1)/n}}\ri) \le(\frac{\diff \s^2}{\s^2} + \diff \y^2 \ri) ,
\end{aligned}\end{equation}
where the second line is obtained by defining new coordinates \((\s,\y)\) through \(z = \s^{1/n} e^{i \y/n}\). The metric in equation~\eqref{eq:D3_higher_pole_metric} is not that of \(\ads[3] \times \sph[1]\) for \(n > 1\), although it becomes locally \(\ads[3] \times \sph[1]\) asymptotically at large \(\s\). Since \(\s \to \infty\) corresponds to \(z \to \infty\) and thus \(r = |y| \to 0\), this \(\ads[3] \times \sph[1]\) regime is in the deep IR.

Now consider more general \(y(z)\), holomorphic in the complex plane minus a set of isolated poles at locations \(z_{*,p}\), and with the D3-branes at arbitrary constant \(\vec{W}\). At a pole in \(y\), the D3-brane touches the boundary of \ads[5], since when \(y\) diverges, so too does \(r^2 = |y|^2 + W^2\). Close to a  pole we have that \(|y| \gg W\), so that we can neglect \(W\) and the worldvolume geometry becomes approximately that of the solutions discussed above. In particular, close to a simple pole the D3-branes' induced metric becomes asymptotically \(\ads[3] \times \sph[1]\) with curvature radius that depends on the residue of the pole, similar to equation~\eqref{eq:D3_gukov_witten_metric}. Likewise, close to a higher-order pole the induced metric takes a form similar to equation~\eqref{eq:D3_higher_pole_metric}. Consequently, we expect a holomorphic D3-brane embedding for which \(y(z)\) has isolated poles to be holographically dual to a state in the presence of surface defects at the locations of the poles; either Gukov--Witten defects at simple poles or the defects of ref.~\cite{Witten:2007td} at higher-order poles.

Let us turn to the IR physics in the QFT, dual to the \(r \to 0\) region of \(\ads[5] \times \sph[5]\). If \(\vec{W} \neq 0\) then the probe D3-branes do not contribute to the IR physics, since \(r^2 = |y|^2 + W^2\) is bounded from below by \(W^2\). On the other hand, for D3-brane embeddings with \(\vec{W} = 0\) the D3-branes reach \(r=0\) at zeros of the holomorphic function \(y(z)\). We will set \(\vec{W} = 0\) in what follows.

Consider a holomorphic D3-brane embedding for which \(\vec{W} = 0\) and with \(y(z)\) having a zero of order \(n\) at some \(z = z_0\), close to which \(y \approx c (z - z_0)^n\) for some complex constant \(c\). Close to \(z = z_0\), \(|\p y|^2/r^4 \propto |z - z_0|^{-2(n+1)}\gg 1\), so that the induced metric in equation~\eqref{eq:d3_metric} becomes approximately
\begin{equation} \begin{aligned}
    \diff s_\mathrm{D3}^2 &\approx \frac{|y|^2}{L^2} \le(- \diff t^2 + \diff x^2\ri)
    + \frac{L^2}{|y|^2} |\p y|^2 \diff z \diff \zb
    \\
    &= \frac{\r^2}{L^2} \le( -\diff t^2 + \diff x^2 \ri) + \frac{L^2}{\r^2} \diff \r^2+ L^2 \diff \y^2 \, ,
\end{aligned}\end{equation}
where the second line is obtained after letting \(y \approx c z = \r e^{i\y}\). This is the metric of \(\ads[3] \times \sph[1]\), where both \ads[3] and \sph[1] have curvature radius \(L\).

Thus, perhaps unsurprisingly, we find superconformal surface defects in the IR, located at zeros of \(y(z)\). What kind of defects? Recall from equation~\eqref{eq:D3_gukov_witten_metric} that probe D3-branes dual to Gukov--Witten defects have \(\ads[3] \times \sph[1]\) worldvolume, where both factors have curvature radius \(L \sqrt{1 + |\k|^2}\), where \(\k \propto \b_1 + i \g_1\). In the IR we find \(\ads[3] \times \sph[1]\) with curvature radius \(L\), so we interpret the defects found in the IR as Gukov--Witten defects in the singular limit \(\b_1, \g_1 \to 0\). This limit is discussed in refs.~\cite{Holguin:2025bfe,Chalabi:2025nbg}.

There is strong evidence that the Coulomb branch of \(\cN=4\) SYM, dual to holomorphic D3-branes with constant \(y(z)\), exhibits integrability, see e.g. ref.~\cite{Ivanovskiy:2024vel} and references therein. Similarly, the singular \(\b_1,\g_1 \to 0\) limit of Gukov--Witten defects that we find at zeros of \(y(z)\) in the IR are integrable~\cite{Holguin:2025bfe,Chalabi:2025nbg}. In both cases, the holographically dual D3-branes provide integrable boundary conditions for strings in \(\ads[5] \times \sph[5]\)~\cite{Dekel:2011ja,Demjaha:2025axy}. It is then natural to wonder whether the QFTs dual to holomorphic embeddings with arbitrary non-constant \(y(z)\) are also integrable. Unfortunately, this cannot be generally so, as the case \(y \propto z^{-1}\) shows: outside of the \(\b_1,\g_1 \to 0\) limit, Gukov--Witten defects are not integrable~\cite{Holguin:2025bfe,Chalabi:2025nbg}.

\section{Summary and outlook}
\label{sec:discussion}

We have generalised the holomorphic probe D7-branes in the D3-brane background described in ref.~\cite{holomorphic_branes} to arbitrary D\(q\)-branes in extremal black D\(p\)-brane backgrounds for \(p < 7\). We have shown that, starting from an intersection between flat D\(p\)- and D\(q\)-branes and then replacing the D\(p\)-branes by the corresponding extremal supergravity background, a complex scalar \(y\) describing the embedding of the D\(q\)-branes may be made a non-trivial holomorphic or antiholomorphic function of a worldvolume coordinate \(z\) if the number of Neumann--Dirichlet directions \(d\) in the original D\(p\)/D\(q\) intersection is a multiple of four. We classified such holomorphic embeddings according to whether \(y\) and \(z\) are formed from directions parallel or perpendicular to the D\(p\)-branes, as summarised in table~\ref{tab:embedding_classification}. Whenever \(d\) is a multiple of four, holomorphic embeddings saturate a BPS bound and preserve a fraction of the supersymmetry of the D\(p\)-brane background --- typically one-half for \(d=0\) or one-quarter for \(d=4\) or \(8\).

We investigated the holography of holomorphic D\(5\)- and D3-branes in the \(\ads[5] \times \sph[5]\) near-horizon limit of the extremal D3-brane background. The holomorphic D5-branes are dual to three-dimensional \(\cN=4\) hypermultiplets coupled to four-dimensional \(\cN=4\) SYM, with a mass that depends holomorphically on position. This mass triggers an RG flow, and we found using holography that in the IR one obtains supersymmetric Wilson lines in an antisymmetric representation of \(\SU(N)\) located at the zeros of the mass. The holomorphic D3-branes are dual to non-trivial translational symmetry breaking states, generically in the presence of Gukov--Witten surface defects located at poles of the embedding scalar. We used holography to show that in the IR one obtains \(\cN=(4,4)\) supersymmetric surface defects, located at the zeros of the holomorphic embedding scalar.

There are many possible directions for future research. For one, our analysis of the holomorphic D5- and D3-branes in \(\ads[5] \times \sph[5]\) and their dual QFTs in section~\ref{sec:d3} is far from complete. A natural next step would be to perform the holographic renormalisation of these probe branes~\cite{deHaro:2000vlm,Karch:2005ms}. There are also several possible further generalisations of the embeddings that we have discussed, with potentially interesting physics to explore. For example, can we find versions of these holomorphic embeddings with non-trivial worldvolume gauge fields? 

We can obtain one immediate generalisation our holomorphic embeddings via a double Wick rotation. Consider the class 1 embeddings discussed in section~\ref{sec:class1_embedding}, for which the complex coordinate \(z\) is built from two directions \(z = x_\parallel^1 + i x_\parallel^2\). Performing the Wick rotations \(t = - i \tilde{x}\) and \(x_\parallel^1 = i \tilde{t}\) to obtain a new spatial coordinate \(\tilde{x}\) and a new time coordinate \(\tilde{t}\), we now have that \(z = i (\tilde{t} + x_\parallel^2)\) and \(\zb = i (\tilde{t} - x_\parallel^2)\). Thus, the result of section~\ref{sec:class1_embedding}, that \(y\) can be any holomorphic or antiholomorphic function of \(z\), becomes the statement that \(y\) can be any function of the lightcone coordinate \(\tilde{t} + x_\parallel^2\) or of \(\tilde{t} - x_\parallel^2\). That this solves the D\(q\)-brane equations of motion in the D\(p\)-brane background can straightforwardly be confirmed by direct calculation.

In appendix~\ref{sec:multiple_coords} we describe another generalisation of class 1 embeddings, for which \(y  = x_\perp^1 + i x_\perp^2\) is a holomorphic function of multiple complex coordinates \(z_n\), each formed from \(x_\parallel^\m\) directions. We show that any such \(y\) solves the D\(q\)-brane equations of motion and preserves a fraction of the supersymmetry of the D\(p\)-brane background. Concretely, we show that with \(M\) complex coordinates \((z_1,\cdots,z_M)\), for \(M > 1\) the possible holomorphic embeddings have \(d=0\) ND directions and preserve a fraction \(1/2^M\) of the supersymmetry of the D\(p\)-brane background.

Another natural direction is to look for holomorphic embeddings of D-branes or other types of extended objects in other supergravity backgrounds. In appendix~\ref{sec:m_brane} we perform a first step in this direction, demonstrating the existence of holomorphic M2- and M5-brane embeddings in the extremal M2- and M5-brane backgrounds of eleven-dimensional supergravity. The holography of these embeddings would also be interesting to explore.

One could also try to go beyond the probe limit and find supergravity solutions that account for the backreaction of holomorphic D-brane embeddings, given the amount of supersymmetry they preserve. A natural place to start may be the class 1 D5-brane embeddings in \(\ads[5] \times \sph[5]\) considered in section~\ref{sec:d5}. The backreacted solutions that would correspond to constant \(y=0\) are known~\cite{DHoker:2007zhm,DHoker:2007hhe}, and one could attempt to find a generalisation of these solutions that would describe non-trivial \(y(z)\). Given the analysis of section~\ref{sec:d5}, we expect that deep in the bulk of this geometry and close to a zero of \(y(z)\), such a solution should approach the solution of type IIB supergravity dual to an antisymmetric-representation Wilson line described in ref.~\cite{DHoker:2007mci}.

We hope that our work can serve as a launchpad for further fruitful research, in the directions we have suggested and others.

\acknowledgments

We thank Adam Chalabi, Konstantin Zarembo, and Neil Lambert for useful discussions. We especially thank Pietro Capuozzo, Jack Holden, Andy O'Bannon, and Benjamin Suzzoni for discussion and collaboration on the related work~\cite{holomorphic_branes}. The work of R.\,R. was supported by the European Union’s Horizon Europe research and innovation program under the Marie Sklodowska-Curie Grant Agreement No.~101104286, and by VR grant 2021-04578. The work of J.\,R. is supported by the STFC consolidated grant ST/X000583/1. J.\,R. would like to thank the Isaac Newton Institute for Mathematical Sciences, Cambridge, for support and hospitality during the programme ``Quantum field theory with boundaries, impurities, and defects'', where work on this paper was undertaken.

\appendix

\section{Multiple holomorphic coordinates}
\label{sec:multiple_coords}

In this appendix, we construct a generalisation of the embeddings described in the main text, for which the embedding scalar \(y\) is a holomorphic function of multiple complex coordinates \(z_j\). We do not aim to fully explore all possibilities for such embeddings. Instead, this appendix serves as a proof of concept, in which we demonstrate the existence of a generalisation of class 1 embeddings in the classification of table~\ref{tab:embedding_classification}, for which \(y\) is built from \(x_\perp^i\) directions, while each of the \(z_j\) are built from \(x_\parallel^\m\) directions.

\subsection{Existence of embeddings}

As in the main text, our aim is to embed \(k\) coincident probe D\(q\)-branes into the extremal black D\(p\)-brane background \eqref{eq:Dp_brane_background} of type IIA or type IIB supergravity. As for the class 1 embeddings described in the main text, we form a complex coordinate \(y\) from two of the \(x_\perp^i\) directions of the background,
\begin{equation}
    y = x_\perp^1 + i x_\perp^2 \, ,
\end{equation}
while we form \(M\) complex coordinates \(z_j\), \(j  = 1,2,\cdots,M\), from \(x_\parallel^\m\) directions,
\begin{equation}
    z_j = x_\parallel^{2j-1} + i x_\parallel^{2j} \, .
\end{equation}
Since in the D\(p\)-brane background the index on \(x_\parallel^\m\) runs from \(\m = 0\) to \(p\), the number of \(z_j\) coordinates we can define is bounded from above: \(M \leq p/2\). The class 1 embeddings constructed in section~\ref{sec:class1_embedding} correspond to \(M = 1\). In this appendix we consider cases with \(M \geq 2\), so we restrict to D\(p\)-brane backgrounds with \(p \geq 4\).\footnote{Ref.~\cite{Koh:2008kt} constructs D3-brane embeddings depending holomorphically on two complex coordinates in the near horizon limit of the D3-brane background by working in Euclidean signature, so that Euclidean time can be used to form one of the complex coordinates.} As in the main text we consider only \(p < 7\).

We embed \(k\) coincident probe D\(q\)-branes into the D\(p\)-brane background, that span \(a\) of the \(x_\parallel^\m\) directions, including time \(t = x_\parallel^0\) and all of the \(z_j\) directions. We adopt the same notation as in table~\ref{tab:notation} for the remaining coordinates: if \(a > 2M+1\) the D\(q\)-branes span more of the \(x_\parallel^\m\) directions which we denote \(\vec{x}\), while any \(x_\parallel^\m\) directions not spanned by the D\(q\)-branes are denoted \(\vec{U}\). Any \(x_\perp^i\) directions spanned by the D\(q\)-branes are denoted by \(\vec{v}\) while, apart from \((y,\yb)\), any \(x_\perp^i\) directions not spanned by the D\(q\)-branes are denoted \(\vec{W}\). In this notation, the metric appearing in the D\(p\)-brane background~\eqref{eq:Dp_brane_background} becomes
\begin{equation}
	\diff s^2 = H(r)^{-1/2}\Bigl(-\diff t^2 + \diff \vec{x}^{\,2} + d\vec{U}^{\,2} + \sum_j \diff z_j \diff \zb_j \Bigr) +  H(r)^{1/2} \le(\diff \vec{v}^{\,2} + \diff \vec{W}^{\,2} + \diff y \diff\yb\ri).
\end{equation}

As in the main text, we make the ansatz that the probe D\(q\)-branes' worldvolume gauge field \(A\) vanishes, while for the worldvolume scalars we make the ansatz that \(y\) depends on all of the complex coordinates \(z_j\), \(y = y(z_1,\zb_1,z_2,\zb_2,\cdots)\) and that \(\vec{U}\) and \(\vec{W}\) are constant. Evaluated on this ansatz, the determinant of the induced metric on the D\(q\)-branes' worldvolume takes the form
\begin{equation} \label{eq:multiple_coordinates_metric_determinant}
    \lvert\det g \rvert = \frac{H(r)^{(q+1-2a)/2}}{4^M} \D \, ,
\end{equation}
where \(\D\) is given by
\begin{equation} \label{eq:multiple_coordinates_Delta}
	\D = \Bigl(1 + \sum_j \cY_4^{(j)} \Bigr)^2 + 4 H(r) \sum_j |\pb_j y|^2 + 4 H(r)^2 \sum_j \sum_{k>j}|\pb_j y \pb_k \yb - \pb_j \yb \pb_k y|^2 \, .
\end{equation} 
In this expression we use the notation \(\p_j = \p/\p z_j\) and \(\pb_j = \p / \p \zb_j\), and we define
\begin{equation} \label{eq:multiple_coordinates_Y}
	\cY_n^{(j)} = H(r)^{n/4} \le(|\p_j y|^2 - |\pb_j y|^2\ri)\,,
\end{equation}
which is a generalisation of the central charge density in equation~\eqref{eq:class1_Y}.

Substituting the metric determinant~\eqref{eq:multiple_coordinates_metric_determinant} into the D\(q\)-brane action~\eqref{eq:Dq_action}, we find
\begin{equation}\begin{aligned} \label{eq:multiple_coordinates_action}
    S &= - \frac{k T_{\mathrm{D}q}}{2^M} \int \diff t \diff z_1 \diff \zb_1 \cdots \diff z_M \diff \zb_M \diff \vec{x} \diff \vec{v} \, \cL \, ,
    \\
    \cL &=H(r)^{(d - 4)/4} \sqrt{\D} - \delta_{d,0} \le[ H(r)^{-1} - 1 \ri],
\end{aligned}\end{equation}
with \(r^2 = |y|^2 + v^2 + W^2\), and where $d$ is again the number of ND directions, given in equation~\eqref{eq:number_ND}. As in section~\ref{sec:class1_embedding}, the term in \(\cL\) proportional to \(\d_{d,0}\) arises from the coupling of the D\(q\)-branes to \(P[C_{p+1}]\), which is non-zero only for \(p=q\) and when the probe branes span all of the \(x_\parallel^\m\) directions, corresponding to \(d=0\). For \(M=1\) equation~\eqref{eq:multiple_coordinates_action} reduces to the action~\eqref{eq:class1_action} for class 1 embeddings depending on a single complex coordinate \(z = z_1\).

It is straightforward to confirm that the equations of motion following from the action~\eqref{eq:multiple_coordinates_action} admit solutions where \(y\) is an arbitrary holomorphic or antiholomorphic function of each of the \(z_j\) when \(d=0\) or \(d=4\), but not for other values of \(d\). For \(M>1\) this includes \(y\) that depends holomorphically on some of the \(z_j\) and antiholomorphically on the others. We refer to any such embedding depending holomorphically or antiholomorphically on each of the \(z_j\) as a holomorphic embedding.

For \(M > 1\) the only possibility admitting holomorphic embeddings is \(d=0\) --- the D\(q\)-branes span at least \((2M+1)\) of the \(x_\parallel^\m\) directions, namely \(t\) and the \((z_j,\zb_j)\), and do not span two of the \(x_\perp^i\) directions \((y,\yb)\). This leaves at most \((7-2M)\) directions that could potentially be ND. Since \(d\) is even, this implies that \(d \leq 2\) for \(M=2\) and \(d=0\) for \(M=3\). Thus, for \(M>1\) holomorphic embeddings exist only for \(d=0\), as claimed.

Since \(d=0\) requires \(p=q=a-1\), and \(p\) is bounded by \(2M \leq p \leq 6\), this greatly limits the possible holomorphic embeddings depending on multiple complex coordinates: they must be probe D\(p\)-brane embeddings for \(4 \leq p \leq 6\), spanning all of the \(x_\parallel^\m\) directions in the D\(p\)-brane background. All three possibilities are shown in table~\ref{tab:class1_holomorphic_multiple}.

\begin{table}[t]
\setlength{\tabcolsep}{2pt}
    \begin{subtable}{0.5\textwidth}
    \begin{tabularx}{0.98\textwidth}{| c | g g g g g Y Y : Y Y Y | c |}
            \hline
            D\(q\) & \(t\) & \(z_1\) & \(\zb_1\) & \(z_2\) & \(\zb_2\) & \(y\) & \(\yb\) &  \(x_\perp^3\) & \(x_\perp^4\) & \(x_\perp^5\)  &\(d\)
            \\ \hline
            D4 & \(\times\) & \(\times\) & \(\times\) & \(\times\) & \(\times\) & & & & & & \(0\)
            \\
            \hline
    \end{tabularx}
    \caption{\(p=4\)}
    \end{subtable}\begin{subtable}{0.5\textwidth}
    \hfill
    \begin{tabularx}{0.98\textwidth}{| c | g g g g g g Y Y : Y Y | c |}
        \hline
        D\(q\) & \(t\) & \(z_1\) & \(\zb_1\) & \(z_2\) & \(\zb_2\) & \(x_\parallel^5\) & \(y\) & \(\yb\) & \(x_\perp^3\) & \(x_\perp^4\)  & \(d\)
        \\ \hline
        D5 & \(\times\) & \(\times\) & \(\times\) & \(\times\) & \(\times\) & \(\times\) & & & & & \(0\)
        \\
        \hline
    \end{tabularx}
    \caption{\(p=5\)}
    \end{subtable}
    \\[1em]
    \begin{subtable}{\textwidth}\centering
    \begin{tabularx}{0.49\textwidth}{| c | g g g g g g g Y Y : Y | c|}
        \hline
        D\(q\) & \(t\) & \(z_1\) & \(\zb_1\) & \(z_2\) & \(\zb_2\) & \(z_3\) & \(\zb_3\) & \(y\) & \(\yb\) & \(x_\perp^3\) & \(d\)
        \\ \hline
        D6 & \(\times\) & \(\times\) & \(\times\) & \(\times\) & \(\times\) & \(\times\) & \(\times\) & & & & \(0\)
        \\
        \hline
    \end{tabularx}
    \caption{\(p=6\)}
    \end{subtable}
    \caption{Holomorphic class 1 D\(q\)-brane embeddings in D\(p\)-backgrounds, for which \(y\) is a holomorphic or antiholomorphic function of \(M>1\) complex coordinates \(z_j\), as constructed in appendix~\ref{sec:multiple_coords}. The shaded columns indicate \(x_\parallel^\m\) directions, while the crosses indicate directions spanned by the D\(q\)-branes. For \(p=4\) and \(p=5\) the only possibility with \(M>1\) is \(M=2\), since there are not enough \(x_\parallel^\m\) directions to form further class 1 embeddings. For \(p=6\) we can go up to \(M=3\) as indicated in the table. The case that \(M=2\) is trivially recovered by taking \(y\) to be independent of \((z_3,\zb_3)\).}
    \label{tab:class1_holomorphic_multiple}
\end{table}

As for the \(M=1\) case discussed in section~\ref{sec:class1_embedding}, the energy of holomorphic embeddings with \(M>1\) satisfies a BPS bound. To see this, we note that by introducing \(M\) uncorrelated signs \(s_j = \pm 1\), \(\D\) in equation~\eqref{eq:multiple_coordinates_Delta} may be written in any of several equivalent forms,
\begin{equation}\begin{aligned} \label{eq:multiple_coordinates_Delta_signs}
    \D &=  \Bigl(1 + \sum_j s_j \cY_4^{(j)} \Bigr)^2 + 4 H(r) \sum_j \le( \frac{1+s_j}{2}|\pb_j y|^2 + \frac{1 - s_j}{2} |\p_j y|^2 \ri)
    \\ &\phantom{=}
    + 4 H(r)^2 \sum_j \sum_{k>j}|\pb_j y \pb_k \yb - \pb_j \yb \pb_k y|^2
    + 2 \sum_j \sum_{k > j} (1 - s_j s_k) \cY_4^{(j)} \cY_4^{(k)} \, .
\end{aligned} \end{equation}
The form in which \(\D\) is written in equation~\eqref{eq:multiple_coordinates_Delta} corresponds to choosing all \(s_j = + 1\). Since each term in equation~\eqref{eq:multiple_coordinates_Delta_signs} is manifestly non-negative, we find that \(\D\) satisfies the inequalities
\begin{equation} \label{eq:multiple_coordinates_Delta_inequality_1}
    \D \geq \Bigl( 1 + \sum_j  s_j \cY_4^{(j)} \Bigr)^2 \, .
\end{equation}
Since this inequality is true for any assignments of the signs \(s_j\), it implies in particular that
\begin{equation} \label{eq:multiple_coordinates_Delta_inequality_2}
    \D \geq \Bigl( 1 + \sum_j  |\cY_4^{(j)}| \Bigr)^2 \, .
\end{equation}
This inequality is saturated for holomorphic embeddings. To see this, for each \(j\) choose \(s_j = +1\) or \(-1\) if \(y\) depends holomorphically or antiholomorphically on \(z_j\), respectively. Then all terms on the right-hand side of equation equation~\eqref{eq:multiple_coordinates_Delta_signs} vanish or cancel apart from the term appearing on the right-hand side of the inequality~\eqref{eq:multiple_coordinates_Delta_inequality_1}, so the inequality is saturated. Moreover, from equation~\eqref{eq:multiple_coordinates_Y} we see that \(\cY_4^{(j)}\) is positive or negative if \(y\) depends holomorphically or antiholomorphically on \(z_j\), respectively. Thus with this choice of the \(s_j\) we have that \(s_j \cY_4^{(j)} = |\cY_4^{(j)}|\), and so equations~\eqref{eq:multiple_coordinates_Delta_inequality_1} and~\eqref{eq:multiple_coordinates_Delta_inequality_2} become equivalent. Hence, equation~\eqref{eq:multiple_coordinates_Delta_inequality_2} is saturated too.

Because of equation~\eqref{eq:multiple_coordinates_Delta_inequality_2}, for \(d=0\) or \(d=4\) the D\(q\)-brane action~\eqref{eq:multiple_coordinates_action} satisfies the inequality
\begin{equation} \label{eq:multiple_coordinates_action_bound}
    S \leq - \int \diff t \, \Bigl( Z + \sum_j Y_d^{(j)} \Bigr) \, , \qquad (d = 0 \text{ or } 4) \, ,
\end{equation}
where
\begin{equation} \begin{aligned}\label{eq:multiple_coordinates_ZY}
    Z &= \frac{k T_q}{2^M} \int \diff z_1 \diff \zb_1 \cdots \diff z_M \diff \zb_M \diff \vec{x} \diff \vec{v} \, ,
    \\
    Y_d^{(j)} &= \frac{k T_q}{2^M} \int \diff z_1 \diff \zb_1 \cdots \diff z_M \diff \zb_M \diff \vec{x} \diff \vec{v} \, | \cY_d^{(j)} | \, .
\end{aligned}\end{equation}
This is a generalisation to \(M \geq 1\) of the \(M=1\) inequality in equations~\eqref{eq:class1_bps_bound} and~\eqref{eq:class1_ZY_integrals}, and the same considerations apply as for the \(M=1\) inequality with regard to regulating the integrals over the D\(q\)-branes' worldvolume. The discussion in the previous paragraph shows that the inequality in equation~\eqref{eq:multiple_coordinates_action_bound} is saturated for \(y\) that depends holomorphically or antiholomorphically on each of the \(z_j\). Equation~\eqref{eq:multiple_coordinates_action_bound} implies a lower bound on the energy \(E\) of the D\(q\)-branes,
\begin{equation} \label{eq:multiple_coordinates_energy_bound}
    E \geq  Z + \sum_j Y_d^{(j)} \, , \qquad (d = 0 \text{ or } 4) \, ,
\end{equation}
which is saturated for holomorphic embeddings.

Just as in the second line of equation~\eqref{eq:class1_ZY_integrals}, we can use the fact that \(\cY_d^{(j)}\) is proportional to the Jacobian for the change of variables \((z_j,\zb_j) \to (y,\yb)\) to exchange the integrals over \((z_j,\zb_j)\) in \(\cY_d^{(j)}\) for integrals over \((y,\yb)\), giving an expression for \(Y_d^{(j)}\) that manifestly depends only on the topological properties of the embedding function \(y\). Thus, the fact that holomorphic embeddings saturate the bounds on the action and energy in equations~\eqref{eq:multiple_coordinates_action_bound} and~\eqref{eq:multiple_coordinates_energy_bound} means that they extremise the action and minimise the energy.

On the other hand, for \(d=2\) equation~\eqref{eq:multiple_coordinates_Delta_inequality_2} implies a bound on the action similar to that in equation~\eqref{eq:class1_d2_bound},
\begin{equation}\begin{aligned}
    S &\leq - \int \diff t \Bigl(\tilde{Z} + \sum_j Y_d^{(j)} \Bigr) \, \qquad (d = 2)\, 
    \\
    \tilde{Z} &\equiv  \frac{k T_q}{2^M}\int \diff z_1 \diff \zb_1 \cdots \diff z_M \diff \zb_M \diff \vec{x} \diff \vec{v} \, H(r)^{-1/2} 
\end{aligned}\end{equation}
This bound is saturated for \(y\) depending holomorphically or antiholomorphically on each of the \(z_j\), but this does not imply that such \(y\) extremises the action, for the same reason as for \(M=1\) in section~\ref{sec:class1_embedding}: we would still need to extremise \(\tilde{Z}\), which requires setting \(y=0\).

\subsection{Supersymmetry}

We now show that the solutions constructed in the previous subsection preserve a fraction of the supersymmetry of the D\(p\)-brane background. We will specialise to holomorphic embeddings with \(M > 1\), the case of \(M=1\) already being covered in section~\ref{sec:class1_kappa}. As shown in the previous subsection and summarised in table~\ref{tab:class1_holomorphic_multiple}, this means we consider only probe D\(p\)-branes  that span all of the \(x_\parallel^\m\) directions of the D\(p\)-brane background (i.e. \(q=p\) and \(a = p+1\)). This will simplify notation somewhat.

As in section~\ref{sec:class1_kappa}, we seek solutions to the kappa symmetry condition \(\G \ve = \ve\), where for \(M\) complex coordinates on the probe D\(p\)-branes the kappa symmetry matrix is
\begin{equation} \label{eq:multiple_coordinates_kappa_matrix}
    \G = \frac{(-i)^M}{\sqrt{\lvert \det g \rvert}}\g_{01 \cdots p} J_{(p)} \, ,
\end{equation}
with \(J_{(p)}\) defined in equation~\eqref{eq:Jp}. Our notation is the same as in section~\ref{sec:class1_kappa}, with the exception of the necessary adaptations to account for multiple complex coordinates \(z_j\).

The probe branes span the \(x_\parallel^\m\) directions. As in the previous subsection we denote, the directions spanned by the probe branes as \(\xi = (t,z_1,\zb_1,\cdots, z_M,\zb_M,\vec{x})\). We will use the following indices to refer to the different components of \(\xi\),
\begin{equation} \label{eq:multiple_coordinates_xi_indices}
    \xi^0 = t \, , \quad
    \xi^{2j-1} = z_j \, , \quad
    \xi^{2j} = \zb_j \, , \quad
    \xi^\a = x^{(\a-2M)} \, ,
\end{equation}
with \(j\) running from \(1\) to \(M\) and \(\a\) running from \(2M+1\) to \(p\). In the event that \(2M+1 > p\), terms involving \(\a\) indices should be ignored as there are no \(\vec{x}\) directions.

We choose vielbeins such that the curved space Dirac matrices on the probe branes' worldvolume take the form
\begin{equation} \label{eq:multiple_coordinates_gamma_matrices}
\begin{aligned}
	\g_0 &= h^{-1} \G_0 \, ,
	\\
	\g_{2j-1} & = \frac{1}{2 h}(\G_{2j-1}  - i \G_{2j}) + \frac{h}{2} \le[ \p_j y (\G_8 - i \G_9) + \p_j \yb ( \G_8 + i \G_9 ) \ri] ,
	\\
	\g_{2j} & = \frac{1}{2 h}(\G_{2j-1}  + i \G_{2j}) + \frac{h}{2} \le[ \pb_j y (\G_8 - i \G_9) + \pb_j \yb ( \G_8 + i \G_9 ) \ri] ,
	\\
    \g_\a &= h^{-1} \G_\a \, .
\end{aligned}
\end{equation}
All of these Dirac matrices anticommute with one-another except for \(\g_{2j-1}\) and \(\g_{2j}\) for each \(j\). The product $\g_{01\cdots p}$ appearing in the kappa symmetry matrix~\eqref{eq:multiple_coordinates_kappa_matrix} thus factorises as
\begin{equation} \label{eq:multiple_coordinates_gamma_product_factorisation}
    \g_{01\cdots p} = \g_{0(2M+1)(2M+2) \cdots p}\g_{12 \cdots (2M)}\,.
\end{equation}
Evaluating the two products on the right-hand side using equation~\eqref{eq:multiple_coordinates_gamma_matrices} and the Clifford algbra satisfied by the \(\G_A\), we find
\begin{equation} \label{eq:multiple_coordinates_gamma_products}
\begin{aligned}
    \g_{0(2M+1)(2M+2) \cdots p} &= \frac{1}{h^{p+1-2M}}\G_{0 (2M+1)(2M+2) \cdots p} \, ,
	\\
    \g_{12 \cdots (2M)} &= \frac{i^M}{2^M h^{2M}} \le( \mathcal{S}_1 - \frac{h^2}{2}  \, \mathcal{S}_2 - \G_{89}  \frac{ h^4}{2} \mathcal{S}_3 \ri) \G_{12 \cdots (2M)}, 
\end{aligned}
\end{equation}
where in the second line we have defined the three combinations
\begin{subequations} \label{eq:multiple_coordinates_S}
\begin{align}
    &\mathcal{S}_1 =  \id - \smash[b]{\sum_{j}} \, \cY_4^{(j)} \, \G_{2j-1} \G_{2j} \G_{89}\, ,
    \label{eq:multiple_coordinates_S1}
    \\[2em]
    &\begin{aligned}
        \mathcal{S}_2 = \smash{\sum_{j}} \Bigl[&
        \le(\p_j y + \pb_j \yb\ri)\le(\G_{2j-1}\G_8 + \G_{2j} \G_9\ri)
        - i \le(\p_j y - \pb_j \yb\ri)\le(\G_{2j-1}\G_9 - \G_{2j} \G_8\ri)
        \\ &
        +\le(\p_j \yb + \pb_j y\ri)\le(\G_{2j-1}\G_8 - \G_{2j} \G_9\ri)
        + i \le(\p_j \yb - \pb_j y\ri)\le(\G_{2j-1}\G_9 + \G_{2j} \G_8\ri)
    \Bigr] \, ,
    \end{aligned}
    \label{eq:multiple_coordinates_S2}
    \\[1em]
    &\begin{aligned}
    \mathcal{S}_3 = \smash{\sum_{j} \sum_{k \neq j}} \Bigl[&
        \le(\p_j y \, \pb_k \yb - \pb_j y \, \p_k \yb\ri) \le(\G_{2j-1} \G_{2k} - \G_{2j} \G_{2k-1} \ri)
        \\ &
        + i \le( \p_j y \, \pb_k \yb + \pb_j y \,\p_k \yb\ri) \le(\G_{2j-1} \G_{2k-1} + \G_{2j} \G_{2k} \ri)
        \\ &
        - \le( \p_j y \, \p_k \yb + \pb_j \yb \, \pb_k y\ri) \le(\G_{2j-1} \G_{2k} + \G_{2j} \G_{2k-1} \ri)
        \\ &
        + i \le(\p_j y \, \p_k \yb  - \pb_j\yb \, \pb_k y\ri) \le(\G_{2j-1} \G_{2k-1} - \G_{2j} \G_{2k} \ri)
    \Bigr] \, .
    \end{aligned}
    \label{eq:multiple_coordinates_S3}
\end{align}
\end{subequations}
The determinant of the induced metric on the probe branes' worldvolume is given in equations~\eqref{eq:multiple_coordinates_metric_determinant} and~\eqref{eq:multiple_coordinates_Delta}. Substituting this (for \(q = p\)) and the expression for \(\g_{01 \cdots p}\) in equations~\eqref{eq:multiple_coordinates_gamma_product_factorisation} and~\eqref{eq:multiple_coordinates_gamma_products} into the kappa symmetry matrix in equation~\eqref{eq:multiple_coordinates_kappa_matrix}, we find
\begin{equation} \label{eq:multiple_coordinates_kappa_matrix_final}
    \G = \G' + \G'' + \G''' \, ,
\end{equation}
where we have defined
\begin{equation} \label{eq:multiple_coordinates_gamma_prime}
    \G' = \frac{1}{\sqrt{\D}} \cS_1  \G_{01 \cdots p}  J_{(p)}\, ,
    \qquad
    \G'' = -\frac{h^2}{2 \sqrt{\D}} \cS_2 \G_{01 \cdots p} J_{(p)} \, ,
    \qquad
    \G''' = \frac{h^4}{2 \sqrt{\D}} \G_{89} \cS_3 \G_{01 \cdots p} J_{(p)} \, .
\end{equation}
As a simple check of this result, for \(M=1\) we have that \(\cS_3 = 0\), so \(\G = \G' + \G''\), and the expressions for \(\G'\) and \(\G''\) can readily be seen to match those in equations~\eqref{eq:class1_kappa_projector} for \(q = p\) after some manipulation with the Clifford algebra satisfied by the \(\G_A\).

Similar to the analysis in section~\ref{sec:class1_kappa}, we now show that holomorphic embeddings admit spinors satisfying \(\G' \ve = \ve\) and \(\G'' \ve = \G''' \ve = 0\), and therefore satisfying the kappa symmetry condition \(\G \ve = \ve\). As noted in the previous subsection, if \(y\) depends holomorphically or antiholomorphically on each of the \(z_j\), then \(\D\) saturates the inequality in equation~\eqref{eq:multiple_coordinates_Delta_inequality_1}, so that
\begin{equation}
    \sqrt{\D} = 1 + \sum_j s_j \cY_4^{(j)} \, ,
\end{equation}
where \(s_j = 1\) or \(-1\) if \(y\) is holomorphic or antiholomorphic in \(z_j\), respectively. Thus, for such \(y\) a spinor \(\ve\) will satisfy \(\G' \ve = \ve\) if it satisfies the conditions
\begin{subequations} \label{eq:multiple_coordinates_kappa_conditions}
\begin{align}
    \G_{01 \cdots p} J_{(p)} \ve &= \ve \, ,
     \label{eq:multiple_coordinates_kappa_conditions_a}
    \\
    \G_{2j-1} \G_{2j} \G_{89} \ve &= - s_j \ve \, .
     \label{eq:multiple_coordinates_kappa_conditions_b}
\end{align}
\end{subequations}
Equation~\eqref{eq:multiple_coordinates_kappa_conditions_a} is the same as equation~\eqref{eq:Dp_brane_projection_conditions_unified}, satisfied by all of the Killing spinors of the D\(p\)-brane background. Thus, the additional constraints from requiring \(\G' \ve = \ve\) are those in equation~\eqref{eq:multiple_coordinates_kappa_conditions_b}.

If \(y\) depends holomorphically on \(z_j\), the second line in the definition of \(\cS_2\) in equation~\eqref{eq:multiple_coordinates_S2} vanishes. Similarly, if \(y\) depends antiholomorphically on \(z_j\) then the first line in the definition of \(\cS_2\) vanishes. Consequently, for \(y\) that depends holomorphically or antiholomorphically on each of the \(z_j\) we will have \(\G'' \ve = 0\) if
\begin{equation} \label{eq:multiple_coordinates_gamma_2_condition}
    \le(\G_{2j-1} \G_8 + s_j \G_{2j} \G_9 \ri) \ve = 0 \, ,
    \qquad
    \le(\G_{2j-1} \G_9 - s_j \G_{2j} \G_8 \ri) \ve = 0 \, .
\end{equation}
These two conditions are equivalent to each other, and to equation~\eqref{eq:multiple_coordinates_kappa_conditions_b}, since the Clifford algebra implies that
\begin{equation}\begin{aligned}
    \G_{2j-1} \G_8 + s_j \G_{2j} \G_9 &= \G_{2j-1} \G_8 \le(\id + s_j \G_{2j-1} \G_j \G_{89}\ri) \, ,
    \\
    \G_{2j-1} \G_9 - s_j \G_{2j} \G_8 &= \G_{2j-1} \G_9 \le(\id + s_j \G_{2j-1} \G_j \G_{89}\ri) \, ,
\end{aligned}\end{equation}
Thus, any \(\ve\) satisfying equation~\eqref{eq:multiple_coordinates_kappa_conditions_b} automatically satisfies equation~\eqref{eq:multiple_coordinates_gamma_2_condition}.

We now check that \(\G'''\ve = 0\) for \(\ve\) satisfying equation~\eqref{eq:multiple_coordinates_kappa_conditions}, which from equation~\eqref{eq:multiple_coordinates_gamma_prime} will happen if \(\cS_3 \ve = 0\). For each \(j\) and \(k\) in the sum in the definition of \(\cS_3\), the derivatives in the bottom two lines of equation~\eqref{eq:multiple_coordinates_S3} vanish if \(y\) is holomorphic or antiholomorphic in both \(z_j\) and \(z_k\), i.e. if \(s_j = s_k\), or equivalently if \(s_j s_k = 1\). We will then have \(\G''' \ve = 0\) if the combinations of Dirac matrices in the top two lines of equation~\eqref{eq:multiple_coordinates_S3} annihilate \(\ve\). Similarly, the derivatives in the top two lines of equation~\eqref{eq:multiple_coordinates_S3} vanish if \(y\) is holomorphic in \(z_j\) and antiholomorphic in \(z_k\), or vice versa, i.e. if \(s_j s_k = -1\), and then we will have \(\G'''\ve = 0\) if the combinations of Dirac matrices in the bottom two lines annihilate \(\ve\). In total, if \(y\) is holomorphic or antiholomorphic in each of the \(z_j\), we will have \(\G'' \ve = 0\) for any \(\ve\) satisfying equation~\eqref{eq:multiple_coordinates_kappa_conditions} that also satisfy
\begin{equation} \label{eq:multiple_coordinates_kappa_conditions_2}
    \le(\G_{2j-1} \G_{2k} - s_j s_k \G_{2j} \G_{2k-1} \ri) \ve = 0 \, ,
    \qquad
    \le(\G_{2j-1} \G_{2k-1} + s_j s_k \G_{2j} \G_{2k} \ri) \ve = 0 \, .
\end{equation}
But these conditions are automatically satisfied for any \(\ve\) satisfying equation~\eqref{eq:multiple_coordinates_kappa_conditions_2}, since the Clifford algebra implies that
\begin{equation}\begin{aligned}
    \le( \G_{2j-1} \G_{2k} - s_j s_k \G_{2j} \G_{2k-1} \ri) \ve &= \G_{2k-1} \G_{2j} \le(s_j s_k - \G_{2j-1} \G_{2j} \G_{89} \G_{2k-1} \G_{2k} \G_{89} \ri) \ve \, ,
    \\
    \le( \G_{2j-1} \G_{2k-1} + s_j s_k \G_{2j} \G_{2k} \ri) \ve &= \G_{2j} \G_{2k} \le(s_j s_k - \G_{2j-1} \G_{2j} \G_{89} \G_{2k-1} \G_{2k} \G_{89} \ri) \ve \, .
\end{aligned}\end{equation}
The right-hand sides of these two expressions manifestly vanish for \(\ve\) satisfying equation~\eqref{eq:multiple_coordinates_kappa_conditions_b}.

In summary, the kappa symmetry condition \(\G \ve = \ve\) will be satisfied if \(y\) is holomorphic or antiholomorphic in each of the \(z_j\), for those Killing spinors \(\ve\) of the D\(p\)-brane background satisfying equation~\eqref{eq:multiple_coordinates_kappa_conditions_b} with \(s_j = +1\) or \(-1\) if \(y\) is holomorphic or antiholomorphic in \(z_j\), respectively. With \(M\) complex coordinates \(z_j\), for each \(j\) equation~\eqref{eq:multiple_coordinates_kappa_conditions_b} eliminates half of the independent components of \(\ve\), so that in total such \(y\) preserves a fraction \(1/2^M\) of the supersymmetry of the D\(p\)-brane background.

\section{Holomorphic M2- and M5-branes}
\label{sec:m_brane}

In this appendix we demonstrate the existence of holomorphic embeddings of M2- and M5-branes in the extremal M2- and M5-brane backgrounds of eleven-dimensional supergravity, analogous the D-brane embeddings in D-brane backgrounds described in the main text. Each of these embeddings is specified by a holomorphic or antiholomorphic embedding function, \(y(z)\) or \(y(\zb)\). As for the D-brane embeddings, we classify the holomorphic embeddings according to whether \(y\) and \(z\) are formed from directions parallel or perpendicular to the M2- or M5-branes sourcing the supergravity background, as summarised in table~\ref{tab:embedding_classification}. The allowed holomorphic embeddings are listed in table~\ref{tab:M_brane_embeddings}.

Some special cases of the holomorphic embeddings that we describe are present in the literature already. For example, both the M2- and M5-brane backgrounds have near horizon limits, in which they become \(\ads[4] \times \sph[7]\) or \(\ads[7] \times \sph[4]\), respectively. Probe M2-brane embeddings in \(\ads[4] \times \sph[7]\) with \(\ads[2] \times \sph[1]\) worldvolume and probe M5-brane embeddings in \(\ads[7] \times \sph[4]\) with \(\ads[5] \times \sph[1]\) worldvolume have both been constructed~\cite{Drukker:2008jm,Gutperle:2020gez}. These embeddings are qualitatively similar to the \(\ads[3] \times \sph[1]\) probe D3-branes in \(\ads[5] \times \sph[5]\) mentioned in section~\ref{sec:d3} which correspond to class 1 holomorphic embeddings with \(y \propto z^{-1}\). Concretely, the \(\ads[2] \times \sph[1]\) M2-brane embedding is, in our language, the near-horizon limit of the class 1 embedding of a probe M2-brane in the M2-brane background, listed in table~\ref{tab:M2_brane_class1}. Similarly, the \(\ads[5] \times \sph[1]\) M5-brane is the near-horizon limit of the class 1 embedding of a probe M5-brane spanning four of the parallel directions in the M5-brane background, listed in in the top row of table~\ref{tab:M5_brane_class1}.

\begin{table}[t]
\setlength{\tabcolsep}{2pt}
    \begin{subtable}{0.5\textwidth}
    \begin{tabularx}{0.98\textwidth}{| c | g g g Y Y : Y Y Y Y Y Y |}
            \hline
            M\(q\) & \(t\) & \(z\) & \(\zb\) & \(y\) & \(\yb\) & \(x_\perp^3\) & \(x_\perp^4\) & \(x_\perp^5\)& \(x_\perp^6\) & \(x_\perp^7\) & \(x_\perp^8\)
            \\ \hline
            M2 & \(\times\) & \(\times\) & \(\times\) & & & & & & & & 
            \\\hline
    \end{tabularx}
    \caption{Class 1, M2-brane background}
    \label{tab:M2_brane_class1}
    \end{subtable}\begin{subtable}{0.5\textwidth}
    \hfill
    \begin{tabularx}{0.98\textwidth}{| c | g g g g g g Y Y : Y Y Y |}
            \hline
            M\(q\) & \(t\) & \(z\) & \(\zb\) & \(x^3_\parallel\) & \(x^4_\parallel\) & \(x^5_\parallel\) & \(y\) & \(\yb\) & \(x_\perp^1\) & \(x_\perp^2\)& \(x_\perp^3\)
            \\ \hline
            M5 & \(\times\) & \(\times\) & \(\times\) & \(\times\) & \(\times\) & \(\times\) & & & & &
            \\
            M5 & \(\times\) & \(\times\) & \(\times\) & \(\times\) & & & & & \(\times\) & \(\times\) &
            \\\hline
    \end{tabularx}
    \caption{Class 1, M5-brane background}
    \label{tab:M5_brane_class1}
    \end{subtable}
    \\[1em]
    \begin{subtable}{0.5\textwidth}
    \begin{tabularx}{0.98\textwidth}{| c | g g g Y Y : Y Y : Y Y Y Y |}
            \hline
            M\(q\) & \(t\) & \(x_\parallel^1\) & \(x_\parallel^2\) & \(z\) & \(\zb\) & \(y\) & \(\yb\)  & \(x_\perp^5\)& \(x_\perp^6\) & \(x_\perp^7\) & \(x_\perp^8\)
            \\ \hline
            M2 & \(\times\) & & & \(\times\) & \(\times\) & & & & & & 
            \\
            M5 & \(\times\) & \(\times\) & & \(\times\) & \(\times\) & & & \(\times\) & \(\times\) & &
            \\\hline
    \end{tabularx}
    \caption{Class 2, M2-brane background}
    \label{tab:M2_brane_class2}
    \end{subtable}\begin{subtable}{0.5\textwidth}
    \hfill
    \begin{tabularx}{0.98\textwidth}{| c | g g g g g g Y Y : Y Y : Y |}
            \hline
            M\(q\) & \(t\) & \(x^1_\parallel\) & \(x^2_\parallel\) & \(x^3_\parallel\) & \(x^4_\parallel\) & \(x^5_\parallel\) & \( z\) & \(\zb\) & \(y\) & \(\yb\) & \(x_\perp^5\)
            \\ \hline
            M5 & \(\times\) & \(\times\) & \(\times\) & \(\times\) & & & \(\times\) & \(\times\) & & &
            \\\hline
    \end{tabularx}
    \caption{Class 2, M5-brane background}
    \label{tab:M5_brane_class2}
    \end{subtable}
    \\[1em]
    \begin{subtable}{\textwidth}\centering
    \begin{tabularx}{0.49\textwidth}{| c | g : g g : g g : g Y Y Y Y Y |}
            \hline
            M\(q\) & \(t\) & \( z\) & \(\zb\) & \(y\) & \(\yb\) & \(x_\parallel^5\) & \(x^1_\perp\) & \(x^2_\perp\) & \(x^3_\perp\) & \(x^4_\perp\) &\(x^5_\perp\)
            \\ \hline
            M5 & \(\times\) & \(\times\) & \(\times\) & & & \(\times\) & \(\times\) & \(\times\) & & &
            \\\hline
    \end{tabularx}
    \caption{Class 3, M5-brane background}
    \label{tab:M5_brane_class3}
    \end{subtable}
    \caption{Holomorphic probe M-brane embeddings in the M2- and M5-brane backgrounds of M-theory. The different classes correspond to whether we form the complex coordinates \(z\) and \(y\) out of \(x_\parallel^\m\) or \(x_\perp^i\) directions, as  indicated in table~\ref{tab:embedding_classification}. The shaded columns in each table indicate the \(x_\parallel^\m\) directions, and the crosses show the directions spanned by the probe branes. As discussed in section~\ref{sec:11d_class3} there are two additional class 3 embeddings in the M5-brane background that are consistent with the M2- or M5-brane equations of motion, but which we do not include in table~\ref{tab:M5_brane_class3} since we do not expect them to be supersymmetric.}
    \label{tab:M_brane_embeddings}
\end{table}

The bosonic fields of eleven-dimensional supergravity are the metric and a three-form gauge field \(C_3\). In the M2-brane background, these fields take the form (see e.g. ref.~\cite{Blumenhagen:2013fgp})
\begin{equation}\begin{aligned} \label{eq:M2_background}
    \diff s^2 &= H(r)^{-2/3} \diff x_\parallel^2 + H(r)^{1/3} \diff x_\perp^2 \, ,
    \\
    C_3 &= \le[H(r)^{-1} -1\ri] \diff x_\parallel^0 \wedge \diff x_\parallel^1 \wedge \diff x_\parallel^2 \, ,
\end{aligned}\end{equation}
where  \(\diff x_\parallel^2 = \h_{\m\n} \diff x_\parallel^\m \diff x_\parallel^\n\), with \(\h_{\m\n}\) the three-dimensional Minkowski metric in mostly-plus signature and \(\diff x_\perp^2 = \d_{ij} \diff x_\perp^i \diff x_\perp^j\). The harmonic function appearing in this solution is
\begin{equation}
    H(r) = 1 + \frac{L^6}{r^6} \, ,
\end{equation}
where \(r^2 = \d_{ij} x_\perp^i x_\perp^j\), and \(L\) is related to the number \(N\) of M2-branes and the eleven-dimensional Planck length \(\ell_P\) by \(L^6 = 2^5 \pi^2 \ell_P^6 N\).

The gauge field of the M5-brane background is most conveniently expressed in terms of its dual, six-form gauge field \(C_6\), defined by \(* \diff C_3 = \diff C_6 - C_3 \wedge \diff C_3\)~\cite{Bandos:1997gd}, where \(*\) is the Hodge star. The metric and six-form of the M5-brane solution take the form
\begin{equation}\begin{aligned} \label{eq:M5_background}
    \diff s^2 &= H(r)^{-1/3} \diff x_\parallel^2 + H(r)^{2/3} \diff x_\perp^2 \, ,
    \\
    C_6 &= \le[H(r)^{-1}-1\ri] \diff x_\parallel^0 \wedge \diff x_\parallel^1 \wedge \cdots \wedge \diff x_\parallel^5 \, ,
\end{aligned}\end{equation}
where now  \(\diff x_\parallel^2 = \h_{\m\n} \diff x_\parallel^\m \diff x_\parallel^\n\), with \(\h_{\m\n}\) the six-dimensional Minkowski metric in mostly-plus signature and \(\diff x_\perp^2 = \d_{ij} \diff x_\perp^i \diff x_\perp^j\). The harmonic function appearing in the M5-brane solution is
\begin{equation}
    H(r) = 1 + \frac{L^3}{r^3} \, ,
\end{equation}
where \(r^2 = \d_{ij} x_\perp^i x_\perp^j\), and \(L\) is related to the number \(N\) of M5-branes and the eleven-dimensional Planck length \(\ell_P\) by \(L^6 = L^3 = \pi \ell_P^3 N\).

We will wish to embed probe M2- and M5-branes into the supergravity backgrounds in equations~\eqref{eq:M2_background} and~\eqref{eq:M5_background}. For this purpose we need the bosonic parts of the M2- and M5-brane actions. The bosonic part of the M2-brane action is 
\begin{equation} \label{eq:M2_action}
    S = - T_\mathrm{M2} \int \, \diff^3 \xi \sqrt{|g|} + T_\mathrm{M2} \int P[C_3] \, ,
\end{equation}
where \(T_\mathrm{M2} = (4\pi^2 \ell_P^{3})^{-1}\) is the M2-brane tension. We will use coordinates \(\xi = (t,z,\zb)\) on the M2-branes, where \(t = x_\parallel^0\). 

The action for a probe M5-brane is complicated by the presence of a two-form gauge field \(A\) with self-dual field strength \(F = \diff A\) on the M5-brane's worldvolume. Several actions exist, which implement the self-duality constraint in different ways~\cite{Aganagic:1997zq,Pasti:1997gx,Bandos:1997ui,Ko:2013dka}. These actions are believed to be classically equivalent~\cite{Bandos:1997gm,Ko:2013dka}. We will follow the approach of refs.~\cite{Pasti:1997gx,Bandos:1997ui}, in which the action contains an auxiliary scalar field \(\vf\). In this approach, the bosonic part of the M5-brane action is
\begin{multline} \label{eq:M5_action}
    S = - T_\mathrm{M5} \int \diff^6 \xi \le[\sqrt{|\det(g + i \tilde{E})|} + \frac{\sqrt{|g|}}{4(\p \vf)^2} \p_m \vf \, E^{*mnl} E_{nlp} \, \p^p \vf \ri]
    \\
    + T_\mathrm{M5} \int \le(P[C_6] + \frac{1}{2} F \wedge P[C_3] \ri),
\end{multline}
where \(E \equiv F + P[C_3]\), \(E^{\star m n l} = \frac{1}{6 \sqrt{\lvert \det g \rvert}} \e^{mnlpqr} E_{pqr}\), and \(\tilde{E}_{mn} = E^*_{mn}{}^l \p_l \vf / \sqrt{(\p \vf)^2}\). The M5-brane tension is \(T_\mathrm{M5} = (2\pi)^{-5} \ell_P^{-6}\). The self-duality constraint follows from a local symmetry of the action in equation~\eqref{eq:M5_action}~\cite{Pasti:1997gx,Bandos:1997ui}.

We now show the existence of the holomorphic embeddings listed in table~\ref{tab:M_brane_embeddings}. Throughout this appendix we use the same notation as in the main text, summarised in table~\ref{tab:notation}. We take our probe M2-branes to span \(\xi = (t,z,\zb)\), and our probe M5-branes to span \(\xi = (t,z,\zb,\vec{x},\vec{v})\), where \(\vec{x}\) and \(\vec{v}\) are formed from \(x_\parallel^\m\) and \(x_\perp^i\) directions, respectively. We denote by \(a\) the total number of \(x_\parallel^\m\) directions spanned by the probe branes. Aside from \((y,\yb)\), the \(x_\parallel^\m\) and \(x_\perp^i\) directions not spanned by the probe branes are denoted \(\vec{U}\) and \(\vec{W}\), respectively.

\subsection{Class 1}
\label{sec:11d_class1}

For class 1 embeddings we form \(z\) from \(x_\parallel^\m\) directions and \(y\) from \(x_\perp^i\) directions of the supergravity backgrounds in equations~\eqref{eq:M2_background} and~\eqref{eq:M5_background}, as in equation~\eqref{eq:class1_z}. We consider probe M2-branes spanning \(\xi = (t,z,\zb)\), which are all \(x_\parallel^\m\) directions and hence \(a = 3\). We also consider probe M5-branes spanning \(\xi = (t,z,\zb,\vec{x},\vec{v})\). In the M2-brane background there are only three \(x_\parallel^\m\) directions, so there are no \(\vec{x}\) directions and hence probe M5-branes also have \(a=3\). In the M5-brane background there are more \(x_\parallel^\m\) directions, so \(a\) can take any value in the range \(3 \leq a \leq 6\).

For both probe M2- and M5-branes, we make the ansatz that \(y = y(z,\zb)\), while the other embedding scalars \(\vec{U}\) and \(\vec{W}\) are constant. For probe M5-branes we make the further ansatz that the worldvolume two-form gauge field \(A\) vanishes and that the auxiliary scalar field takes the form \(\vf = \vf(t)\). We substitute this ansatz into the M2- and M5- brane actions~\eqref{eq:M2_action} and~\eqref{eq:M5_action}, evaluated in the M2- and M5-brane backgrounds~\eqref{eq:M2_background} and~\eqref{eq:M5_background}. The result is that the action for a probe M\(q\)-brane in the M\(p\)-brane background, evaluated on our ansatz, may be written in the unified form
\begin{equation}\begin{aligned} \label{eq:class1_M_brane_action}
    S_1 &= - \frac{T_{\mathrm{M}q}}{2} \int \diff t \diff z \diff \zb \diff \vec{x} \diff \vec{v}\, \cL_1 \, ,
    \\
    \cL_1 &= H(r)^{(d-4)/4}\sqrt{\le[1 + H(r) \le(|\p y|^2 + |\pb y|^2 \ri) \ri]^2    - 4 H(r)^2 |\p y|^2 |\pb y|^2} - \d_{d,0} \le[H(r)^{-1} - 1\ri] ,
\end{aligned}\end{equation}
where \(r^2 = |y|^2 + W^2\) for a probe M2-brane, \(r^2 = |y|^2 + v^2 + W^2\) for a probe M5-brane, and for a probe M2-brane \(\diff \vec{x}\) and \(\diff \vec{v}\) should be dropped from the above expression. In equation~\eqref{eq:class1_M_brane_action} we have defined
\begin{equation} \label{eq:class1_M_brane_d}
    d = \frac{2}{9}(p+1)(q+1) + 4 - 2 a \, .
\end{equation}
The different possibilities for the numbers \((p,q,a)\) and the resulting values of \(d\) are given in table~\ref{tab:class1_pqa}.

\begin{table}[t]
    \centering
    \begin{tabular}{|c:c:c|c|}
     \hline
     \(p\) & \(q\) & \(a\) & \(d\)  \\ \hline
     \(2\) & \(2\) & \(3\) & \(0\)  \\
     \(2\) & \(5\) & \(3\) & \(2\)  \\
     \(5\) & \(2\) & \(3\) & \(2\)  \\
     \(5\) & \(5\) & \(3\) & \(6\)  \\
     \(5\) & \(5\) & \(4\) & \(4\) \\
     \(5\) & \(5\) & \(5\) & \(2\) \\
     \(5\) & \(5\) & \(6\) & \(0\) \\
     \hline
\end{tabular}
\caption{All possible assignments of \(p\) , \(q\), and \(a\) for a class 1 embedding of a probe M\(p\)-brane in the M\(q\)-brane background of eleven-dimensional supergravity. The final column is the resulting value of \(d\), defined in equation~\eqref{eq:class1_M_brane_d}.}
\label{tab:class1_pqa}
\end{table}

The term proportional to \(\d_{d,0}\) in equation~\eqref{eq:class1_M_brane_action} arises from a probe M2-branes' coupling to \(C_3\) in the M2-brane background, or a probe M5-brane's coupling to \(C_6\) in the M5-brane background --- from table~\ref{tab:class1_pqa} we see that \(d=0\) for \((p,q,a) = (2,2,3)\) or \((5,5,6)\), i.e. when a probe M\(p\)-brane spans all of the \(x_\parallel^\m\) directions in the M\(p\)-brane background, in which case the pullback of \(C_{p+1}\) is trivial.

The action in equation~\eqref{eq:class1_M_brane_action} takes exactly the same form as that for class 1 D-brane embeddings in equation~\eqref{eq:class1_action}. Thus, the analysis of section~\ref{sec:class1_embedding} immediately implies that the action~\eqref{eq:class1_M_brane_action} admits solutions where \(y\) is an arbitrary holomorphic or antiholomorphic function of \(z\) if and only if \(d\) is a multiple of four. From table~\ref{tab:class1_pqa} we see that there are three possible assignments of \((p,q,a)\) for which this is the case, namely the two \(d=0\) configurations already mentioned, and the M5-brane in M5-brane background embedding with \((p,q,a) = (5,5,4)\), for which \(d=4\). These three cases correspond to the three class 1 embeddings in table~\ref{tab:M_brane_embeddings}.

The analysis of section~\ref{sec:class1_embedding} also implies that the energy of holomorphic M2- and M5-brane embeddings saturates a BPS bound similar to equation~\eqref{eq:class1_bps_bound}. Although we do not check the kappa symmetry of the embeddings here, by analogy to the analysis for D-branes in section~\ref{sec:class1_kappa} we expect that holomorphic M2- and M5-brane embeddings will preserve a fraction of the supersymmetry of their supergravity backgrounds, one-half for \(d=0\) and one-quarter for \(d=4\). As a consistency check, when \(y\) is constant the two \(d=0\) examples correspond to parallel M2- or M5-brane pairs, which preserve supersymmetry, and the \(d=4\) example corresponds to two stacks of M5-branes with a (3+1)-dimensional intersection, the only dimensionality of an M5-brane intersection consistent with supersymmetry~\cite{Papadopoulos:1996uq}.

For completeness, we note that for \((p,q,a) = (5,2,3)\), i.e. a probe M5-brane in the M2-brane background, our ansatz that the M5-brane's worldvolume gauge field vanishes is inconsistent, by the same argument as made for D-branes around equation~\eqref{eq:extra_wz_term}. Concretely, for such a configuration, the term in the M5-brane action~\eqref{eq:M5_action} containing \(F \wedge P[C_3]\) acts as a source for worldvolume gauge field. However, this configuration is not one of the holomorphic embeddings listed in table~\ref{tab:M_brane_embeddings}, since from table~\ref{tab:class1_pqa} this configuration has \(d=2\).

\subsection{Class 2}
\label{sec:11d_class2}

For class 2 embeddings, we form both \(z\) and \(y\) from \(x_\perp^\m\) directions, as in equation~\eqref{eq:class2_z}. Thus, a probe M2-brane spanning \(\xi = (t,z,\zb)\) spans only one \(x_\parallel^\m\) direction, i.e. has \(a=1\). A probe M5-brane spanning \(\xi = (t,z,\zb,\vec{x},\vec{v})\) has \((a-1)\) \(\vec{x}\) directions and consequently \((4-a)\) \(\vec{v}\) directions. In the M2-brane background, since there are only three \(x_\parallel^\m\) directions, \(a\) for a probe M5-brane takes values in the range \(1 \leq a \leq 3\). In the M5-brane background, four of the five \(x_\perp^i\) directions have been used to form \(z\) and \(y\), leaving only one \(x_\perp^i\) direction that could be a \(\vec{v}\) direction. Hence for a probe M5-brane in the M5-brane background, \(3 \leq a \leq 4\).

As in the previous subsection, we make the ansatz \(y = y(z,\zb)\) and constant \(\vec{U}\) and \(\vec{W}\), as well as for a probe M5-brane \(\vf = \vf(t)\) and \(A=0\). With this ansatz, the action for a probe M\(q\)-brane in the M\(p\)-brane background evaluates to 
\begin{equation} \begin{aligned} \label{eq:class2_M_brane_action}
    S_2 &= - \frac{T_{\mathrm{M}q}}{2} \int \diff t \diff z \diff \zb \, \cL_2 \, ,
    \\
    \cL_2 &= H(r)^{(d-4)/4} \sqrt{\le(1 +|\p y|^2 + |\pb y|^2 \ri)^2   - 4 |\p y|^2 |\pb y|^2} \, ,
\end{aligned} \end{equation}
where \(r^2 = |z|^2 + |y|^2 + W^2\) for a probe M2-brane and \(r^2 = |z|^2 + |y|^2 + v^2 + W^2\) for a probe M5-brane, and where \(d\) is again given by equation~\eqref{eq:class1_M_brane_d}.

The action in equation~\eqref{eq:class2_M_brane_action} takes the same form as the action for class 2 D-brane embeddings in equation~\eqref{eq:class2_action}. Thus, the analysis of section~\ref{sec:class2_embedding} implies that holomorphic embeddings of M2- and M5-branes exist for those combinations of \((p,q,a)\) such that \(d=4\) in equation~\eqref{eq:class1_M_brane_d}. It also implies that the energy of such a holomorphic embedding saturates a BPS bound similar to that in equation~\eqref{eq:class2_bps_bound}.

Of the various values of \((p,q,a)\) consistent with the considerations in the opening paragraph of this subsection, we find from equation~\eqref{eq:class1_M_brane_d} that \(d=4\) for \((p,q,a) = (2,2,1)\), \((2,5,2)\), and \((5,5,4)\). These three combinations correspond to the three class 2 embeddings listed in tables~\ref{tab:M2_brane_class2} and~\ref{tab:M5_brane_class2}.

\subsection{Class 3}
\label{sec:11d_class3}

For class 3 embeddings we form both \(z\) and \(y\) from \(x_\parallel^\m\) directions, as in equation~\eqref{eq:class3_z}. This means that we cannot construct class 3 embeddings in the M2-brane background, as this background does not have enough \(x_\parallel^\m\) directions. Therefore, in this subsection we restrict to embeddings in the M5-brane background. A probe M2-brane spanning \(\xi = (t,z,\zb)\) spans \(a = 3\) of the \(x_\parallel^\m\) directions. A probe M5-brane spanning \(\xi = (t,z,\zb,\vec{x},\vec{v})\) spans \(a = 3\) or \(a=4\) of the \(x_\parallel^\m\) directions. The upper bound on \(a\) arises because two of the six \(x_\parallel^\m\) directions are used to form the complex coordinate \(y\), which is not spanned by the probe branes.

Substituting the same ansatz as in the previous sections, \(y = y(z,\zb)\) and for a probe M5-brane \(A = 0\) and \(\vf = \vf(t)\), into the M2- and M5-brane actions~\eqref{eq:M2_action} and~\eqref{eq:M5_action}, we find that the action for a probe M\(q\)-brane in the M5-brane background takes the form
\begin{equation} \begin{aligned} \label{eq:class3_M_brane_action}
    S_3 &= - \frac{T_{\mathrm{M}q}}{2} \int \diff t \diff z \diff \zb \, \cL_3 \, ,
    \\
    \cL_3 &= H(r)^{(d-4)/4} \sqrt{\le(1 +|\p y|^2 + |\pb y|^2 \ri)^2   - 4 |\p y|^2 |\pb y|^2} \, ,
\end{aligned} \end{equation}
with \(r^2 = W^2\) for a probe M2-brane and \(r^2 = v^2 + W^2\) for a probe M5-brane, and where \(d\) is given again by equation~\eqref{eq:class1_M_brane_d}.

The action in equation~\eqref{eq:class3_M_brane_action} takes the same form as the action for class 3 D-brane embeddings in equation~\eqref{eq:class3_action}. Thus, the analysis of that section implies that the equations of motion following from the action in equation~\eqref{eq:class3_M_brane_action} admit solutions with arbitrary holomorphic or antiholomorphic \(y\) for any \(d\), and that the action of such embeddings saturates a bound similar to that in equation~\eqref{eq:class_3_action_bound}.

There are only three combinations of \((p,q,a)\) compatible with the considerations in the opening paragraph of this subsection. They are \((p,q,a) = (5,2,3)\), \((5,5,3)\), and \((5,5,4)\). The corresponding values of \(d(p,q,a)\) are
\begin{equation}
    d(5,2,3) = 2 \, , \qquad
    d(5,5,3) = 6\, , \qquad 
    d(5,5,4) = 4 \, .
\end{equation}
By analogy to the D-brane embeddings discussed in the main text, we expect that only for \(d=4\) does the probe brane preserve a fraction of the supersymmetry of the background. More concretely, \((p,q,a) = (5,2,3)\) describes an M2-brane and M5-brane intersecting over a 2-brane, while the \((p,q,a) = (5,5,3)\) describes M5-branes intersecting over a 2-brane. Neither of these intersections is compatible with supersymmetry~\cite{Strominger:1995ac,Papadopoulos:1996uq,Tseytlin:1996bh}. On the other hand, \((p,q,a) = (5,5,4)\) corresponds to M5-branes intersecting over a 3-brane, which \emph{is} compatible with supersymmetry~\cite{Papadopoulos:1996uq}. This is the case with \(d=4\), and the only one we show in table~\ref{tab:M5_brane_class3}.

\bibliographystyle{JHEP}
\bibliography{holoprobe}

\providecommand{\href}[2]{#2}\begingroup\raggedright\begin{thebibliography}{10}

\bibitem{holomorphic_branes}
P.~Capuozzo, J.~Holden, A.~O'Bannon, J.~Ratcliffe, R.~Rodgers and B.~Suzzoni, \emph{{Supersymmetric Holomorphic Masses in AdS/CFT with Flavour}},  \href{https://arxiv.org/abs/2512.19688}{{\ttfamily 2512.19688}}.

\bibitem{Gauntlett:1997ss}
J.P.~Gauntlett, J.~Gomis and P.K.~Townsend, \emph{{BPS bounds for world volume branes}}, \href{https://doi.org/10.1088/1126-6708/1998/01/003}{\emph{JHEP} {\bfseries 01} (1998) 003} [\href{https://arxiv.org/abs/hep-th/9711205}{{\ttfamily hep-th/9711205}}].

\bibitem{Maldacena:1997re}
J.M.~Maldacena, \emph{{The Large N limit of superconformal field theories and supergravity}}, \href{https://doi.org/10.1023/A:1026654312961, 10.4310/ATMP.1998.v2.n2.a1}{\emph{Int. J. Theor. Phys.} {\bfseries 38} (1999) 1113} [\href{https://arxiv.org/abs/hep-th/9711200}{{\ttfamily hep-th/9711200}}].

\bibitem{Gubser:1998bc}
S.~Gubser, I.R.~Klebanov and A.M.~Polyakov, \emph{{Gauge theory correlators from noncritical string theory}}, \href{https://doi.org/10.1016/S0370-2693(98)00377-3}{\emph{Phys. Lett. B} {\bfseries 428} (1998) 105} [\href{https://arxiv.org/abs/hep-th/9802109}{{\ttfamily hep-th/9802109}}].

\bibitem{Witten:1998qj}
E.~Witten, \emph{{Anti-de Sitter space and holography}}, \href{https://doi.org/10.4310/ATMP.1998.v2.n2.a2}{\emph{Adv. Theor. Math. Phys.} {\bfseries 2} (1998) 253} [\href{https://arxiv.org/abs/hep-th/9802150}{{\ttfamily hep-th/9802150}}].

\bibitem{Karch:2002sh}
A.~Karch and E.~Katz, \emph{{Adding flavor to AdS / CFT}}, \href{https://doi.org/10.1088/1126-6708/2002/06/043}{\emph{JHEP} {\bfseries 06} (2002) 043} [\href{https://arxiv.org/abs/hep-th/0205236}{{\ttfamily hep-th/0205236}}].

\bibitem{Blumenhagen:2013fgp}
R.~Blumenhagen, D.~L{\"u}st and S.~Theisen, \emph{{Basic concepts of string theory}}, Theoretical and Mathematical Physics, Springer, Heidelberg, Germany (2013), \href{https://doi.org/10.1007/978-3-642-29497-6}{10.1007/978-3-642-29497-6}.

\bibitem{Polchinski:1998rr}
J.~Polchinski, \emph{{String theory. Vol. 2: Superstring theory and beyond}}, Cambridge Monographs on Mathematical Physics, Cambridge University Press (12, 2007), \href{https://doi.org/10.1017/CBO9780511618123}{10.1017/CBO9780511618123}.

\bibitem{Skenderis:2002vf}
K.~Skenderis and M.~Taylor, \emph{{Branes in AdS and p p wave space-times}}, \href{https://doi.org/10.1088/1126-6708/2002/06/025}{\emph{JHEP} {\bfseries 06} (2002) 025} [\href{https://arxiv.org/abs/hep-th/0204054}{{\ttfamily hep-th/0204054}}].

\bibitem{Itzhaki:1998dd}
N.~Itzhaki, J.M.~Maldacena, J.~Sonnenschein and S.~Yankielowicz, \emph{{Supergravity and the large N limit of theories with sixteen supercharges}}, \href{https://doi.org/10.1103/PhysRevD.58.046004}{\emph{Phys. Rev. D} {\bfseries 58} (1998) 046004} [\href{https://arxiv.org/abs/hep-th/9802042}{{\ttfamily hep-th/9802042}}].

\bibitem{Boonstra:1998mp}
H.J.~Boonstra, K.~Skenderis and P.K.~Townsend, \emph{{The domain wall / QFT correspondence}}, \href{https://doi.org/10.1088/1126-6708/1999/01/003}{\emph{JHEP} {\bfseries 01} (1999) 003} [\href{https://arxiv.org/abs/hep-th/9807137}{{\ttfamily hep-th/9807137}}].

\bibitem{Skenderis:1998dq}
K.~Skenderis, \emph{{Field theory limit of branes and gauged supergravities}}, \href{https://doi.org/10.1002/(SICI)1521-3978(20001)48:1/3<205::AID-PROP205>3.0.CO;2-F}{\emph{Fortsch. Phys.} {\bfseries 48} (2000) 205} [\href{https://arxiv.org/abs/hep-th/9903003}{{\ttfamily hep-th/9903003}}].

\bibitem{Kanitscheider:2008kd}
I.~Kanitscheider, K.~Skenderis and M.~Taylor, \emph{{Precision holography for non-conformal branes}}, \href{https://doi.org/10.1088/1126-6708/2008/09/094}{\emph{JHEP} {\bfseries 09} (2008) 094} [\href{https://arxiv.org/abs/0807.3324}{{\ttfamily 0807.3324}}].

\bibitem{Wiseman:2008qa}
T.~Wiseman and B.~Withers, \emph{{Holographic renormalization for coincident Dp-branes}}, \href{https://doi.org/10.1088/1126-6708/2008/10/037}{\emph{JHEP} {\bfseries 10} (2008) 037} [\href{https://arxiv.org/abs/0807.0755}{{\ttfamily 0807.0755}}].

\bibitem{Karch:2000ct}
A.~Karch and L.~Randall, \emph{{Locally localized gravity}}, \href{https://doi.org/10.1088/1126-6708/2001/05/008}{\emph{JHEP} {\bfseries 05} (2001) 008} [\href{https://arxiv.org/abs/hep-th/0011156}{{\ttfamily hep-th/0011156}}].

\bibitem{Karch_2001a}
A.~Karch and L.~Randall, \emph{Open and closed string interpretation of susy cft’s on branes with boundaries}, \href{https://doi.org/10.1088/1126-6708/2001/06/063}{\emph{Journal of High Energy Physics} {\bfseries 2001} (2001) 063–063}.

\bibitem{Gukov:2006jk}
S.~Gukov and E.~Witten, \emph{{Gauge Theory, Ramification, And The Geometric Langlands Program}},  \href{https://arxiv.org/abs/hep-th/0612073}{{\ttfamily hep-th/0612073}}.

\bibitem{Witten:2007td}
E.~Witten, \emph{{Gauge theory and wild ramification}},  \href{https://arxiv.org/abs/0710.0631}{{\ttfamily 0710.0631}}.

\bibitem{Horowitz:1991cd}
G.T.~Horowitz and A.~Strominger, \emph{{Black strings and P-branes}}, \href{https://doi.org/10.1016/0550-3213(91)90440-9}{\emph{Nucl. Phys. B} {\bfseries 360} (1991) 197}.

\bibitem{deAzcarraga:1989mza}
J.A.~de~Azcarraga, J.P.~Gauntlett, J.M.~Izquierdo and P.K.~Townsend, \emph{{Topological Extensions of the Supersymmetry Algebra for Extended Objects}}, \href{https://doi.org/10.1103/PhysRevLett.63.2443}{\emph{Phys. Rev. Lett.} {\bfseries 63} (1989) 2443}.

\bibitem{Sato:1998ax}
T.~Sato, \emph{{The Space-time superalgebras in a massive background via brane probes}}, \href{https://doi.org/10.1016/S0370-2693(98)01177-0}{\emph{Phys. Lett. B} {\bfseries 441} (1998) 105} [\href{https://arxiv.org/abs/hep-th/9805209}{{\ttfamily hep-th/9805209}}].

\bibitem{Sato:1998yu}
T.~Sato, \emph{{Various supersymmetric brane configurations from superalgebras in many types of M-brane backgrounds}}, \href{https://doi.org/10.1016/S0550-3213(99)00123-6}{\emph{Nucl. Phys. B} {\bfseries 548} (1999) 231} [\href{https://arxiv.org/abs/hep-th/9812014}{{\ttfamily hep-th/9812014}}].

\bibitem{Callister:2007jy}
A.K.~Callister and D.J.~Smith, \emph{{Topological BPS charges in 10-dimensional and 11-dimensional supergravity}}, \href{https://doi.org/10.1103/PhysRevD.78.065042}{\emph{Phys. Rev. D} {\bfseries 78} (2008) 065042} [\href{https://arxiv.org/abs/0712.3235}{{\ttfamily 0712.3235}}].

\bibitem{Harvey:1982xk}
R.~Harvey and H.B.~Lawson, Jr., \emph{{Calibrated geometries}}, \href{https://doi.org/10.1007/BF02392726}{\emph{Acta Math.} {\bfseries 148} (1982) 47}.

\bibitem{Gutowski_1999}
J.~Gutowski, G.~Papadopoulos and P.K.~Townsend, \emph{Supersymmetry and generalized calibrations}, \href{https://doi.org/10.1103/physrevd.60.106006}{\emph{Physical Review D} {\bfseries 60} (1999) }.

\bibitem{Gutowski_1999b}
J.~Gutowski, G.~Papadopoulos and P.K.~Townsend, \emph{Supersymmetry and generalized calibrations}, \href{https://doi.org/10.1103/physrevd.60.106006}{\emph{Physical Review D} {\bfseries 60} (1999) }.

\bibitem{gauntlett2003branescalibrationssupergravity}
J.P.~Gauntlett, \emph{Branes, calibrations and supergravity},  2003.

\bibitem{Sim_n_2012}
J.~Simón, \emph{Brane effective actions, kappa-symmetry and applications}, \href{https://doi.org/10.12942/lrr-2012-3}{\emph{Living Reviews in Relativity} {\bfseries 15} (2012) }.

\bibitem{townsend2000branetheorysolitons}
P.K.~Townsend, \emph{Brane theory solitons},  2000.

\bibitem{Townsend_2000}
P.K.~Townsend, \emph{Phremology: calibrating m-branes}, \href{https://doi.org/10.1088/0264-9381/17/5/336}{\emph{Classical and Quantum Gravity} {\bfseries 17} (2000) 1267–1276}.

\bibitem{Li:1998ce}
M.~Li, \emph{{'t Hooft vortices on D-branes}}, \href{https://doi.org/10.1088/1126-6708/1998/07/003}{\emph{JHEP} {\bfseries 07} (1998) 003} [\href{https://arxiv.org/abs/hep-th/9803252}{{\ttfamily hep-th/9803252}}].

\bibitem{Freedman_Van_Proeyen_2012}
D.Z.~Freedman and A.~Van~Proeyen, \emph{Supergravity}, Cambridge University Press (2012).

\bibitem{Bergshoeff:1997kr}
E.~Bergshoeff, R.~Kallosh, T.~Ortin and G.~Papadopoulos, \emph{{Kappa symmetry, supersymmetry and intersecting branes}}, \href{https://doi.org/10.1016/S0550-3213(97)00470-7}{\emph{Nucl. Phys. B} {\bfseries 502} (1997) 149} [\href{https://arxiv.org/abs/hep-th/9705040}{{\ttfamily hep-th/9705040}}].

\bibitem{Bergshoeff:1996wk}
E.~Bergshoeff, \emph{{p-branes, D-branes and M-branes}},  in \emph{{Strings 96: Current Trends in String Theory}}, pp.~210--217, 7, 1996, \href{https://doi.org/10.1142/9781848160927_0010}{DOI} [\href{https://arxiv.org/abs/hep-th/9611099}{{\ttfamily hep-th/9611099}}].

\bibitem{Kehagias:1998gn}
A.~Kehagias, \emph{{New type IIB vacua and their F theory interpretation}}, \href{https://doi.org/10.1016/S0370-2693(98)00809-0}{\emph{Phys. Lett. B} {\bfseries 435} (1998) 337} [\href{https://arxiv.org/abs/hep-th/9805131}{{\ttfamily hep-th/9805131}}].

\bibitem{Grana:2000jj}
M.~Grana and J.~Polchinski, \emph{{Supersymmetric three form flux perturbations on AdS(5)}}, \href{https://doi.org/10.1103/PhysRevD.63.026001}{\emph{Phys. Rev. D} {\bfseries 63} (2001) 026001} [\href{https://arxiv.org/abs/hep-th/0009211}{{\ttfamily hep-th/0009211}}].

\bibitem{weinberg_1981}
E.J.~Weinberg, \emph{Index calculations for the fermion-vortex system}, \href{https://doi.org/10.1103/PhysRevD.24.2669}{\emph{Phys. Rev. D} {\bfseries 24} (1981) 2669}.

\bibitem{Harvey:2007ab}
J.A.~Harvey and A.B.~Royston, \emph{{Localized modes at a D-brane-O-plane intersection and heterotic Alice atrings}}, \href{https://doi.org/10.1088/1126-6708/2008/04/018}{\emph{JHEP} {\bfseries 04} (2008) 018} [\href{https://arxiv.org/abs/0709.1482}{{\ttfamily 0709.1482}}].

\bibitem{Buchbinder:2007ar}
E.I.~Buchbinder, J.~Gomis and F.~Passerini, \emph{{Holographic gauge theories in background fields and surface operators}}, \href{https://doi.org/10.1088/1126-6708/2007/12/101}{\emph{JHEP} {\bfseries 12} (2007) 101} [\href{https://arxiv.org/abs/0710.5170}{{\ttfamily 0710.5170}}].

\bibitem{Harvey:2008zz}
J.A.~Harvey and A.B.~Royston, \emph{{Gauge/Gravity duality with a chiral N=(0,8) string defect}}, \href{https://doi.org/10.1088/1126-6708/2008/08/006}{\emph{JHEP} {\bfseries 08} (2008) 006} [\href{https://arxiv.org/abs/0804.2854}{{\ttfamily 0804.2854}}].

\bibitem{DeWolfe:2001pq}
O.~DeWolfe, D.Z.~Freedman and H.~Ooguri, \emph{{Holography and defect conformal field theories}}, \href{https://doi.org/10.1103/PhysRevD.66.025009}{\emph{Phys. Rev. D} {\bfseries 66} (2002) 025009} [\href{https://arxiv.org/abs/hep-th/0111135}{{\ttfamily hep-th/0111135}}].

\bibitem{Erdmenger_2002}
J.~Erdmenger, Z.~Guralnik and I.~Kirsch, \emph{Four-dimensional superconformal theories with interacting boundaries or defects}, \href{https://doi.org/10.1103/physrevd.66.025020}{\emph{Physical Review D} {\bfseries 66} (2002) }.

\bibitem{Yamaguchi:2006tq}
S.~Yamaguchi, \emph{{Wilson loops of anti-symmetric representation and D5-branes}}, \href{https://doi.org/10.1088/1126-6708/2006/05/037}{\emph{JHEP} {\bfseries 05} (2006) 037} [\href{https://arxiv.org/abs/hep-th/0603208}{{\ttfamily hep-th/0603208}}].

\bibitem{Gomis:2006sb}
J.~Gomis and F.~Passerini, \emph{{Holographic Wilson Loops}}, \href{https://doi.org/10.1088/1126-6708/2006/08/074}{\emph{JHEP} {\bfseries 08} (2006) 074} [\href{https://arxiv.org/abs/hep-th/0604007}{{\ttfamily hep-th/0604007}}].

\bibitem{Camino:2001at}
J.M.~Camino, A.~Paredes and A.V.~Ramallo, \emph{{Stable wrapped branes}}, \href{https://doi.org/10.1088/1126-6708/2001/05/011}{\emph{JHEP} {\bfseries 05} (2001) 011} [\href{https://arxiv.org/abs/hep-th/0104082}{{\ttfamily hep-th/0104082}}].

\bibitem{ahlfors}
L.~Ahlfors, \emph{{Complex analysis}}, McGraw-Hill, Inc., New York (1979).

\bibitem{Constable:2002xt}
N.R.~Constable, J.~Erdmenger, Z.~Guralnik and I.~Kirsch, \emph{{Intersecting D-3 branes and holography}}, \href{https://doi.org/10.1103/PhysRevD.68.106007}{\emph{Phys. Rev. D} {\bfseries 68} (2003) 106007} [\href{https://arxiv.org/abs/hep-th/0211222}{{\ttfamily hep-th/0211222}}].

\bibitem{Drukker:2008wr}
N.~Drukker, J.~Gomis and S.~Matsuura, \emph{{Probing N=4 SYM With Surface Operators}}, \href{https://doi.org/10.1088/1126-6708/2008/10/048}{\emph{JHEP} {\bfseries 10} (2008) 048} [\href{https://arxiv.org/abs/0805.4199}{{\ttfamily 0805.4199}}].

\bibitem{Koh:2008kt}
E.~Koh and S.~Yamaguchi, \emph{{Holography of BPS surface operators}}, \href{https://doi.org/10.1088/1126-6708/2009/02/012}{\emph{JHEP} {\bfseries 02} (2009) 012} [\href{https://arxiv.org/abs/0812.1420}{{\ttfamily 0812.1420}}].

\bibitem{DHoker:2002nbb}
E.~D'Hoker and D.Z.~Freedman, \emph{{Supersymmetric gauge theories and the AdS / CFT correspondence}},  in \emph{{Theoretical Advanced Study Institute in Elementary Particle Physics (TASI 2001): Strings, Branes and EXTRA Dimensions}}, pp.~3--158, 1, 2002 [\href{https://arxiv.org/abs/hep-th/0201253}{{\ttfamily hep-th/0201253}}].

\bibitem{Gomis:2007fi}
J.~Gomis and S.~Matsuura, \emph{{Bubbling surface operators and S-duality}}, \href{https://doi.org/10.1088/1126-6708/2007/06/025}{\emph{JHEP} {\bfseries 06} (2007) 025} [\href{https://arxiv.org/abs/0704.1657}{{\ttfamily 0704.1657}}].

\bibitem{Klebanov:1999tb}
I.R.~Klebanov and E.~Witten, \emph{{AdS / CFT correspondence and symmetry breaking}}, \href{https://doi.org/10.1016/S0550-3213(99)00387-9}{\emph{Nucl. Phys.} {\bfseries B556} (1999) 89} [\href{https://arxiv.org/abs/hep-th/9905104}{{\ttfamily hep-th/9905104}}].

\bibitem{Holguin:2025bfe}
A.~Holguin and H.~Kawai, \emph{{Integrability and conformal blocks for surface defects in $\mathcal{N}=4$ SYM}}, \href{https://doi.org/10.1007/JHEP11(2025)043}{\emph{JHEP} {\bfseries 11} (2025) 043} [\href{https://arxiv.org/abs/2503.09944}{{\ttfamily 2503.09944}}].

\bibitem{Chalabi:2025nbg}
A.~Chalabi, C.~Kristjansen and C.~Su, \emph{{Integrable corners in the space of Gukov-Witten surface defects}}, \href{https://doi.org/10.1016/j.physletb.2025.139512}{\emph{Phys. Lett. B} {\bfseries 866} (2025) 139512} [\href{https://arxiv.org/abs/2503.22598}{{\ttfamily 2503.22598}}].

\bibitem{Ivanovskiy:2024vel}
V.~Ivanovskiy, S.~Komatsu, V.~Mishnyakov, N.~Terziev, N.~Zaigraev and K.~Zarembo, \emph{{Vacuum Condensates on the Coulomb Branch}},  \href{https://arxiv.org/abs/2405.19043}{{\ttfamily 2405.19043}}.

\bibitem{Dekel:2011ja}
A.~Dekel and Y.~Oz, \emph{{Integrability of Green-Schwarz Sigma Models with Boundaries}}, \href{https://doi.org/10.1007/JHEP08(2011)004}{\emph{JHEP} {\bfseries 08} (2011) 004} [\href{https://arxiv.org/abs/1106.3446}{{\ttfamily 1106.3446}}].

\bibitem{Demjaha:2025axy}
R.~Demjaha and K.~Zarembo, \emph{{String integrability on the Coulomb branch}}, \href{https://doi.org/10.1007/JHEP09(2025)154}{\emph{JHEP} {\bfseries 09} (2025) 154} [\href{https://arxiv.org/abs/2506.17955}{{\ttfamily 2506.17955}}].

\bibitem{deHaro:2000vlm}
S.~de~Haro, S.N.~Solodukhin and K.~Skenderis, \emph{{Holographic reconstruction of space-time and renormalization in the AdS / CFT correspondence}}, \href{https://doi.org/10.1007/s002200100381}{\emph{Commun. Math. Phys.} {\bfseries 217} (2001) 595} [\href{https://arxiv.org/abs/hep-th/0002230}{{\ttfamily hep-th/0002230}}].

\bibitem{Karch:2005ms}
A.~Karch, A.~O'Bannon and K.~Skenderis, \emph{{Holographic renormalization of probe D-branes in AdS/CFT}}, \href{https://doi.org/10.1088/1126-6708/2006/04/015}{\emph{JHEP} {\bfseries 04} (2006) 015} [\href{https://arxiv.org/abs/hep-th/0512125}{{\ttfamily hep-th/0512125}}].

\bibitem{DHoker:2007zhm}
E.~D'Hoker, J.~Estes and M.~Gutperle, \emph{{Exact half-BPS Type IIB interface solutions. I. Local solution and supersymmetric Janus}}, \href{https://doi.org/10.1088/1126-6708/2007/06/021}{\emph{JHEP} {\bfseries 06} (2007) 021} [\href{https://arxiv.org/abs/0705.0022}{{\ttfamily 0705.0022}}].

\bibitem{DHoker:2007hhe}
E.~D'Hoker, J.~Estes and M.~Gutperle, \emph{{Exact half-BPS Type IIB interface solutions. II. Flux solutions and multi-Janus}}, \href{https://doi.org/10.1088/1126-6708/2007/06/022}{\emph{JHEP} {\bfseries 06} (2007) 022} [\href{https://arxiv.org/abs/0705.0024}{{\ttfamily 0705.0024}}].

\bibitem{DHoker:2007mci}
E.~D'Hoker, J.~Estes and M.~Gutperle, \emph{{Gravity duals of half-BPS Wilson loops}}, \href{https://doi.org/10.1088/1126-6708/2007/06/063}{\emph{JHEP} {\bfseries 06} (2007) 063} [\href{https://arxiv.org/abs/0705.1004}{{\ttfamily 0705.1004}}].

\bibitem{Drukker:2008jm}
N.~Drukker, J.~Gomis and D.~Young, \emph{{Vortex Loop Operators, M2-branes and Holography}}, \href{https://doi.org/10.1088/1126-6708/2009/03/004}{\emph{JHEP} {\bfseries 03} (2009) 004} [\href{https://arxiv.org/abs/0810.4344}{{\ttfamily 0810.4344}}].

\bibitem{Gutperle:2020gez}
M.~Gutperle and C.F.~Uhlemann, \emph{{Janus on the Brane}}, \href{https://doi.org/10.1007/JHEP07(2020)243}{\emph{JHEP} {\bfseries 07} (2020) 243} [\href{https://arxiv.org/abs/2003.12080}{{\ttfamily 2003.12080}}].

\bibitem{Bandos:1997gd}
I.A.~Bandos, N.~Berkovits and D.P.~Sorokin, \emph{{Duality symmetric eleven-dimensional supergravity and its coupling to M-branes}}, \href{https://doi.org/10.1016/S0550-3213(98)00102-3}{\emph{Nucl. Phys. B} {\bfseries 522} (1998) 214} [\href{https://arxiv.org/abs/hep-th/9711055}{{\ttfamily hep-th/9711055}}].

\bibitem{Aganagic:1997zq}
M.~Aganagic, J.~Park, C.~Popescu and J.H.~Schwarz, \emph{{World volume action of the M theory five-brane}}, \href{https://doi.org/10.1016/S0550-3213(97)00227-7}{\emph{Nucl. Phys. B} {\bfseries 496} (1997) 191} [\href{https://arxiv.org/abs/hep-th/9701166}{{\ttfamily hep-th/9701166}}].

\bibitem{Pasti:1997gx}
P.~Pasti, D.P.~Sorokin and M.~Tonin, \emph{{Covariant action for a D = 11 five-brane with the chiral field}}, \href{https://doi.org/10.1016/S0370-2693(97)00188-3}{\emph{Phys. Lett. B} {\bfseries 398} (1997) 41} [\href{https://arxiv.org/abs/hep-th/9701037}{{\ttfamily hep-th/9701037}}].

\bibitem{Bandos:1997ui}
I.A.~Bandos, K.~Lechner, A.~Nurmagambetov, P.~Pasti, D.P.~Sorokin and M.~Tonin, \emph{{Covariant action for the superfive-brane of M theory}}, \href{https://doi.org/10.1103/PhysRevLett.78.4332}{\emph{Phys. Rev. Lett.} {\bfseries 78} (1997) 4332} [\href{https://arxiv.org/abs/hep-th/9701149}{{\ttfamily hep-th/9701149}}].

\bibitem{Ko:2013dka}
S.-L.~Ko, D.~Sorokin and P.~Vanichchapongjaroen, \emph{{The M5-brane action revisited}}, \href{https://doi.org/10.1007/JHEP11(2013)072}{\emph{JHEP} {\bfseries 11} (2013) 072} [\href{https://arxiv.org/abs/1308.2231}{{\ttfamily 1308.2231}}].

\bibitem{Bandos:1997gm}
I.A.~Bandos, K.~Lechner, A.~Nurmagambetov, P.~Pasti, D.P.~Sorokin and M.~Tonin, \emph{{On the equivalence of different formulations of the M theory five-brane}}, \href{https://doi.org/10.1016/S0370-2693(97)00784-3}{\emph{Phys. Lett. B} {\bfseries 408} (1997) 135} [\href{https://arxiv.org/abs/hep-th/9703127}{{\ttfamily hep-th/9703127}}].

\bibitem{Papadopoulos:1996uq}
G.~Papadopoulos and P.K.~Townsend, \emph{{Intersecting M-branes}}, \href{https://doi.org/10.1201/9781482268737-27}{\emph{Phys. Lett. B} {\bfseries 380} (1996) 273} [\href{https://arxiv.org/abs/hep-th/9603087}{{\ttfamily hep-th/9603087}}].

\bibitem{Strominger:1995ac}
A.~Strominger and M.~Dine, \emph{{Open p-branes}}, \href{https://doi.org/10.1201/9781482268737-13}{\emph{Phys. Lett. B} {\bfseries 383} (1996) 44} [\href{https://arxiv.org/abs/hep-th/9512059}{{\ttfamily hep-th/9512059}}].

\bibitem{Tseytlin:1996bh}
A.A.~Tseytlin, \emph{{Harmonic superpositions of M-branes}}, \href{https://doi.org/10.1201/9781482268737-28}{\emph{Nucl. Phys. B} {\bfseries 475} (1996) 149} [\href{https://arxiv.org/abs/hep-th/9604035}{{\ttfamily hep-th/9604035}}].

\end{thebibliography}\endgroup

\end{document}